\documentclass[%
 reprint,
 amsmath,amssymb,
 aps,showpacs,superscriptaddress
]{revtex4-1}

\usepackage{graphicx}
\usepackage{dcolumn}
\usepackage{bm}
\usepackage{color}
\usepackage{dsfont}

\usepackage{hyperref}

\newcommand{\beqar}{\begin{eqnarray}}
\newcommand{\eeqar}{\end{eqnarray}}
\newcommand{\bea}{\begin{eqnarray}}
\newcommand{\eea}{\end{eqnarray}}
\newcommand{\bcen}{\begin{center}}
\newcommand{\ecen}{\end{center}}
\newcommand{\ra}{\rightarrow}
\newcommand{\bra}[1]{\left< #1 \right|}
\newcommand{\ket}[1]{\left| #1 \right>}

\newcommand{\tr}{\mathrm{tr}}

\newcommand{\eps}{\varepsilon}
\newcommand{\lam}{\lambda}

\newcommand{\f}[2]{\frac{#1}{#2}}
\renewcommand{\b}[1]{\left({#1}\right)}
\renewcommand{\v}[1]{\vec{#1}}
\newcommand{\pd}[2]{\frac {\partial #1}{\partial #2}}

\renewcommand{\sb}[1]{\left[{#1}\right]}
\newcommand{\mean}[1]{\langle {#1} \rangle}

\newcommand{\trr}{\textcolor{black}}

\usepackage[normalem]{ulem}
\newcommand{\stkout}[1]{\ifmmode\text{\sout{\ensuremath{#1}}}\else\sout{#1}\fi}
\newcommand{\cyan}[1]{\textcolor{cyan}{#1}}

\newcommand{\nm}[1]{\cyan{#1}}

\begin{document}

\preprint{APS/123-QED}
\title{Non-Markovian dynamics under time-translation symmetry}


\author{Roie Dann}
\email{roie.dann@mail.huji.ac.il}
\affiliation{The Institute of Chemistry, The Hebrew University of Jerusalem, Jerusalem 9190401, Israel}
\author{Nina Megier}
\email{nina.megier@mi.infn.it}
\affiliation{Dipartimento di Fisica “Aldo Pontremoli”, Università degli Studi di Milano, via Celoria 16, 20133 Milan, Italy}
\affiliation{Istituto Nazionale di Fisica Nucleare, Sezione di Milano, via Celoria 16, 20133 Milan, Italy}
\author{Ronnie Kosloff}%
\email{kosloff1948@gmail.com}
\affiliation{The Institute of Chemistry, The Hebrew University of Jerusalem, Jerusalem 9190401, Israel}%

\date{\today}
\begin{abstract}
A dynamical symmetry is employed to determine the structure of the quantum non-Markovian time-local master equation. Such a structure is composed from two components: scalar kinetic coefficients and the standard quantum Markovian operator form. The kinetic coefficients are generally time-dependent and incorporate information on the kinematics and memory effects, while the operators manifest the dynamical symmetry. 
Specifically, we focus on time-translation symmetric dynamics, where the Lindblad jump operators constitute the eigenoperators of the free dynamics. This 
symmetry is motivated by thermodynamic microscopic considerations, where strict energy conservation between system and environment imposes the time-translation symmetry. The construction is generalized to other symmetries, and to driven quantum systems. 
The formalism is illustrated by three exactly solvable non-Markovian models,
where
the exact reduced description
exhibits a
dynamical symmetric structure. 
The formal structure of the master equation leads to a first principle calculation of the exact kinetic coefficients. This opens the possibility to simulate in a modular fashion non-Markovian dynamics.
\end{abstract}

\maketitle

 

\section{Introduction}
\label{sec:intro}

Quantum systems exhibit a wide range of characteristic dynamical behaviour. In small isolated systems the fundamental time-reversal symmetry manifests itself by quasi-periodic evolution. However, with increasing system size and reduction in symmetry, the periodicity becomes harder to witness and the characteristic local behavior becomes increasingly irreversible. Underlying this typical transition are emerging correlations between the system under study and its surrounding. Essentially, any interaction, even asymptotically small, leads to leakage of information to the environment and formation of joint correlations. In turn, the fragile nature of quantum information, high dimensionality of the environment and the limited access the observer has to the environmental degrees of freedom leads to local irreversibility \cite{zurek1991quantum,zurek2003decoherence}.

The materializing system-environment correlations and their influence on the system dynamics are related to the concept of ``memory". In the framework of open quantum systems, memory quantifies the extent information on the system's past state influences the future system dynamics. Under memoryless dynamics  only a one-directional flow of information occurs from the reduced state to the environment. However, in practice the information flow rate is 
limited
by the Lieb-Robinson bound \cite{lieb1972finite}, which defines an associated timescale of decay of system and environment correlations. For shorter timescales the back-flow from the environment is inevitable. Consequently, correlations forming in the past influence the system's future evolution. 
Between the two extremes there is a wide range of possible dynamical phenomena. 

In practice, neither of the extremes is completely accurate \cite{davies1976quantum,huppert1990long,breuer2002theory,fain2000irreversibilities,rivas2012open}. Any quantum system has some residual interaction with its surrounding, which inevitably includes a large number of degrees of freedom. As a result, information on the system's past state gets decoded in highly global correlations (including many degrees of freedom), which only slightly affects the system present evolution. Conversely, memoryless dynamics relies on  the negligible role of the system and environment correlations on the reduced dynamics. The description therefore includes an implicit effective coarse-graining in time, leading to deviations in short times spans \cite{chapter1998radiation, Rivas_2010}.

      

The aspiration for an accurate description of quantum dynamics is motivated by the recent advancements in quantum technology, which rely on the reduction of the environmental impact on the quantum system \cite{unruh1995maintaining,nielsen2002quantum,ladd2010quantum}. To reduce the detrimental environmental influence, one first needs to precisely model its effect on the quantum dynamics, which include memory effects. For example,  the development of error mitigation schemes
relies on an accurate dynamical description \cite{shor1995scheme,ekert1996quantum,gottesman1997stabilizer,cory1998experimental,knill2000theory,aharonov2008fault,lidar2013quantum}.  It has also been shown that non-Markovianity can be utilized to assist tasks for quantum information processing \cite{vasile2011continuous,huelga2012non,bylicka2013non,laine2014nonlocal,xiang2014entanglement,reich2015exploiting,dong2018non,anand2019quantifying} and quantum metrology \cite{chin2012quantum,matsuzaki2011magnetic}. In addition, the inevitable leakage of information and decoherence of the quantum state motivates rapid operations on the quantum system \cite{zurek2003decoherence}. Consequently, an accurate description in the short time regime is required, where information back-flow cannot be ignored.




In the present study, we analyze the non-Markovian dynamics of open quantum systems and construct a master equation which includes memory effects. 
This issue is tackled by adopting a first principle axiomatic approach. We first introduce two thermodynamically motivated postulates which manifest a time-translation dynamical symmetry. The symmetry of the map enables conducting a spectral analysis and leads to the general form of the master equation that complies with the initial postulates. The obtained master equation is of the GKLS form with time dependent and possibly negative kinetic coefficients. This form is coined the {\emph{dynamical symmetric structure}}. Interestingly, the kinetic coefficients include all the information regarding the dissipation rates, details of environmental properties and coupling strength. They can be determined by employing a perturbative treatment, while taking advantage of the master equation's defined operatorial structure. In principle, this approach is valid for strong system-environment coupling as well as highly non-Markovian environments.

The dynamical symmetric structure complies with the form of the Davies master equation \cite{davies1974markovian}. Nevertheless,  the obtained master equation can be non-Markovian. We compare the two master equations and the initial assumptions involved in their construction. The proposed approach is very general, allowing a straightforward generalization of the construction to other dynamical symmetries. We demonstrate this by analyzing dynamics which conserve the total number of excitations in the system and environment. Under this symmetry, the dynamics does exhibit a Markovian limit in the long time regime. Moreover, by building upon the case of a stationary Hamiltonian, we extend the description to time-dependent Hamiltonians. 



We begin by setting the framework and discuss the basic postulates of the theory in Sec. \ref{sec:framework}. Following, we shortly review the prime known results on dynamical generators. Building upon the postulates we then prove the general form of the generator of the dynamical map (Sec. \ref{sec:gen_form}). 
In sections  \ref{sec:kinetic coefficients} and \ref{sec:symmetry_constraints} we describe a perturbative treatment to calculate the accurate kinetic coefficients, and discuss symmetry imposed restrictions on the coefficients. Following, we discuss the subtle relation between strict energy conservation and non-Markovianity and compare the present approach  to the Davies construction \ref{sec:davies_compare} and \ref{sec:sec_and_markovianity}. The framework is then extended to other symmetries, analyze dynamics under conservation of the total number of excitations Sec. \ref{sec:generalization to other symmetries}, and time-dependent Hamiltonians, Sec. \ref{sec:td_Hamiltonians}. 
Finally, in Sec. \ref{sec:example} we demonstrate the theory by analyzing the exact dynamical solutions of the Jaynes-Cummings, spin-star models under time-translation symmetry and a spin-boson bath model for dynamics conserving the number of excitations, and conclude.



\section{Quantum Markovianity and beyond}
\label{sec:background}

In the classical theory of stochastic processes, lack of memory is formalized in terms of Markovianity. This property infers that the present system evolution, which is described in terms of a time-dependent probability distribution $p\b{x,t}$, is independent of the state history \cite{norris1998markov,ross1996stochastic}. Generalizing this natural concept to the quantum regime is not straight-forward, as classical Markovianity is a property of probabilities conditioned on the history of the process: 
$p\b{x_n,t_n|x_{n-1},t_{n-1},\dots,x_0,} = p\b{x_n,t_n|x_{n-1},t_{n-1}}$ (meaning that the random variable $X$ takes the value $x_i$ at time $t_i$ for $t_n\geq t_{n-1}\geq\dots \geq t_0$). In the quantum setting, the conditional probabilities depend not only on the dynamics but also on the chosen measurement procedure. This a definition which may differ from one experiment to another. To circumvent this complication, we adopt an approach that
identifies quantum Markovianity 
as a property of the evolution of the reduced density operator $\hat{\rho}_S\b t$, with no reference to any measurement scheme.

We associate quantum
Markovianity 
with the property of complete positive (CP)-divisibility. The identification is motivated by the fact that on the level of the one-point probabilities $p\b{x,t}$, (classical) divisible and Markovian processes are equivalent. That is, when the experimenter only has access to one-point probabilities she cannot distinguish between classical Markovian and divisible processes \cite{rivas2014quantum}. Note that alternative definitions of quantum Markovianity exist \cite{li2018concepts}, among the best known are the approaches based on P-divisibility \cite{PhysRevLett.112.120404,HelstromPdiv}, the monotonicity of the distinguishability quantifiers between two distinct reduced states \cite{BLP,HelstromPdiv,megier2021entropic} and the change of the volume of accessible reduced states \cite{PhysRevA.88.020102}. All of these characterisations are based solely on the properties of the dynamics of the open quantum system. An alternative approach termed process tensor formalism requires a specific environmental realization \cite{PhysRevLett.123.040401,PhysRevA.97.012127,PhysRevA.100.012120}.

A common restrictive case of CP-divisibility includes dynamical maps which form a dynamical semi-group. In the present analysis we categorize such maps as ``strictly Markovian".
Strict Markovianity is abundant in the analysis of open quantum systems, and is frequently employed in quantum optics, solid state physics, quantum information and quantum thermodynamics  \cite{carmichael2009open,scully1999quantum,childs2000quantum,kosloff2013quantum,vinjanampathy2016quantum}. This assumption is supported by a vast number of experiments, which exhibit a typical exponential decay towards equilibrium  \footnote{The semi-group property implies that the map can be expressed in terms of a time-independent generator ${\cal L}$: $\Lambda_{t,t_0}=e^{{\cal L}t}$, leading to an exponential decay.}. Moreover, the Markovian assumption highly simplifies the theoretical description, allowing to derive time-local equations of motion of a specified form. 

In their seminal papers,  Gorini, Kossakowski and Sudarshan and separately Lindblad (GKLS) \cite{gorini1976completely,lindblad1976generators} supplied the general structure of the generator of a
strictly Markovian dynamical map. 
This result was later generalized to  CP-divisible dynamics, which can be described with the generalized GKLS master equation, where the kinetic coefficients are still positive, but time dependent (at least for finite-dimensional systems) \cite{chruscinski2012markovianity,rivas2012open,PhysRevA.89.042120}.
 Despite the simplified theoretical structure, we stress that Markovian evolution can only be an approximate description, which implicitly includes a coarse-graining of the dynamics over some timescale $\Delta t_{\text{c.g}}$. Rapid changes occurring within the timescale of $\Delta t_{\text{c.g}}$ are averaged over and one obtains a smoothed theoretical description. 
Beyond these regimes the general structure of the generator is unknown, apart from some restrictive cases \cite{PhysRevA.59.3290,PhysRevA.94.022118}.

\section{Dynamical map, symmetry and the master equation}
\label{sec:framework}
The theory of open quantum systems considers a composite system consisting of a primary system coupled to an external environment. The total system is isolated, leading to joint unitary dynamics which are generated by the joint Hamiltonian
\begin{equation}
    \hat{H}=\hat{H}_S+\hat{H}_{SE}+\hat{H}_E~~,
    \label{eq:hamil}
\end{equation}
where $\hat{H}_S$ and $\hat{H}_E$ are the bare system and environment Hamiltonians and $\hat{H}_{SE}$ is the interaction term. Under a broader treatment, the description may include an external controller, which  leads to an explicit time-dependence (in a semi-classical description \cite{glauber1963coherent,allen1987optical,scully1999quantum,vsindelka2010derivation,dann2020thermodynamically}), nonetheless, we will first address the simplest case including a static Hamiltonian $\hat{H}_S$ of dimension $N$. 

Formally, the evolution of the composite state $\hat{\rho}\b t$  is determined by the Liouville von-Neumann equation. However, in practice this relation is not much of use, as a general accurate solution is intractable due to the vast number of environmental degrees of freedom. Luckily, a general solution is not essential for most purposes, since only the reduced system dynamics are typically of interest. Under the standard assumption of a separable initial state (no correlations initially) the reduced system dynamics are given by a linear completely positive trace preserving (CPTP) dynamical map \cite{kraus1971general}
\begin{multline}
    \hat{\rho}_S\b t \equiv  \Lambda{\b{t,t_0}}\sb{\hat{\rho}_S\b {{t_0}}}\\= \text{tr}_E\b{\hat{U}\b{t,t_0}\hat{\rho}_S\b {{t_0}}\otimes\hat{\rho}_E\b {{t_0}}\hat{U}^\dagger\b{t,t_0}}~~,
\label{eq:dynamical_map}
\end{multline}
where $\hat{\rho}_E\b{t_0}$ is the environment initial state, $\text{tr}_i$ denotes the partial trace, with $i=S,E$ and $\hat{U}\b{t,t_0}=e^{-i\hat{H}\b{t-t_0}/\hbar}$ is the total system propagator.

In the present study, the dynamical map is (quantum) Markovian if and only if it satisfies CP-divisibility. This property implies that the map can be expressed as
\begin{equation}
    \Lambda{\b{t,t_0}} = V{\b{t,s}}\Lambda{\b{s,t_0}},
\label{eq:divisibility}
\end{equation}
 where $V{\b{t,s}}$ is also a CPTP map which satisfies the composition law
\begin{equation}
    V\b{t,s}V\b{s,u}=V\b{t,u}~~.
\end{equation}
In case of a semi-group (strictly Markovian dynamics)
the composition law is 
replaced by \cite{gorini1976completely,lindblad1976generators,davies1976quantum}
\begin{align}
\label{eq:semigroup}
    \Lambda{\b{t,t_0}} = \Lambda{\b{t-s+t_0,t_0}}\Lambda{\b{s,t_0}}~~.
\end{align}
Note that 
the second property, relation \eqref{eq:semigroup}, implies the former, Eq. \eqref{eq:divisibility}, but not conversely.

 

The composite propagator $\hat{U}\b{t,t_0}$ in Eq.  \eqref{eq:dynamical_map} has only a formal role, since an exact solution is unfeasible even for simple environments. As a consequence, Eq. \eqref{eq:dynamical_map} serves as a starting point for approximate solutions of the reduced dynamics.
The customary derivation begins with a complete description $\hat{H}$, defined in Eq. \eqref{eq:hamil}, and employs the Born-Markov-secular approximation \cite{davies1974markovian,breuer2002theory,rivas2012open,dann2018time}. This construction leads to a GKLS master equation satisfying Eq. \eqref{eq:semigroup}, corresponding to strictly Markovian dynamics.
Many variants of the construction exist, nonetheless, they are characterized by three main features: weak system-environment coupling, rapid decay of environmental correlations and a coarse-graining in time. These assumptions effectively discard the memory effects 
and therefore lead to a Markovian dynamical map \cite{Rivas_2010}.
If some of these assumptions are not satisfied, one can try to describe the dynamics with a generalized GKLS equation, which form guarantees the CPTP property of the associated dynamical map and satisfaction of Eq. \eqref{eq:divisibility}.



Our present analysis adopts an alternative methodology, we introduce two additional thermodynamically motivated postulates to the standard CPTP framework \footnote{The CPTP framework assumes an initial separable state at initial time and composite unitary dynamics.}. We prove the general structure of the master equation within this axiomatic framework. 
We consider the following two postulates:
\begin{enumerate}
    \item Strict energy conservation - The system environment interaction term commutes with the sum of bare Hamiltonians  $\sb{\hat{H}_S+\hat{H}_E,\hat{H}_{SE}}=0$.
    \item The initial environment state is stationary with respect to the bare environment Hamiltonian $\hat{H}_E$.
\end{enumerate}

Strict energy conservation implies that the interface between the system and environment does not accumulate any energy. This assumption is motivated by the classical thermodynamic idealization which neglects the properties of the interface between subsystems, and analyses only energy currents and changes within the subsystems. The second postulate includes the common case, where the environment is a thermal reservoir, and allows for generalizations of multiple reservoirs which are diagonal in the energy basis of $\hat{H}_E$. Nonetheless, it excludes any coherent bath such as squeezed states. This restriction essentially serves as a strict distinction between what we consider as a control system or environment.  According to our characterization, quantum control systems include non-stationary dynamics (with respect to their bare Hamiltonian), while the environment must be stationary \cite{janzing2000thermodynamic,brandao2013resource,horodecki2013fundamental,dann2020thermodynamically,landi2021irreversible}.

These two postulates along with an assumption that the map satisfy the semi-group property (strictly Markovian dynamics) were introduced in Ref. \cite{dann2021open}. There they served as a mathematical basis to prove the general form of thermodynamically consistent strictly Markovian master equation. In the following, we relax the Markovian assumption and analyze the structure of the generator of a dynamical map $\Lambda{\b{t,t_0}}$, Eq. \eqref{eq:dynamical_map}, that only complies with postulates 1 and 2.
\\

\subsection{The thermodynamic postulates and time-translation symmetry}

The two thermodynamic postulates, 1. and 2. introduced above,
imply that dynamical map commutes with the isolated map \cite{marvian2012symmetry,marvian2014modes,dann2021open} (the proof is presented in Appendix \ref{apsec:proofA} for completeness)
\begin{equation}
    \Lambda\sb{{\cal{U}}_{S}\sb{\hat{\rho}_S}}={\cal{U}}_{S}\sb{\Lambda\sb{\hat{\rho}_S}}~~,
    \label{eq:commutivatity_maps}
\end{equation}
or equivalently $\Lambda\circ{\cal{U}}_{S}={\cal{U}}_{S}\circ\Lambda$,
where the time dependence was removed for brevity and ${\cal U}_{S}\b{t,t_0}\sb{\bullet}\equiv \hat{U}_S\b{t,t_0}\bullet\hat{U}_S^\dagger\b{t,t_0}$, with $\hat{U}_S\b{t,t_0}=e^{-i\hat{H}_S\b{t-t_0}/\hbar}$. 
 This relation is known in the literature under the name of {\emph{time-translation symmetry}} or {\emph{phase covariance}}. A quantum evolution obeying this symmetry emerges in many different contexts, as nicely summarized in \cite{PhysRevA.96.032109}. In particular, in the field of quantum optics such dynamics is related to the rotating wave approximation (RWA), i.e. neglection of the counter-rotating terms in the interaction Hamiltonian (cf. Jaynes-Cummings model on resonance, in Sec. \ref{sec:2}).

In the following analysis, the two thermodynamic postulates can be replaced by the time-translation condition on the map, Eq. \eqref{eq:commutivatity_maps}, leading to identical (and somewhat more general) conclusions. To be precise, the two thermodynamic postulates define a set of dynamical maps $\{\Lambda\}$. These constitute a subset of the set of maps which are symmetric under time-translation, hence, Eq. \eqref{eq:commutivatity_maps} serves as a weaker condition on the dynamics. In the present study we chose to introduce the framework in terms of the thermodynamic postulates in order to allow a clear physical picture.  

A different point of view on the relation between the thermodynamics postulates and time-translation symmetry is obtained by embedding  the symmetric dynamics within a larger Hilbert space. It has been shown that any time-translation symmetric maps can be cast as strict energy conserving dynamics of the system, environment and an additional auxiliary system \cite{keyl1999optimal,marvian2012symmetry,lostaglio2019introductory} (a result termed as Stinespring dilation for time-translation symmetric maps). Therefore, one can always view the dynamics satisfying time-translation as arising from a larger total system, which satisfies the thermodynamics postulates.

The relation between strict energy conservation and time-translation symmetry has been previously studied in the context of thermodynamic resource theories. This framework establishes a set of allowed ``free" operations and characterizes the possible state transformation, enabled by these operations. These theories focus on possible transformation and not the explicit dynamics, which is the subject of the current study. By comparing the possible state transitions one obtains insight on the operational implications of each property. 
The conditions of strict energy conservation and an initial thermal environment define the free operations of the resource theory of thermal operation \cite{janzing2000thermodynamic,horodecki2013fundamental,ng2018resource,lostaglio2019introductory}.
Similarly, an initial thermal environment and a dynamical map satisfying time-translation symmetry defines the free operations in the resource theory of thermal processes (or enhanced thermal operations) \cite{cwiklinski2015limitations,gour2018quantum}. 
Comparing the two resources theories, the characterized possible transition are identical for both theories, therefore it is not yet clear 
whether
thermal processes have any operative advantage over thermal operations.

\subsection{Generators of dynamical maps}

The generator of a general quantum dynamical map is a function of the joint Hamiltonian $\hat{H}$, which contains all the information regarding both the system and environment. The structure of such a generator is unknown, nevertheless, a number of formal approaches have been developed which allow analyzing specific cases. 

The standard Nakajima-Zwanzig projection operator technique shows that the dynamical properties of the reduced system can be expressed accurately in terms of the memory kernel  ${\cal K}$ \cite{nakajima1958quantum,zwanzig1960ensemble,breuer2002theory}
\begin{equation}\label{eq:kernel}
   \f{d}{dt} \Lambda{\b {t,t_0}} = \int_{t_0}^t {\cal K}\b{t-s}\Lambda{\b{s,t_0}}ds ~~\text{with}~~\Lambda{\b{t_0,t_0}}=\mathbb{I}~~.
\end{equation}
The time non-local structure of this equation is computationally demanding as the right hand side depends on the whole history of the process.
Alternatively, when the dynamical map is invertible \footnote{Actually, the time-local master equation exists when some weaker conditions are satisfied \cite{doi:10.1080/09500340701352581}.},  the dynamics can be equivalently described by a time-local equation \cite{chaturvedi1979time,shibata1977generalized}
\begin{equation}
\f{d}{dt}\Lambda{\b{t,t_0}}={\cal L}\b{\tau}\Lambda{\b{t,t_0}}~~,    
\label{eq:2}
\end{equation}
where ${\cal L}$ is known as the time-local generator or time-convolution-less generator (referred to as dynamical generator from here on) and $\tau =t-t_0$. The simple form of this time-local equation may be misleading as ${\cal L}\b{\tau}$ contains memory and is effectively non-local in time due to the dependence on $t_0$ \cite{chruscinski2010non}. Similarly to the generator, the dynamical map is also only a function of the time-difference, allowing to replace the notation, $\Lambda\b{t,t_0}\ra\Lambda\b{\tau}$. This property is a consequence of the fact that any non-Markovian dynamics can be embedded within Markovian dynamics of a larger Hilbert space, see Ref. \cite{chruscinski2010non} for further details. 

The non-local character of the generator is hidden within the notation, since the initial time is frequently taken to be zero. Nevertheless, for the sake of brevity we also set $t_0=0$ in the following analysis. Therefore $\tau$ is replaced by $t$, and the two times dependence in the propagators $\b{t,t_0}$ is replaced by a single time. Despite the notation it should be clear that ${\cal L}\b{t}$ is non-local in time and depends on the whole history of the state ($\tau=t-0$), not solely on time $t$.   

By utilizing Eq. \eqref{eq:2}  the generator can be expressed as 
\begin{equation}
    {\cal L}\b \tau = \b{\pd{\Lambda}{\tau}}\Lambda^{-1}\b{{\tau}}~~.
    \label{eq:lambda_diff}
\end{equation}
This relation directly implies the linearity of the generator whenever an inverse $\Lambda^{-1}\b{t}$ exists \footnote{The dynamical map is linear as we assume factorized initial condition, therefore its inverse is as well. Moreover, differentiation is also linear which leads to the linearity of ${\cal L}\b t$}. 

If the dynamical generator is time-independent  the dynamics acquire the form $\Lambda\b{t}=e^{{\cal L}t}$.  It is then straight forward to check that such a map satisfies the semi-group property Eq. \eqref{eq:semigroup}, and therefore governs strictly Markovian dynamics. In their pioneering work GKLS
 proved the general form of $\cal L$.
 This result was later generalized for time-dependent generators ${\cal L}\b{t}$ (generalized GKLS form) of CP-divisible dynamical maps in \cite{chruscinski2012markovianity,rivas2012open}
 \begin{multline}
     {\cal{L}}\b t\sb{\bullet} = -i\sb{\hat{H}'\b t,\bullet}+\\
     \sum_\alpha \gamma_{\alpha}\b t\b{\hat{V}_{\alpha}\b t\bullet\hat{V}_{\alpha}^\dagger\b t-\f{1}{2}\{\hat{V}_{\alpha}^\dagger\b t\hat{V}_{\alpha}\b t,\bullet\}}~~,
      \label{eq:master_equation}
 \end{multline}
where $\hat{H}'\b t$ is Hermitian, $\gamma_{\alpha}\b t\geq 0$ for every $\alpha$ and time $t$, and $\bullet$ denotes any operator in the $C^{*}$ algebra of the $N\times N$ complex matrices. 
It is important to stress that the mentioned results rely on the CP-divisibility  property and therefore their validity is guaranteed only under Markovian dynamics.  

 
A specific example of non-Markovian generators can be obtained by considering a set of mutually commutative strictly Markovian generators $\{{\cal L}_k\}$ and real scalar functions $\{l_k\b t\}$, satisfying $\int_0^{t}l_k\b{t'}dt'\geq 0$ at any time \cite{chruscinski2010non,chruscinski2012markovianity,megier2020interplay}. Taking a linear combination of these generators
\begin{equation}
{\cal L}\b t=l_1\b t{\cal L}_1+\cdots+l_{K}\b t{\cal L}_{K}~~,
\label{eq:commut_gen}
\end{equation}
generates a dynamical map which is guaranteed to be a valid quantum channel (CPTP). Such a channel necessarily exhibits memory effects if the coefficients obtain negative values for some time $t$, $l_k\b t<0$. 
Relation \eqref{eq:commut_gen} serves as a specific case of the class of dynamical maps which self commute at different times, $\sb{\Lambda\b{t},\Lambda\b{s}}=0$ for all times $t$ and $s$ \cite{megier2020interplay}. 


\subsection{Generators of invertible dynamical maps}

The linearity of ${\cal L}\b t$, cf. Eq. \eqref{eq:lambda_diff}, allows utilizing Lemma 2.2 of Ref.  \cite{gorini1976completely} (see Appendix \ref{apsec:dynamical_map_properties}) to uniquely express the generator in terms of a complete orthonormal set ${\hat{\sigma}_\alpha}$ (satisfying $\b{\hat \sigma_{\alpha},\hat \sigma_{\beta}}=\text{tr}\b{\hat \sigma_{\alpha}^\dagger\hat \sigma_{\beta}}=\delta_{\alpha\beta}$) 
\begin{equation}
    {\cal L}\b{t}\sb{\bullet}=\sum_{\alpha,\beta = 1}^{N^2} c_{\alpha \beta }\b{t}\hat{\sigma}_{\alpha} \bullet\hat{\sigma}_{\beta}^\dagger~~.
    \label{eq:linear_decomp}
\end{equation}
A further restriction is imposed by 
demanding that the map will preserve the Hermiticity property:
\begin{equation}
    {\cal L}\b t\sb{\bullet^\dagger}=\b{{\cal{L}}\b t\sb{\bullet}}^\dagger~~.
    \label{eq:hermitiacy}
\end{equation}
As a consequence  the coefficients form a Hermitian matrix $c_{\alpha\beta}=c_{ \beta\alpha}^*$. Note, that the same form of the time-local generator can also be obtained under even weaker conditions \cite{doi:10.1080/09500340701352581}.  By enforcing condition \eqref{eq:hermitiacy} and the trace-preserving property, $\text{tr}\b{{\cal L}\b{t}\sb{\bullet}}=0$, on the linear structure, Eq. \eqref{eq:linear_decomp},  Gorini et al. showed that the linear generator acquires the general form  (Theorem 2.2 of \cite{gorini1976completely})
\begin{multline}
        {\cal{L}}\b{t}\sb{\bullet} = -i\sb{\hat{H}''\b t,\bullet}+\\
     \sum_{\alpha\beta=1}^{N^2-1} c_{\alpha\beta}\b{t}\b{\hat{\sigma}_\alpha\bullet\hat{\sigma}_\beta^\dagger-\f{1}{2}\{\hat{\sigma}_\beta^\dagger\hat{\sigma}_\alpha,\bullet\}}~~.
     \label{eq:linear_struct}
\end{multline}
Here, the complete orthonormal set is chosen such that $\{\sigma_i\}$ for $i=1,\dots,N^2-1$ are traceless, where  $N$ is the dimension of the system's Hilbert space, and $\hat{H}''\b{t}$ is Hermitian and satisfies $\text{tr}\b{\hat{H}''\b{t}}=0$. 
We emphasise, that generally the coefficients $c_{\alpha\beta}\b t$ may be complex for $\alpha\neq\beta$, but are restricted by the Hermiticity condition. For such coefficients, the structure does not guarantee the CPTP of the associated map.  
Nevertheless,  Eq. \eqref{eq:linear_struct}, will serve as a template to analyse the general structure which complies with the thermodynamic postulates.

\section{General structure of the master equation beyond the Markovian regime}
\label{sec:gen_form}

The thermodynamic postulates impose constraints on the structure of the dynamical equation. Mathematically, we employ a spectral analysis to translate the postulates into conditions on the generator.  
Both $\Lambda\b{t}$ and ${\cal{U}}_{S}\b t$ are linear (super)-operators on the space of system operators, hence, their commutativity, Eq. \eqref{eq:commutivatity_maps}, implies that they share a common set of eigenoperators $\{\hat{S}\}$. 
Accordingly, these operators satisfy an eigenvalue type equation  \footnote{Note that similar considerations could be conducted in Heisenberg picture. The eigenoperators of the original map and dual map built the bi-orthogonal bases \cite{megier2020interplay}. In our case this implies that the unitary non-invariant eigenoperators are also eigenoperators of the dual map, but the unitary invariant operators under action of the dual map are in general mapped to different mixtures of invariant operators than in the case of the map in Schr{\"o}dinger picture. Accordingly, the independence of the unitary invariant and  non-invariant subspaces is still given.}
\begin{equation}
    {\cal U}_{S}\b t \sb{\hat{S}_\alpha}=\hat{U}_S\b{t}\hat{S}_\alpha\hat{U}_S^\dagger\b{t} = e^{i\theta_\alpha\b t} \hat{S}_\alpha~~,
    \label{eq:eigen_val}
\end{equation}
 with $\theta_\alpha\in \mathbb{R}$ and similarly for 
 $\Lambda\b{t}$
 with eigenvalues $\lam_\alpha\in \mathbb{C}$. 

For a time-independent Hamiltonian with a non-degenerate spectrum, the eigenoperators can be categorized into two sets: {\emph{unitary invariant}} and {\emph{unitary non-invariant}} operators.  The unitary invariant operators have unity eigenvalues ($\theta_k\b t=0$) and are spanned by the energy  projection operators of $\hat{H}_S$, $\{\hat{\Pi}_j = \ket{j}\bra{j}\}$, where $\hat{H}_S\ket{n} = \eps_n\ket{n}$. The non-invariant set includes all the transition operators between energy states $\{\hat{F}_{nm}=\ket{n}\bra{m}\}$ for which $n\neq m$.
For conciseness of the analysis, when convenient we use a single index instead of the double index $nm$ ($\hat{F}_{\alpha}=\hat{F}_{nm}$). Throughout the paper the single index are denoted by Greek letters $\alpha,\beta=1,\dots,N^2$, where $N^2$ is the dimension of the space of system operators, while English letters indices run over states of the system's Hilbert space $i,j,n,m=1,\dots,N$.

If we assume that the Bohr frequencies are non-degenerate, that is $\eps_n-\eps_m\neq\eps_k-\eps_l$ for $n\neq k$ or $m\neq l$, the unitary non-invariant eigenoperators also constitute eigenoperators of $\Lambda\b{t}$. Meaning that the transition operator satisfy
\begin{equation}
     \Lambda\b{t}\sb{\hat{F}_{\alpha}} = \lam_{\alpha}\b t \hat{F}_{\alpha}~~.
    \label{eq:non-invariant_eig}
\end{equation}
In addition, commutativity of the maps (time-translation symmetry) implies that the unitary invariant and non-invariant subspaces are independent, that is, the unitary invariant operators are mapped to invariant operators
\begin{equation}
      \Lambda\b{t}\sb{\hat{\Pi}_j} = \sum_{i=1}^{N} \mu_{ji}\b{t} \hat{\Pi}_{i}~~.
     \label{eq:invariant_eig}
\end{equation}
Note, that due to the time dependency of $\mu_{ji}\b{t}$, the corresponding dynamical map is in general non-commutative and the associated eigenoperators are generally time-dependent \cite{megier2020interplay}.

\begin{figure}
\centering
\includegraphics[width=6.cm]{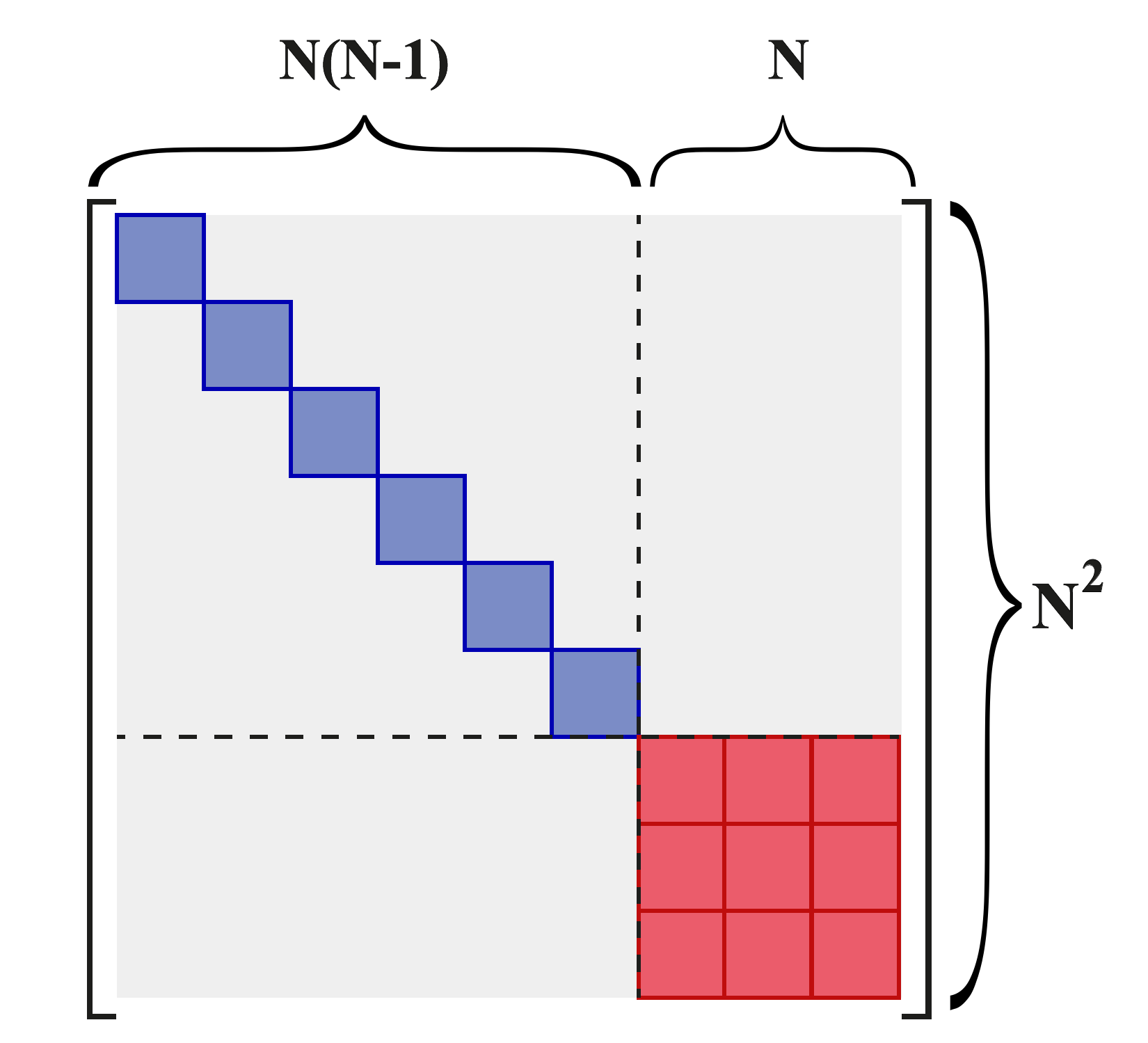}
\caption{The schematic structure of $\Lambda\b t$ and its corresponding dynamical generator ${\cal L}\b{t}$, in Hilbert-Schmidt space, displayed in the operator basis $\{\hat S\}$, Eq. \eqref{eq:S_basis}. The matrix  representation of the superoperators is block diagonal: the upper left part is a $N\b{N-1}$ real diagonal matrix (blue), corresponding to the unitary non-invariant part $\{\hat{F}_{nm}\}$, while the lower right part is an $N$ by $N$ Hermitian matrix (red) which governs the dynamics of the unitary invariant subspace (spanned by the energy projection operators $\{\hat{\Pi}_j\}$). The grey colored elements vanish.  
}
\label{fig:scheme}
\end{figure}

The above relations, Eqs. \eqref{eq:non-invariant_eig} and \eqref{eq:invariant_eig}, can be rationalized by representing the dynamical map as a matrix in the Hilbert-Schmidt space of operators (also known as Liouville space). Such a space is the vector space of system operators $\{\hat{X}\}$ endowed with an inner 
product $\b{\hat{X}_i,\hat{X}_j}=\text{tr}\b{\hat{X}_i^\dagger\hat{X}_j}$. In this framework, by choosing an operator basis  
\begin{equation}
\{\hat{S}\}\equiv\{\hat{F}_1,\dots, \hat{F}_{N\b{N-1}},\hat{\Pi}_{1},\dots,\hat{\Pi}_N\}~~, 
\label{eq:S_basis}
\end{equation}
the dynamical map obtains a block diagonal form. The upper block is diagonal and contains the eigenvalues $\lam_\alpha \b t$, while the lower block is generally a full time-dependent matrix , see Fig. \ref{fig:scheme}. 

Relations \eqref{eq:eigen_val}, and \eqref{eq:non-invariant_eig} along with Eq. \eqref{eq:lambda_diff} imply that the generator and dynamical map share a similar  structure in Hilbert-Schmidt space , for details see Appendix \ref{apsec:dynamical_map_properties}.
This structure implies similar conditions on the dynamical generator:
\begin{equation}
 {\cal L}\b{t}\sb{\hat{F}_{\alpha}} =a_{\alpha}\b{t} \hat{F}_{\alpha}~~,
    \label{eq:non-invariant_eig_gen}
\end{equation}
and
\begin{equation}
      {\cal L}\b{t}\sb{\hat{\Pi}_j} = \sum_{n=1}^N b_{ji}\b{t} \hat{\Pi}_{i}~~,
         \label{eq:invariant_eig_gen}
\end{equation}
By enforcing these conditions on Eq. \eqref{eq:linear_decomp}, we obtain the restrictive structure of the dynamical generator.

We begin the derivation by expressing  the linear structure of the dynamical map in terms of the operator basis $\{\hat{S}\}$. In this basis Eq. \eqref{eq:linear_decomp} becomes 
\begin{equation}
    {\cal L}\b{t}\sb{\bullet}=\sum_{\alpha,\beta = 1}^{N^2} c_{\alpha \beta }\b{t}\hat{S}_{\alpha} \bullet\hat{S}_{\beta}^\dagger~~.
    \label{eq:linear_decomp_2}
\end{equation}
Next, we enforce conditions Eqs. \eqref{eq:hermitiacy},  \eqref{eq:non-invariant_eig_gen} and \eqref{eq:invariant_eig_gen} on the linear structure 
(see Appendix \ref{apsec:conditions} for an explicit derivation)
\begin{multline}
    {\cal L}\b{t} \sb{\bullet} = \\
   \sum_{\alpha=1}^{N\b{N-1}} c_{\alpha\alpha}\b{t} \hat{F}_{\alpha} \bullet\hat{F}_{\alpha}^\dagger 
    +\sum_{i,j=1}^{N}  p_{ij}\b{t} \hat{\Pi}_{i} \bullet\hat{\Pi}_{j}~~.
   \label{eq:17}
\end{multline}
Here, the coefficients of the first term are real, while the coefficients of the second term $p_{ij}$ correspond to $c_{\alpha\beta}$ with $\alpha = N\b{N-1}+i
$ and $\beta = N\b{N-1}+j$.  These form an $N$-dimensional complex Hermitian matrix. 
The first term of Eq. \eqref{eq:17} generates transition between the system's energy levels, transferring heat between the system and environment. The second term serves as source or drain term, generating or destroying coherence in the energy basis (dependent on the sign of real part of the associated coefficient)

Uniqueness of \eqref{eq:17} follows from a dimensional analysis. This is achieved by counting the number of independent variables under the Hermiticity preserving property and relations \eqref{eq:eigen_val}, and \eqref{eq:non-invariant_eig}. The Hermiticity preserving property of ${\cal L}\b t$, Eq. \eqref{eq:hermitiacy}, implies that the linear structure has $N^4$ independent degrees of freedom (DOF) (number of DOF in an Hermitian matrix of dimension $N^2$). In addition,
relations \eqref{eq:eigen_val}, and \eqref{eq:non-invariant_eig} enforce $N^4-N\b{N-1}-N^2$ constraints on this linear structure, leaving $N\b{2N-1}$ free variables. Alternatively, the DOF in the generator can be counted by the number of undetermined DOF in the associated superoperator (see scheme \ref{fig:scheme}): the diagonal  non-invariant part contribute $N\b{N-1}$ DOF and the coefficients of the invariant subspace contribute $N^2$ DOF (Hermitian matrix of dimension $N$). In comparison, the coefficients of the transition terms introduce $N\b{N-1}$ DOF, while the source-drain term contribute $N^2$ DOF, giving a total of $N\b{2N-1}$ independent DOF in Eq. \eqref{eq:17}. 
Since the number of independent DOF coincides with the number of free variables, the resulting form of the dynamical generator is unique.  

Equation \eqref{eq:17} can be simplified further by enforcing the trace-preserving property. Following the seminal work of Gorini et al. \cite{gorini1976completely}, we introduce a new operator basis $\{\hat{P}\}$ for the invariant subspace (linear combinations of $\{\hat{\Pi}\}$), satisfying $\hat{P}_N=\hat{I}/N$, while the rest of the operators are traceless operators.
These operators define a new operator basis for the system
\begin{equation}
\{\hat{T}\}\equiv\{\hat{F}_1,\dots, \hat{F}_{N\b{N-1}},\hat{P}_{1},\dots,
\hat{P}_N\}.    
\label{eq:S_basis_2}
\end{equation}
A possible choice is the diagonal matrices of the $SU\b{N}$ generalized Gell-Mann matrices $\hat{P}_j=\sqrt{\f{2}{j\b{j+1}}}\b{\sum_{l=1}^{j}\ket{l}\bra{l}-j\ket{j+1}\bra{j+1}}$, for $j=1,\dots,N-1$ \cite{bertlmann2008bloch}. By demanding that the mapping preserves the trace of the operators, the standard derivation leads to the final form of the dynamical generator (see details in Appendix \ref{apsec:invariant_term})
\begin{multline}
    {\cal L}\b{t} \sb{\bullet} = -\f{i}{\hbar}\sb{\bar{H}(t),\bullet}\\
   +\sum_{\alpha=1}^{N\b{N-1}} c_{\alpha\alpha}\b{t} \left( \hat{F}_{\alpha} \bullet\hat{F}_{\alpha}^\dagger -\f{1}{2}\{\hat{F}_{\alpha}^\dagger\hat{F}_{\alpha},\bullet\}\right)\\
    +\sum_{i,j=1}^{N-1}d_{ij}\b{t}\b{\hat{P}_{i}\bullet\hat{P}_{j}^{\dagger}-\f 12\{\hat{P}_{j}^{\dagger}\hat{P}_{i},\bullet\}}~~,
   \label{eq:L_time_independent}
\end{multline}
where $\bar{H}\b{t}=\f {\hbar}{2i}\b{\hat{P}^{\dagger}\b{t}-\hat{P}\b{t}}$ is a Hermitian operator, with $\hat{P}\b{t}=\f 1N\sum_{i=1}^{N-1}d_{iN}\b{t}\hat{P}_{i}$, and $d_{ij}\b{t}=d_{ji}^*\b{t}$. The kinetic coefficients $c_{\alpha \alpha}\b{t}$ must be real but may be negative, while $d_{ij}\b{t}$ are generally complex. The form of Eq. \eqref{eq:L_time_independent} is termed the {\emph{open system dynamical symmetric structure}}. When $c_{\alpha \alpha}$ obtain negative value the dynamics necessarily is non-Markovian and include memory effects.

The initial assumption concerning the non-degeneracy of the spectrum seems limiting. However, in Sec. \ref{sec:degeneracy}  we show that this assumption does not impose a practical limitation on the studied systems, as a free propagator with a degenerate spectrum can be well-approximated by a suitable non-degenerate propagator.

Overall, Eq. \eqref{eq:L_time_independent} serves as the general structure of a valid time-translation symmetric dynamical generator. We emphasise, that this form does not guarantee complete positivity of the associated dynamical map. The main advantage of Eq. \eqref{eq:L_time_independent} is that all the correct Lindblad jump operators are determined, leading to a generator which satisfies the desired symmetries and a minimal number of undetermined kinetic coefficients. 
We show in Sec. \ref{sec:kinetic coefficients} that by utilizing a polynomial expansion these coefficients can be determined up to the desired accuracy, thus restoring complete positivity within the associated error range.

We would like to point out, that time-translation symmetry and the form of the corresponding master equations was mainly addressed in the literature  for the case of strictly Markovian dynamics \cite{HOLEVO199395,10.1007/BFb0106777,Vacchini2010,chruscinski2020universal}. An exception is the case of the time-translation symmetry for a qubit (mostly called phase covariance in this context), for which a large number of publications have appeared recently, motivated by the phase estimation problems \cite{Teittinen_2018,Haase2019,Filippov_2020,PhysRevLett.116.120801,Haase_2018}. These results  reproduce our findings for $N=2$, see Sec. \ref{sec:2} for an example in terms of the Jaynes-Cummings model.

\section{Determining the kinetic coefficients}
\label{sec:kinetic coefficients}

When the coupling to the environment is restricted, the kinetic coefficients of the dynamical generator $\{\{c_{\alpha\alpha}\},\{d_{ij}\}\}$ can be determined by comparing Eq. \eqref{eq:L_time_independent} to a perturbative expansion of the exact master equation. We propose a general procedure to determine these coefficients up to a chosen accuracy.

In the interaction picture with respect to the bare system and environment Hamiltonian ($\hat{H}_S+\hat{H}_E$), the joint system dynamics are given by
\begin{equation}
    \f{d}{dt}\tilde{\rho}\b t =\tilde{\cal L}^{\b{SE}}\b t\sb{\tilde{\rho}\b t} =-\f{i}{\hbar}\sb{\tilde{H}_{SE}\b t,\tilde{\rho}\b t}~~,
    \label{eq:interaction_dyn}
\end{equation}
where the overscript tilde designates operators in the interaction picture, $\tilde{X}=e^{i\b{\hat{H}_S+\hat{H}_E}t/\hbar}\hat{X}e^{-i\b{\hat{H}_S+\hat{H}_E}t/\hbar}$ and $\tilde{\cal L}^{\b{SE}}$ is the generator of the total system in the interaction picture. For a generic interaction the solution of Eq. \eqref{eq:interaction_dyn} is involved due to the time-dependence in the generator. However, when the interaction satisfies strict energy conservation (Postulate 1) the interaction term in the interaction picture is time-independent $\tilde{H}_{SE}\b t=\hat{H}_{SE}$. As a result, the joint dynamical generator is also time-independent $\tilde{\cal L}^{\b{SE}}\b t =\tilde{\cal L}^{\b{SE}}$, leading to a simplified solution. 
Equation \eqref{eq:interaction_dyn} can be integrated to obtain the formal solution 
\begin{equation}
    \tilde{\rho}_S\b{t} =\text{tr}_E\b{e^{\tilde{\cal L}^{\b{SE}} t}\sb{\hat{\rho}\b{0}}}~~.
\end{equation}
We proceed by expressing the joint dynamical map $e^{\tilde{\cal L}^{\b{SE}} t}$ in terms of a suitable polynomial expansion. Consequently, this leads to an expansion for the reduced system generator 
\begin{equation}
    \tilde{\cal L}\b t \sb{\hat{\rho}_S\b{t}}=\sum_n w_n\b{t}\varphi_n\b{\tilde{\cal L}^{\b{SE}}{\sb{\hat{\rho}\b{0}}}}~~,
\end{equation}
where $\{w_n\}$ are expansion coefficients and $\{\varphi_n\}$ are operator valued functions, which depend on the chosen polynomial.

We focus on two specific polynomial series, the Maclaurin (Taylor expansion around zero)  and Chebychev series. The Maclaurin series is chosen due to its simple form and widespread use. In addition, its expansion point is the origin, therefore, it faithfully captures the dynamics in the short-time regime, which commonly exhibits non-Markovian behavior.
In comparison, the Chebychev series captures the global character of the approximated function, and is constructed so to minimize the maximal error for the chosen time interval. Hence, the Chebychev series is advantageous for intermediate and long timescales \cite{kosloff1994propagation}.

Other polynomials can be chosen, specifically tailored to approximate the dynamics at different timescales. The solutions for different time-regimes of the kinetic coefficients can then be stitched together, leading to a combined accurate description.

\subsection{Solution based on the Maclaurin series}

Expanding the joint map in terms of a Maclaurin series gives 
\begin{multline}
    \tilde{\cal L}\b t\sb{\tilde{\rho}_S\b t}=\sum_{n=1}^{\infty}\b{-\f i{\hbar}}^{n}\f{t^{n-1}}{\b{n-1}!}\\\times\text{tr}_E\b{{\sb{\hat H_{SE},\sb{...\sb{\hat H_{SE},\hat{\rho}\b{0}}}}}}~~,
    \label{eq:maclaurin}
\end{multline}
where the last term includes $n$ commutation relations.
We now truncate the series in the desired order and introduce an explicit initial state for the joint system. We denote the $M$'th order approximate dynamical generator by $\tilde{\cal{L}}^{\b{M}}\b t$ (discarding all terms $n>M$).
By utilizing the orthonormality  the operator basis $\{\hat{S}\}$, Eq. \eqref{eq:S_basis}, and the structure of the master equation, Eq. \eqref{eq:17}, we obtain a set of the linear equations for the coefficients. In terms of the double indices notation ($\alpha\ra nm$) the set of equation are expressed as (see Eq. \eqref{eq:Cond1} in Appendix \ref{apsec:conditions}), 
\begin{equation}
c_{nnmm}-\f 12\sum_{i=1}^{N}\b{c_{inin}+c_{imim}}=    \text{tr}\b{\hat{F}_{nm}^\dagger\tilde{\cal L}^{\b{M}}\b t\sb{\hat{F}_{nm}}}~~,
\label{eq:28}
\end{equation}
for $n\neq m$, and (Eq. \eqref{eq:Cond2})
\begin{equation}
   c_{inin}=\text{tr}\b{\hat{\Pi}_{i}^\dagger\tilde{\cal L}^{\b{M}}\b {t}\sb{\hat{\Pi}_{n}}}~~,
   \label{eq:29}
\end{equation}
for $i\neq m$, and
\begin{equation}
    {c_{nnnn}-\sum_{i=1}^{N}c_{inin}}=\text{tr}\b{\hat{\Pi}_{n}^\dagger\tilde{\cal L}^{\b{M}}\b{t}\sb{\hat{\Pi}_{n}}}~~.
    \label{eq:30}
\end{equation}
Here $c_{nnmm}$ corresponds to the coefficient $p_{nm}$ of Eq. \eqref{eq:17} (single index notation of the source-drain term) and the coefficients of the form $c_{inin}$ are the coefficients of  $\hat{F}_{\alpha}=\ket{i}\bra{n}$.

The set of linear coupled equations can be solved by standard numerical techniques, leading to a complete characterization of the coefficients. Finally, the non-Markovian generator, Eq. \eqref{eq:L_time_independent}, is completely determined by employing the unitary transformation that relates $\{\hat{S}\}$ and $\{\hat T\}$, Eq. \eqref{eq:S_basis_2}, operator bases.

\subsection{Solution based on a Chebychev series}
\label{subsec:chebychev}

Expansion of the joint dynamical map in terms of the Chebychev polynomials leads to 
\begin{equation}
    \tilde{\cal L}\b t\sb{\tilde{\rho}_S\b{t}} = e^{s}\sum_{n=1}^{\infty}w_{n}\b{r\b t} \text{tr}_E\b{T_{n}\b{\tilde{{\cal{O}}}\sb{\hat{\rho}\b{0}}}}
    \label{eq:Chebychev}
\end{equation}
where $T_n\b{x}=\cos\b{n \arccos\b{x}}$ is the $n\in\mathbb{N}$ Cheybychev polynomial and $\{w_n\}$ are the associated coefficients. The superoperator $\tilde{\cal O}$ is a normalized version of the joint dynamical generator $\tilde{\cal L}^{\b{SE}}\b t$ and $s$ and $r$ are suitable normalization constants (cf. Appendix \ref{apsec:chebychev} for further details).
By following an analogous treatment as in the Maclaurin series procedure, we truncate the series, Eq. \eqref{eq:Chebychev}, and evaluate the kinetic coefficients (Eq. \eqref{eq:28}, \eqref{eq:29} and \eqref{eq:30}). In Sec. \ref{sec:spin_star} we demonstrate this method by calculating the dynamics of a spin star, see Fig. \ref{fig:spin_star}.
\\

Overall,  for time-translation symmetric dynamics, the perturbative treatment leads to a master equation with the correct operator structure and kinetic coefficients within the desired accuracy.   
This procedure does not guarantee that the associated dynamical map is completely positive, since the kinetic coefficients are determined up to a certain error. Nevertheless, if the error in the coefficients is taken into account, the generated map will be completely positive, as the exact CPTP dynamical map $\Lambda\b{t}$, defined in Eq.  \eqref{eq:dynamical_map}, resides with in the error range.

For a certain order of the polynomial series and time $t$, the accuracy improves with a decrease of the interaction strength. Nevertheless, the proposed method can be employed in the strong coupling case if a sufficiently high order polynomial is chosen. The high order polynomial terms include high order environment correlation functions. These can be calculated efficiently utilizing Wick's theorem and graphical tools, such as Feynmann diagrams \cite{PhysRev.76.749,dyson1949radiation,wick1950evaluation}.

The computational resources can be reduced, if the environment's  memory decay rate is taken into consideration. Typically, the decay rate of environmental correlations increases with the order of the correlations. For example, for a bath with  Gaussian spectrum, the $n$'th order correlations decay $n$ times faster \cite{gardiner2004quantum}. This behaviour motivates performing a `higher order' Markovian approximation, which involves choosing a course-graining time $\Delta t_{\text{c.g}}$, and then truncating the series in orders for which the associated correlations decay faster then $\Delta t_{\text{c.g}}$.

\section{Dynamical Symmetry constraints}
\label{sec:symmetry_constraints}

The time-translation symmetry of the dynamical map not only determines the Lindblad jump operators and structure of the dynamical generator, but also enforces restrictions on the kinetic coefficients. Essentially, these additional constraints on the dynamical generator are a consequence of the fact that {\emph{the asymmetry of a state cannot increase under symmetric dynamics}}. 
These symmetry considerations can supplement the perturbative treatment, employed to determine the kinetic coefficients.
Such analysis may be crucial when a detailed description of the environment 
spectrum is not possible, therefore prohibiting a perturbative treatment. 


The concept of asymmetry under symmetric dynamic was introduced and formalized by Marvian et al. \cite{marvian2012symmetry,marvian2014extending,marvian2014modes}, highlighting the fact that the conservation laws arising from Noether's theorem are neither necessary nor sufficient conditions to characterize the possible transitions of open quantum systems \footnote{Under closed system symmetric dynamics of mixed states, Noether's conservation laws serve as necessary, but not sufficient conditions, on the possible state transformations. Only for pure states, under going symmetric reversible dynamics, do Noether's conservation laws provide both necessary and sufficient condition for an achievable state transformations.}. In open systems, dynamical symmetries are, instead, manifested by a monotonic behaviour of certain information-theoretic functions, termed asymmetry monotones. These can be utilized to introduce additional necessary conditions on the structure of the non-Markovian master equation.

We begin the analysis with a  brief description of the theory of symmetric dynamics and asymmetry of states.  A dynamical symmetry is defined with respect to a set of symmetry transformations ${\cal U}_g\sb{\bullet}=\hat{U}\b{g}\bullet\hat{U}^\dagger\b{g}$, where $\hat{U}\b{g}$ are unitary operators, associated with group elements $g\in G$. A symmetric dynamical map, $\Lambda_G$ with respect to group $G$, known also as a {\emph{G-covariant}} map, satisfies the property (commutation relation)
\begin{equation}
   { \cal U}_g\circ\Lambda_G = \Lambda_G \circ{\cal U}_g~~,
\label{eq:general_group_com}
\end{equation} 
for all $g\in G$. For example, in the present case, the time-translation symmetry Eq. \eqref{eq:commutivatity_maps}, is associated with the group $U\b{1}$. This is a Lie group which is generated by the system Hamiltonian $\hat{H}_S$, $\hat{U}\b{g=t}\equiv\hat{U}_{S}\b t= e^{-i\hat{H}_S t/\hbar}$. 


The asymmetry of a state is a measure of the extent it breaks the associated symmetry. For example, for time-translation symmetry, symmetric states are those which are invariant under ${\cal{U}}_{S}\b t\sb{\bullet}=\hat{U}_{S}\b t\bullet\hat{U}_{S}^{\dagger}\b t$, i.e., all incoherent mixtures states in the distinct system's energy states. In contrast, states with coherences between eigenstates with different energies are asymmetric.  

Under symmetric dynamics, asymmetry can be viewed as a resource, as the extent of asymmetry of the initial state dictates which transformations are possible \cite{marvian2014extending,lostaglio2015quantum}. The asymmetry of a state can be quantified in terms of asymmetry monotones.
Formally, an asymmetry monotone is a function $\cal A$ from the space of states to real numbers for which the existence of a G-covariant
channel ${\Lambda}\b{t}$
implies that ${\cal A}\b{\hat{\rho}_S}\geq {\cal A}\b{{\Lambda}\b{t}\sb{\hat{\rho}_S}}$. For time-translation symmetry, asymmetry coincides with coherences in the energy basis of the generator $\hat{H}_S$ (denoted just as coherences). As asymmetry cannot increase under symmetric dynamics, the connection infers that time-translation symmetric dynamics degrade coherences.  

Another important tool in the analysis of symmetric dynamics is the decomposition of a state to asymmetry modes \cite{marvian2014modes}. 
A state $\hat{\rho}_S$ can be expressed as a sum of asymmetry modes
\begin{equation}
    \hat{\rho}_S = \sum_k \hat{\rho}^{\b k}_S
\end{equation}
where each mode $\hat{\rho}^{\b k}_S$, is an eigenoperator of the symmetry transformation ${\cal{U}}_g$.
For the studied symmetry, the asymmetry modes correspond to the non-invariant eigenoperators $\{\hat{F}_{\alpha}\}$, which satisfy ${\cal U}_{S}\b t\sb{\hat{F}_\alpha}=e^{-i\omega_{\alpha} t}\hat{F}_\alpha$, with corresponding Bohr frequencies $\{\omega_\alpha\}$.



In order to impose constraints on the master equation we focus on asymmetry monotones which only measure the degree of asymmetry associated with a certain mode. The trace-norm $||\hat{O}||=\text{tr}\b{\sqrt{\hat{O}\hat{O}^\dagger}}$ constitutes such a monotone, which is crucially non-increasing under the operation of any quantum channel.  This property along with Eq. \eqref{eq:non-invariant_eig} then implies that \cite{marvian2014modes} 
\begin{equation}
    ||\hat{F}_\alpha||\geq\Big|\Big|\Lambda\b t\sb{\hat{F}_\alpha}\Big|\Big|~~.
\end{equation}
This infers that the absolute values of the associated eigenvalues of $\Lambda\b t$ are smaller than one. 
Combining this inequality with Eq. \eqref{eq:2} translates to an integral condition on the eigenvalues of the dynamical generator (Eq. \eqref{eq:non-invariant_eig_gen})
\begin{equation}
    \int_{0}^t a_{\alpha}\b{s}ds\leq 0~~
\end{equation}
for all $\alpha$. 

Finally, the connection to the kinetic coefficients is obtained by following the scheme in Sec. \ref{sec:kinetic coefficients}. We utilize Eq. \eqref{eq:28} and take into account the entire infinite series ($M\ra \infty$) to obtain a constraint on the exact kinetic coefficients of the source-drain term (the term including the unitary non-invariant eigenoperators)
\begin{equation}
\int_{0}^{t} \b{c_{nnmm}\b{s}-\f 12\sum_{i=1}^{N}\b{c_{inin}\b{s}+c_{imim}\b{s}}}ds\leq 0 ~~.
\label{eq:37}
\end{equation}

Further restrictions on the kinetic coefficients depend on the properties of both the unitary non-invariant and invariant eigenoperators of $\Lambda\b{t}$. In Ref. \cite{cwiklinski2015limitations} {{\' C}}wikli{{\' n}}ski et al. point out that the CPTP property and conditions \eqref{eq:invariant_eig_gen}
and \eqref{eq:non-invariant_eig_gen} imply that the damping matrix ${\cal M}$ must be positive \cite{roga2010davies}. The diagonal of ${\cal M}$ is constructed from the eigenvalues of the invariant eigenoperators  $\text{diag}\b{{\cal M}}=\{b_{11},\dots,b_{NN}\}$, while the off-diagonal elements coincide with the eigenvalues of the non-invariant operators ${\cal M}_{nm}=a_{nm}$, corresponding to eigenoperator $\hat{F}_{nm}$ (in the double index notation $\alpha= nm$). The positivity of the damping matrix does not translate to straight forward restrictions on the kinetic  coefficients (for a system with dimension $d>2$ \cite{cwiklinski2015limitations}), nevertheless, it can be used to verify the validity of the master equation.

Restrictions on the kinetic coefficients of the energy transfer term, $\{c_{\alpha\alpha}\}$, arise from the limitations on the possible transitions of energy population.  
This issue has been studied thoroughly for the case of a thermal reservoir, in the context of the resource theory of thermal operations \cite{horodecki2003reversible,horodecki2013fundamental}. Specifically, the allowed transitions must be such that the initial state thermomajorizes the final state (for completeness we provide a complete definition in Appendix \ref{apsec:thermomaj}, for further details see Ref. \cite{lostaglio2019introductory}). Meaning that for a thermal reservoir, the kinetic coefficients cannot lead to a state which violates the thermomajorization condition with respect to the initial state. On the level of transformations between states, the thermomajorization condition has a clear graphical interpretation utilizing Lorentz curves, however, this restriction does not translate to concise closed form conditions on the kinetic coefficients. Nevertheless, as the positivity of the damping matrix, the thermomajorization condition can be utilized as an additional validation check of the master equation of symmetric dynamics.

\section{Strict energy conservation and non-Markovianity}
\label{sec:sec_and_markovianity}

The derivation of the reduced dynamics of an open quantum system, for a Markovian environment, relies on the rapid decay of the environment's correlation functions. These represent the ``memory'' of the environment. Such a rapid decay of correlations implies that the environment effectively remains in its initial stationary state, and allows simplifying the exact dynamical equation to obtain the quantum Markovian master equation \cite{breuer2002theory}.
Interestingly, for an environment initially in a stationary state, the environment correlation functions do not decay under the strict energy conservation condition. This property illuminates a basic relation between processes that violate strict energy conservation and Markovian behaviour.

The basic connection can be understood by studying the reduced system dynamics in the interaction picture relative to the free dynamics. In this picture the reduced dynamics can be expressed as
\begin{equation}\label{eq:Markov}
  \f{d}{dt} \tilde{\rho}_S\b t =-\f{1}{\hbar^2}\int_0^tds\,\text{tr}_E\b{\sb{\tilde{H}_{SE}\b t,\sb{\tilde{H}_{SE}\b{s},\tilde{\rho}\b{s}}}}~~.
\end{equation}
This expression depends on the two-time correlation functions $\tr_E\b{\tilde{B}_k\b{t}\tilde{B}_l\b{s} \tilde{\rho}_E\b{s}}$, where $\hat{B}_{k}$ and $\hat{B}_l$ are environment operators. Here we have approximated $\tilde\rho(s)\approx \tilde\rho_S(s) \otimes \tilde\rho_E(s)$, which makes the right hand side of Eq. \eqref{eq:Markov} accurate up to the second order in interaction strength and is well justified in a try to obtain Markov limit.
Under strict energy conservation the interaction term commutes with the free dynamics leading to a time-independent term $\tilde{H}_{SE}\b{t}=\hat{H}_{SE}$. In addition, due to the size of the environment the environment is modified only slightly by the interaction with the system. As a result, the correlations $\mean{\hat{B}_k\hat{B}_l}_E$ are only slightly modified, and do not decay in time.
The environmental correlations represent the memory. Since they do not decay the resulting dynamic are non-Markovianity.

This issue can be bypassed by adopting a practical approach; assuming weak coupling and taking into account the basic limit of time-energy uncertainty, one can relax the strict commutative condition and replace it by an approximate condition. This allows incorporating an effective strict energy conservation while allowing for Markovian dynamics. Such an analysis leads to the standard form of the master equation which complies with the Davies construction \cite{davies1974markovian}.

\section{Comparison to the Davies construction}
\label{sec:davies_compare}
The symmetric structure coincides with the master equation obtained from the Davies construction \cite{davies1974markovian}. This property implies that the Davies map obeys time-translation symmetry, however there are key differences in the two approaches.

The Davies construction includes first a weak coupling limit assumption (Born), justifying the second order perturbation treatment. Followed, by a renormalization of the time, associated with the Markovian assumption, and finally imposes the secular approximation, leading the GKLS structure. It is illuminating to compare these well studied approximations to the thermodynamically motivated postulates, Sec. \ref{sec:framework}.

Time-translation symmetry is closely related to the secular approximation \cite{Haase_2018}. In the secular approximation one neglects in the master equation all mixed terms containing eigenoperators of the free evolution ${\cal{U}}_{S}(t)$ (isolated map) corresponding to different eigenvalues. This leads to decoupling of the dynamics of the unitary invariant and non-invariant operators and, additionally, to mutually independent evolution of the non-invariant operators, exactly as in Eqs. \eqref{eq:non-invariant_eig} and \eqref{eq:invariant_eig_gen}.
 The secular approximation is often characterized as a procedure which neglects the terms which ``violate" energy conservation.  However, this justification is misleading since the joint dynamics are unitary, therefore, they cannot violate energy conservation. The neglected terms correspond to accumulation of energy in the interface between system and environment.  The approximation shares similar traits as the rotating wave approximations (RWA). Both of them can be
 justified by the fact that the free evolution of the open system occurs on a much shorter timescale than the relaxation dynamics. However, the secular approximation is conducted on the level of evolution equation, while the RWA is employed on the level of the interaction Hamiltonian itself, which makes the RWA nonphysical in some respect \cite{FORD1997377,Fleming_2010}.
 In comparison, strict energy conservation prevents any change in the interface energy, which leads to a master equation without mixed terms. 
Hence, the crucial difference between the Davies approach and the current analysis is the hierarchy of assumptions. In the present approach, the symmetry restriction is imposed at the outset. This allows going beyond the Born-Markov approximation.


The Markovian dynamics and weak system-environment coupling of the Davies construction, allows neglecting the change in environment, leading to $\hat{\rho}_E\b{t}=\hat{\rho}_E\b 0$ at all times. In contrast, the dynamical symmetric structure only limits the initial environmental state and incorporates the dynamical changes of $\hat{\rho}_E\b t$ within the kinetic coefficients. Note, that recently a generalization of Davies master equation beyond the secular approximation was introduced \cite{majenz2013coarse,PhysRevA.103.062226,PhysRevA.95.042104,farina2019open,winczewski2021bypassing}.

An additional, important difference between the two constructions concerns the system-environment coupling. The Davies construction relies on weak coupling, while in principle, the symmetric structure allows for arbitrary coupling strength. Under symmetric dynamics the operatorial form is solely dictated by symmetry considerations, and as long as these consideration are satisfied, the structure remains unaffected by the coupling strength. In practice, the assumption of strict energy conservation may be scrutinized as nonphysical in the strong coupling regime. Nevertheless, certain processes such as scattering phenomena and the collision model in the low density regime satisfy strict energy conservation under strong coupling  \cite{karplus1948note,dumcke1985low}. For a critical analysis regarding this issue see Ref. \cite{dann2021open} Sec. III. In addition, the practical task of calculating accurate kinetic coefficients becomes computationally demanding with the increase of the coupling strength and the non-Markovian behaviour, cf. Sec. \ref{sec:kinetic coefficients}.

\section{Bypassing the non-degeneracy condition}
\label{sec:degeneracy}
The spectral analysis of Sec. \ref{sec:gen_form} relied on the condition that the spectrum of the free propagator $\hat{U}_S\b{t}$ (system's Bohr frequencies) is non-degenerate. This restriction can be somewhat overcome by the following reasoning.
For simplicity consider a free propagator with a single degeneracy  $\hat{U}_S^{deg}\b{t}$, we can introduce a new propagator $\hat{U}_S^{\eps}\b{t}$ for which one of the degenerate Bohr frequencies is modified by a gap $\eps>0$, effectively removing the degeneracy in the spectrum of $\hat{U}_S^{deg}\b{t}$. Such modification enables employing the spectral analysis of Sec. \ref{sec:gen_form} to obtain the  ($\eps$-exact) master equation and the associated dynamical map $\Lambda^{\eps}\b t$. The question arises: What is the difference between the ``degenerate" map and the ``non-degenerate" map? 
The difference can be evaluated by analysing the the difference in the probabilities $P^{deg}$ and $P^{\eps}$ (associated with $\Lambda^{deg}\b {t}$ and $\Lambda^{\eps}\b{t}$, correspondingly) of obtaining a certain outcome, related to an element $\hat M$ of an arbitrary POVM. The difference is bounded by (see Appendix \ref{apsec:bound})   
\begin{multline}
    |P^{deg}-P^{\eps}|\\=\bigg|\text{tr}\b{\b{\Lambda^{deg}\b t-\Lambda^{\eps}\b t}\sb{\hat{\rho}\b{0}}\hat{M}}\bigg|\leq\eps t+O\b{\eps^{2}}~~.
    \label{eq:error}
\end{multline}
which for sufficiently small $\eps$ becomes negligible.  
This result implies that the two maps are practically indistinguishable with respect to any observable, for an appropriate choice of $\eps$ and a time duration of interest \trr{($\eps$ can be chosen to be time-dependent, e.g, $\eps=\delta/t$, for some small $\delta$. This enables removing the degeneracy for any finite time.)}.

A similar procedure can be done for the case where the free propagator has multiple degeneracies by introducing $\eps$-small changes to the spectrum of $\hat{U}_S^{deg}\b{t}$. 
Such analysis leads to analogous conclusions as Eq. \eqref{eq:error}.

\trr{The removal of the degeneracy is also motivated by practical  physical considerations. External interactions typically break the symmetries of physical systems. Therefore, excluding degeneracy which arise from nature's fundamental symmetries, tiny perturbations will remove the degeneracy between eigenstates. Interestingly, if there is an inherent symmetry thermalization is not guaranteed due to the existence of multiple fixed points of the dynamical map \cite{albert2019asymptotics}.}

\section{Generalization to other symmetries}
\label{sec:generalization to other symmetries}

The symmetry analysis, that was conducted for the case of time-translation symmetry, can be generalized to other symmetry classes. An analogous analysis to Sec. \ref{sec:gen_form} can be performed for any dynamical symmetry, represented by a finite or Lie groups $G$ \cite{HOLEVO199395,marvian2012symmetry}. Generally, condition Eq. \eqref{eq:commutivatity_maps} is replaced by Eq. \eqref{eq:general_group_com}, where the initial environmental state must now be stationary with respect to the generator of the symmetry.
This relation implies that $\Lambda\b{t}$ shares an eigenoperator basis with ${\cal{U}}_g$. Moreover, if the spectrum of $\hat{U}\b{g}$ is non-degenerate, the representation of $\Lambda\b{t}$ and ${\cal{L}}\b{t}$ in Hilbert-Schmidt space are block diagonal (as shown in Fig. \ref{fig:scheme}) in the eigenoperator basis of ${\cal{U}}_g$. This leads to a master equation of the form of Eq. \eqref{eq:L_time_independent}, where $\{\hat{F}_\alpha\}$ and $\{\hat{\Pi}_j\}$ are replaced by the transition and invariant operator of $\hat{U}\b g$, respectively.

Finally, when ${\cal U}_g$ represents a symmetry transformation belonging to a Lie group, the kinetic coefficients can be determined by a similar perturbative treatment as described in Sec. \ref{sec:kinetic coefficients}.
In this procedure, instead of transforming to the interaction picture with respect to the free dynamics, the transformation should be conducted with respect to the generators of $\hat U\b g$. An appropriate transformation will then lead an  interaction Hamiltonian relative to the symmetry transformation, and analogous relations to Eq. \eqref{eq:28}, \eqref{eq:29} and \eqref{eq:30}.

\subsection{Conservation of the number of excitations}
\label{sec:excitationNumber}
Noether's theorem relates the symmetry of global gauge invariance to the conservation of the number of excitations (particles). In thermodynamics this conservation law is associated with the grand canonical ensemble.  
We next study the consequences of such a symmetry on the form of the dynamical generator.
We emphasis that such a conservation law, does not require the conservation of the total (free) system and environment energies, therefore does not generally satisfy strict energy conservation. For instance, when the excitations are not on resonance. In reverse, strict energy conservation can hold even  when the total number of excitation changes. For example, when the multiple excitations in the environment correspond to a single energy quanta of the system. 

Conservation of the total number of excitations is represented governed by the total Hamiltonian \begin{equation}
    \hat{H}^{\b{N}} = \hat{H}_S+\hat{H}_E +\hat{H}_I~~,
    \label{eq:Ham_num}
\end{equation}
with an interaction of the form
\begin{equation}
    \hat{H}_I = \sum_{k,\omega,\omega':\omega,\omega'>0} \hat{A}_k\b{\omega}\otimes \hat{B}_k^{\dagger}\b{\omega'}~~,
\end{equation}
where $\hat{A}_k\b{\omega}$ and $\hat{B}_k\b{\omega'}$ are eigenoperators of the system and environment. These satisfy $\sb{\hat{H}_S,\hat{A}_k\b{\omega}} =-\omega\hat{A}_k\b{\omega}$ and $\sb{\hat{H}_E,\hat{B}_k\b{\omega'}} =-\omega'\hat{B}_k\b{\omega'}$ and $\hat{A}_k^\dagger\b{\omega}=\hat{A}_k\b{-\omega}$ and similarly for $\hat{B}_k\b{\omega'}$. In addition, $\hat{A}_k\b{\omega}$ and $\hat{B}_k\b{\omega '}$ must generate a creation and annihilation of the same number of excitations. For example, for a system Hamiltonian $\hat{H}_S = \sum_k \eps_k \ket{\eps_k}\bra{\eps_k}$ coupled to a bosonic bath, $\hat{H}_I$ may include terms of the form $\ket{n}\bra{n+1}\otimes\hat{b}^\dagger\b{\omega'}+\text{h.c}$ and $\ket{n}\bra{n+2}\otimes\b{\hat{b}^\dagger\b{\omega'}}^2+\text{h.c}$, but not $\ket{n}\bra{n+2}\otimes\b{\hat{b}^\dagger\b{\omega'}}+\text{h.c}$\,. The super-script in Eq. \eqref{eq:Ham_num} signifies that the Hamiltonian is associated with the conservation of the total number of excitations. 

As expected, the total Hamiltonian, $\hat{H}^{\b{N}}$,  commutes with the total number operators $\hat{N} = \hat{N}_S+\hat{N}_E$
\begin{equation}
    \sb{\hat{N},\hat{H}^{\b{N}}} = 0~~,
    \label{eq:46}
\end{equation} 
where $\hat{N}_S$ and $\hat{N}_E$ are the system and environment number operators. These operators can be written explicitly by enumerating the free energystates of the system $\{\ket{k}\}$ and environment $\{\ket{\chi_j}\}$ in increasing order, giving $\hat{N}_S =\sum\limits_k k\ket{k}\bra{k}$ and $\hat{N}_E =\sum\limits_j j\ket{\chi_j}\bra{\chi_j}$.
Relation \eqref{eq:46} motivates defining two associated unitary operators $\hat{U}_{N,S}=e^{i\hat{N}_S}$ and $\hat{U}_{N,E}=e^{i\hat{N}_E}$, along with their associated propagators ${\cal{U}}_{N,i}\sb{\bullet}=\hat{U}_{N,i}\bullet\hat{U}_{N,i}^\dagger$, where $i=S,E$. A straightforward generalization of the strict energy case leads to an analogous dynamical symmetry relation (see Appendix \ref{apsec:number_conservation})
\begin{equation}
    {\cal U}_{N,S}\circ\Lambda=\Lambda\circ{\cal U}_{N,S}~~.
    \label{eq:num_conserv_comm}
\end{equation}
This relation allows determining the general form of the master equation that complies with the conservation of the total number of excitations.

The eigenoperators of ${\cal{U}}_{N,S}$ are composed of several sets of degenerate operators. 
All the non-invariant creation operators, associated with the same {\emph{excitation number}} $l$, are degenerate. Formally, the operators of the set $\{\ket{n+l}\bra{n}\}$, with a fixed $l$, all satisfy ${\cal U}_{N,S}\sb{\ket{n+l}\bra{{n}}}=e^{il}$, and are therefore degenerate. A similar relation holds for the annihilation operators. The degeneracy along with Eq. \eqref{eq:num_conserv_comm} implies that the Liouville representation of the open system map obtains a block diagonal form in the operators basis $\{\ket{n}\bra{m}\}$. Each block corresponds to a different excitation number and can be labelled by it (positive for creation operators, negative for annihilation operators and $0$ for the invariant operators).

An analogous analysis as performed for the time-translation symmetry, Sec. \ref{sec:gen_form}, implies that the master equation is of the following form 
\begin{multline}
    {\cal L}^{\b{N}}\b{t}\sb{\bullet}= -\f{i}{\hbar}\sb{\bar{H}^{\b{N}}\b{t},\bullet} \\+\sum_{n,m,l}\gamma_{nml}\b t\bigg(\hat{G}_{n,n+l}\bullet\hat{G}_{m,m+l}^{\dagger}-\f 12\{\hat{G}_{m,m+l}^{\dagger}\hat{G}_{n,n+l},\bullet\}\bigg)~~,
    \label{eq:num_gen}
\end{multline}
where $\bar{H}^{\b{N}}\b{t}$ is generally a linear combination of invariant operators and $\hat{G}_{ab} = \ket{a}\bra{b}$ and $\gamma_{nml}$ are complex coefficients (there are additional restrictions concerning their values due to  the Hermitiacy preservation property of the map).

The obtained master equation, Eq. \eqref{eq:num_gen}, contains terms mixing the transition operators. This contrasts with time-translation symmetric dynamics, described by Eq. \eqref{eq:L_time_independent}, which is a result of the assumption of non-degeneracy of the Bohr frequencies. Note, that in the case of reduced system being a two-level system, the both configurations coincide.

In section \ref{sec:num_cons_example} we analyse the non-Markovian master equation dynamics of a spin coupled to a bosonic bath under the conservation of the total number of particles condition. Unlike the strict energy case (Sec. \ref{sec:sec_and_markovianity}), the current symmetry exhibits a Markovian limit at long time regimes.

\section{Generalization to time-dependent Hamiltonians}
\label{sec:td_Hamiltonians}

In laboratory experiments quantum systems are frequently manipulated by external ``control" fields, which are typically described in terms of an explicit time-dependent Hamiltonian $\hat{H}_S\b t$. Such a description is essentially semi-classical, as the field itself is a quantum system including an infinite number of modes. In the semi-classical regime the energy of the field is much larger then the energy stored in the system control interaction. As a result, the effect of the field can be reliably captured by a time-dependent scalar function, denoted as the drive or control.

This realization allows extending the dynamical symmetry based framework to driven time-dependent systems. It suggests the following approach to solve the open system dynamics: (i) Incorporate the field within the complete quantum description. (ii) Deduce the structure of the master equation utilizing symmetry considerations. (iii) Take the semi-classical limit to obtain the master equation for a driven quantum system. 

We next analyze this procedure for non-Markovian dynamics under time-translation symmetry.
 The Markovian case was developed in  Ref. \cite{dann2020thermodynamically} for a certain time-translation symmetry. 

The composite system, including the primary system, control and environment, is represented by the Hamiltonian
\begin{equation}
    \hat{H}=\hat{H}_S+\hat{H}_C+\hat{H}_{SC}+\hat{H}_I+\hat{H}_E~~,
    \label{eq:Ham_complete}
\end{equation}
where $\hat{H}_C$ is the control Hamiltonian and $\hat{H}_{SC}$ and $\hat{H}_I$ are the system-control and environmental interaction terms.

The  semi-classical description is obtained by taking an asymptotic limit. This limit is defined by two conditions: (i) The control field state is only slightly affected by the interaction with the system. (ii) The correlations between the control field and system are negligible.
In this limit the semi-classical Hamiltonian reads
\begin{equation}
    \hat{H}^{s.c}\b t=\hat{H}_S^{s.c}\b t+\hat{H}_I+\hat{H}_E~~.
\end{equation}
Physically, the semi-classical regime occurs when the control system is initially prepared in a very energetic state with respect to the system's energy scale, $||\hat{H}_C||\gg||\hat{H}_{SC}||\sim||\hat{H}_S||$. In this regime, the evolution of the  control state is dominated by the free dynamics 
$\hat{\rho}_C\b t\approx \hat{U}_C\hat{\rho}_C\b 0\hat{U}_C^\dagger$. As a consequence, if the system and control are initially uncorrelated they will remain approximately separable $\hat{\rho}_D\b t\approx\hat{\rho}_S^{s.c}\b t\otimes\hat{\rho}_C\b t$, with (condition (ii)) \cite{dann2020thermodynamically}.

\begin{figure}
    \centering
        \includegraphics[width=8cm]{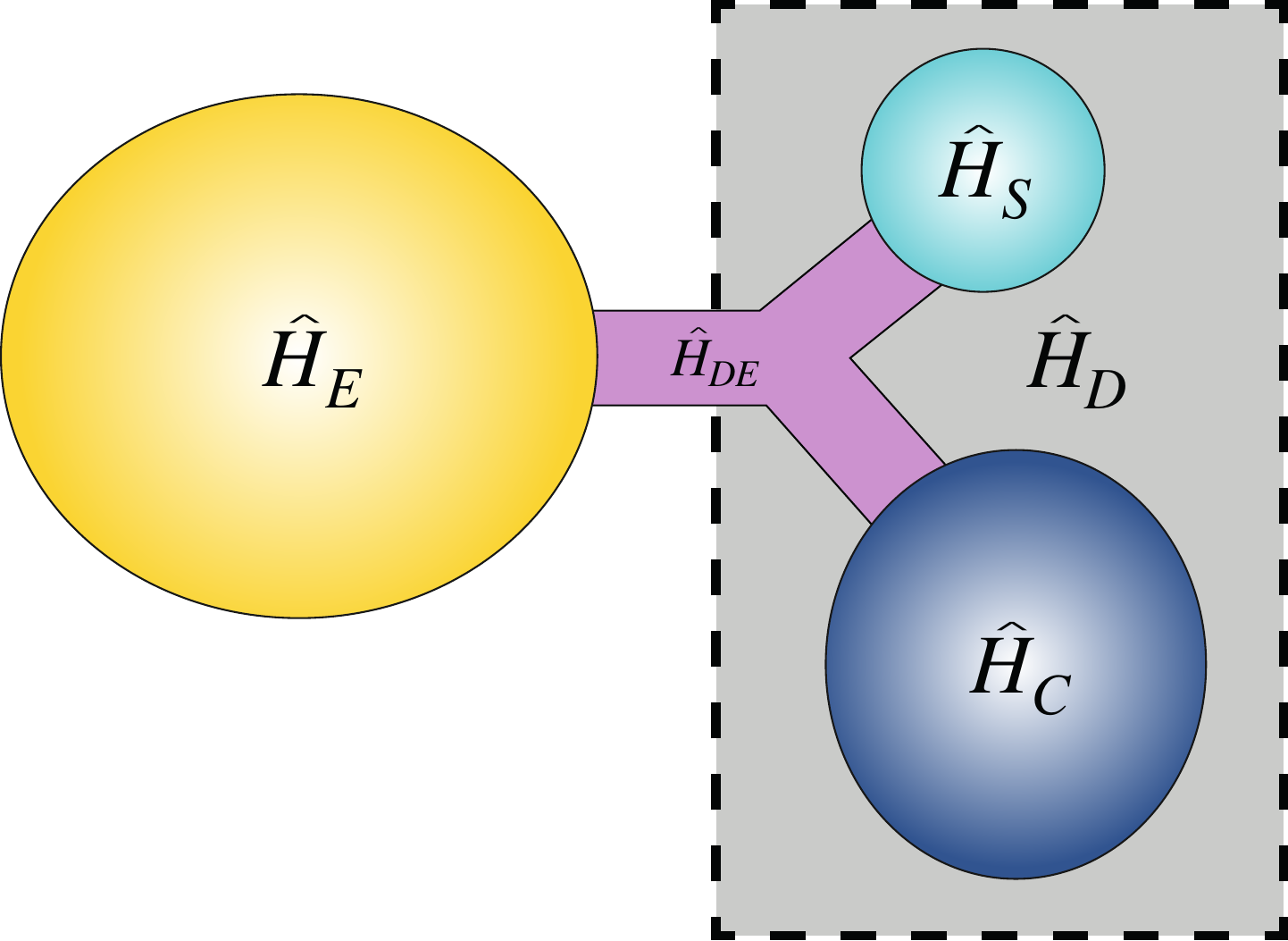}
    \includegraphics[width=8cm]{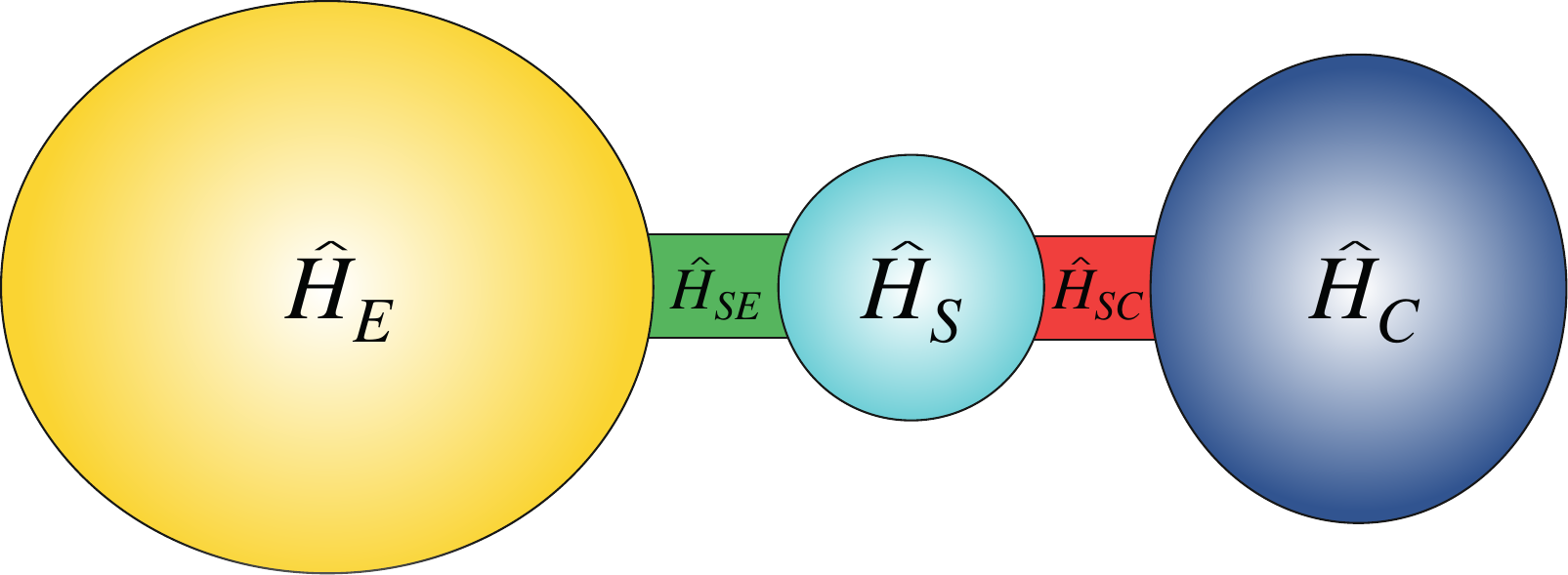}
    \caption{Interaction set ups between system, environment and  control. Top panel - Device set up: System and control form a composite system called the device (gray background) which is embedded in the environment. Strict energy conservation is satisfied between the device and the environment. Bottom panel - Tandem set up: System coupled independently to the environment  and the control. The interface between the system and environment
    and the system and control obey strict energy conservation.}
    \label{fig:scheme_2}
\end{figure}

For open system dynamics which are symmetric under time-translation, we identify two relevant interaction set ups (see Fig. \ref{fig:scheme_2}):
\begin{enumerate}
    \item The {\emph{device}} set up - the system and control constitute a ``device", which interacts with the environment via a strict energy conserving coupling. That is, the device Hamiltonian is identified as $\hat{H}_D=\hat{H}_S+\hat{H}_{SC}+\hat{H}_C$, and the interaction satisfies $\sb{\hat{H}_E+\hat{H}_D,\hat{H}_I}=0$. This set up corresponds to the so-called ``global'' approach towards constructing master equations \cite{geva1995relaxation}. 
    \item \emph{Tandem} set up - The system is coupled to both the control and environment, which do not interact with each other. In this case, the time-translation symmetry is achieved by conditions $\sb{\hat{H}_S+\hat{H}_C,\hat{H}_{SC}}=0$ and $\sb{\hat{H}_S+\hat{H}_E,\hat{H}_I}=0$. This set up is related to the ``local'' approach towards constructing master equations.
\end{enumerate}

The scenario are characterized by different time-translation symmetries. In the device set up, the dynamical map of the composite system is symmetric under time-translation: $\Lambda_{D}\circ{\cal{U}}_{D}={\cal{U}}_{D}\circ\Lambda_{D}$, where ${\cal{U}}_{D}\b t\sb{\bullet}=e^{-i\hat{H}_Dt/\hbar}\bullet e^{i\hat{H}_Dt/\hbar}$. 
In comparison, the interactions of the tandem set up imply that dynamical map of the device (including the primary system, control and the interaction between them) is time-translation symmetric with respect to the free dynamics: $\Lambda_{D}\circ{\cal{U}}_{0}={\cal{U}}_{0}\circ\Lambda_{D}$ \cite{dann2021unification}, where ${\cal U}_0\b{t}\sb{\bullet}=\hat{U}_0\b{t}\bullet\hat{U}^\dagger_0\b t$ is the free propagator (excluding all the interactions) with $\hat{U}_0\b{t}=\exp\b{-i\b{\hat{H}_S+\hat{H}_C}t/\hbar}$.
The distinct dynamical symmetries lead to different master equations. 

The connection to the semi-classical regime and a time-dependent Hamiltonian is obtained by taking the semi-classical limit \cite{dann2020thermodynamically}. This procedure includes tracing over all control degrees of freedom. 

In the device setup, the Lindblad jump operators are associated with the propagator related to $\text{tr}_C\b{\hat{U}_D}$. In the semi-classical limit, the device's free propagator can be written as a product of the free control and system's semi-classical propagators  \cite{dann2020thermodynamically} 
\begin{equation}
    \hat{U}_{D}\b t=\hat{U}_C\b t\otimes\hat{U}_{S}^{s.c}\b t~~,
    \label{eq:prop}
\end{equation}
where 
\begin{equation}
    \hat{U}_{S}^{s.c}\b t={\cal T}\exp\b{-\f{i}{\hbar}\int_{0}^t \hat{H}^{s.c}_S\b{s}ds}~~,
\end{equation}
which is generated by the semi-classical Hamiltonian $\hat{H}_S^{s.c}\b{t}=\text{tr}_C\b{\hat{H}_D\hat{\rho}_C\b t}$.
Relation \eqref{eq:prop} then leads to a master equation, which Lindblad jump operators are eigenoperators of ${\cal{U}}_S^{s.c}\sb{\bullet}=\hat{U}_S^{s.c }\bullet\hat{U}_S^{s.c\dagger }$. For further details, see Ref. \cite{dann2020thermodynamically}

The tandem set up is characterized by time-translation symmetry of the device system with respect to the free dynamics. This symmetry implies that the Lindblad jump operators of the device's master equation are eigenoperators of the free propagator. Due to the commutativity between the free Hamiltonians of the different constituents, the eigenoperators of the free propagator are composed of a product of primary system and control eigenoperators. Once the trace over the control is performed, only the eigenoperators of ${\cal{U}}_S\sb{\bullet}=\hat{U}_S\bullet \hat{U}_S^\dagger$ remain \cite{dann2021unification}.
This means that in the tandem interaction scenario, the controlled system dynamics is characterized by the same symmetry as the case with no control Eq. \eqref{eq:commutivatity_maps}. The identical symmetry infers that the master equation shares the same structure as the case of a time-independent Hamiltonian, Eq. \eqref{eq:L_time_independent}. 
Hence, the energetic transitions and dephasing occur in the system's local energy basis, and the effect of the control system is only incorporated within the kinetic coefficients.


\trr{\section{Discussion on the validity of the strict energy conservation}}

Strict energy conservation between the system and environment constitutes an idealized mathematical condition, which is associated with the neglection of any change in the interface energy. When the system is large, as is the case in traditional thermodynamics, the interface energy is discarded due to its relative minuscule contribution to the energy flows. However, in microscopic systems the interface energy (and the change in the interface energy) is comparable with the bulk energy, and therefore cannot be discarded. Hence, we do not expect strict energy conservation to hold under generic conditions, or even to be strictly valid for real physical systems. Nevertheless, when the system interacts weakly with the environment, the strict energy conservation condition effectively holds in the timescales of interest. In this regime, the condition can be viewed as an emerging effective symmetry of the dynamics.  

This claim can be understood by studying the reduced system dynamic in the weak coupling limit. The composite dynamics are governed by the Hamiltonian of Eq. \eqref{eq:hamil}, and the interaction term is of the order $||\hat{H}_I||\sim\hbar g$, where $\hbar g\ll ||\hat{H}_S||$. Up to second order in the system-environment coupling, and when the environment initially resides in a stationary state (with respect to its bare Hamiltonian) the reduced system dynamics in the interaction picture relative to the bare Hamiltonian can be expressed as 
\begin{equation}
    \f{d}{dt}\tilde{\rho}_S\b t = -\f{1}{\hbar^2}\int_0^t ds \sb{\tilde{H}_I\b{t},\sb{\tilde{H}_I\b{s},\tilde{\rho}\b s}}. 
\end{equation}
Next, we decompose the interaction Hamiltonian into the system (or device) and environment eigenoperators (transition operators between the different energy eigenstates)
$\hat{H}_I = \sum_k \hat{S}_k\otimes \hat{B}_k$. Following the standard microscopic derivation \cite{breuer2002theory}, we substitute this expression into the master equation, which leads to terms of the form $\sim g^2 e^{i\b{\omega-\omega'}s}$. Such terms contribute to the dissipative dynamics corrections which oscillate with an amplitude $\propto 1/\b{\b{\omega-\omega'}t}$, relative to the resonant terms that comply with the strict energy conservation, see for instance Ref. \cite{rivas2017refined} or \cite{winczewski2021bypassing}.
Since the coupling with the environment is weak, such oscillating corrections add small rapid oscillations on top of the dominant free evolution.  
In this regime, the rapid oscillations can be averaged over, producing an effective evolution which complies with strict energy conservation.

\section{Examples}
\label{sec:example}
\subsection{Jaynes-Cummings model in resonance}
\label{sec:2}
We start with a simple model of a qubit coupled to a single bosonic mode, 
where the Hamiltonian is given by 
\begin{align}
\hat{H}= \frac{\omega}{2} \hat{\sigma}_z\otimes \hat{I}_E + g (\hat{\sigma}_+ \otimes \hat{b} + \hat{\sigma}_- \otimes \hat{b}^\dagger ) +\omega \hat{I}_S \otimes \hat n~~,
\end{align}
with the number operator $\hat n= \hat{b}^\dagger \hat{b}$. The model is exactly solvable \cite{TN_libero_mab2)61772}. It serves as an extreme example of non-Markovian dynamics, where the combined evolution is quasi-periodic, while the reduced description can still be cast in the format of the open system dynamical symmetric structure, Eq. \eqref{eq:L_time_independent}.

For this model, the strict energy conservation (Postulate 1) requires both resonance condition and the RWA (absence of the terms $\hat{\sigma}_-\otimes \hat{b}$ and $\hat{\sigma}_+\otimes \hat{b}^\dagger$). 
If we additionally assume that the initial environmental operator commutes with the number operator $[\hat{\rho}_E(0),\hat n]$ (Postulate 2), the corresponding exact master equation takes 
the phase covariant form (interaction picture) \cite{smirne2010nakajima}
\begin{align}\label{eq:phcov}
\frac{d}{dt}\tilde\rho_S(t)=
&\gamma_+(t)\left(\hat\sigma_+\tilde\rho_S(t)\hat\sigma_--\frac{1}{2}\{\tilde\rho_S(t),\hat\sigma_-\hat\sigma_+\}\right) \nonumber\\+
&\gamma_-(t)\left(\hat\sigma_-\tilde\rho_S(t)\hat\sigma_+-\frac{1}{2}\{\tilde\rho_S(t),\hat\sigma_+\hat\sigma_-\}\right)\nonumber\\+
&\gamma_z(t)\left(\hat\sigma_z\tilde\rho_S(t)\hat\sigma_z-\tilde\rho_S(t)\right)~~,
\end{align}
where the time dependent rates read
\begin{align*}
\gamma_{\pm}(t)&=\frac{\eta_{\parallel }(t)}{2}\frac{d}{dt}\frac{1\pm r(t)}{\eta_{\parallel }(t)},\nonumber \\
\gamma_z(t)&=\frac{1}{4}\left(\frac{\dot\eta_{\parallel }(t)}{\eta_{\parallel }(t)}-2\frac{\dot\eta_{\bot }(t)}{\eta_{\bot }(t)}\right),
\end{align*}
which are defined in terms of expectation values over the environment's state
\begin{align}\label{eq:FctJC}
\eta_{\parallel }(t)&=\langle w (\hat n,t)w(\hat n,t) + w (\hat n+1,t)w(\hat n+1,t) \rangle_E-1,  \nonumber\\
r(t)&=\langle w (\hat n+1,t)w(\hat n+1,t) - w (\hat n,t)w(\hat n,t) \rangle_E, \nonumber \\
\eta_{\bot }(t)&=\langle w (\hat n,t)w(\hat n+1,t) \rangle_E, \nonumber \\
w(\hat n,t)&= \cos(g\sqrt{\hat n}t)~~,
\end{align}
with $\mean{\bullet}_E\equiv\tr_E\b{\bullet\hat{\rho}_E\b 0}$.
The quantities $|\eta_{\bot }(t)|$ and  $|\eta_{\parallel }(t)|$ describe shrinking of the Bloch ball in the $x$-$y$ plane and in the $z$-direction, respectively, and the $r(t)$ is responsible for its translation along the $z$-axis \cite{Filippov_2020}. 


Note, that the resonance condition is not necessary for a master equation of a phase covariant form \cite{smirne2010nakajima}. It can be seen by the fact that the off-resonant Hamiltonian conserves the number of excitations, Sec. \ref{sec:excitationNumber}. In the qubit case under this symmetry, the master equation obtains the same operator structure as the time-translation symmetry.

\begin{figure}
\centering
\includegraphics[width=8.cm]{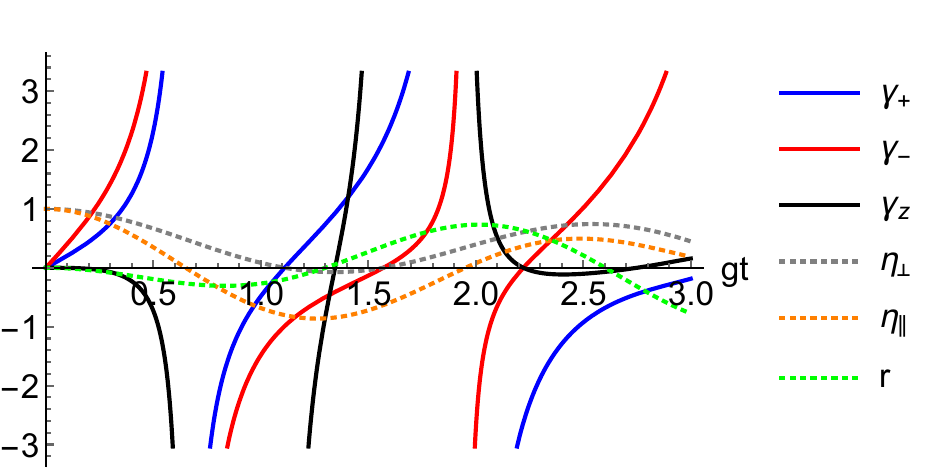}
\caption{ The time behavior of rates corresponding to the Jaynes-Cummings master equation, Eq. \eqref{eq:phcov}, and the functions $\eta_{\bot }(t)$, $\eta_{\parallel }(t)$ and $r(t)$, Eq. \eqref{eq:FctJC}, for the environment initially in the number state $\ket{1}$. All three rates $\gamma_+(t)$, $\gamma_-(t)$ and $\gamma_z(t)$ are negative for some time intervals and diverge at isolated points in time.
}
\label{fig:JC}
\end{figure}

The master equation, Eq. \eqref{eq:phcov}, is of the expected form, Eq. \eqref{eq:L_time_independent}, where the coherent part is absent due to the resonance condition. Only in the situation when $\gamma_{\nm{+}}(t)=\nu\gamma_{\nm{-}}(t)$ for a constant $\nu$,  the generator is commutative {at different times} and the associated eigenoperators are time-independent. They generally read
\begin{align}
\frac{1}{\sqrt{2}}\left\{ \hat{I} -\frac{\gamma_-(t)-\gamma_+(t)}{\gamma_-(t)+\gamma_+(t)} \hat\sigma_z, \hat\sigma_x,\hat \sigma_y, \hat\sigma_z \right\}.
\end{align}

The commutativity of the generator implies the commutativity of the dynamical map and in this case their eigenoperators coincide \cite{Teittinen_2018, megier2020interplay}. In particular, this is true for unital dynamics, i.e. when no translation along the $z$-axis takes place, $r(t)=0$. In this situation the rates corresponding to dissipation and heating are the same, $\gamma_-(t)=\gamma_+(t)$.

Additionally,
the constant ratio between $\gamma_+$ and $\gamma_-$ also occurs for an environment initially in a vacuum state, for which it follows $\nu=0$. Remarkably, in this case we get $\eta_{\bot }^2(t)=\eta_{\parallel }(t)$ which implies that also $\gamma_z(t)=0$ and only the term associated with decay of excitations to the bath remains. Note, that $\gamma_z$ does not vanish in the corresponding time non-local master equation, Eq. \eqref{eq:kernel}, where the term still appears. As a result, the operator structure is not preserved in the transition between the time-local and time non-local master equations, and one should be cautious in the interpretation of the disappearance of a term in one of the pictures \cite{smirne2010nakajima,megier2020interplay}.
The non-Markovianity is clearly
manifested by negative values $\gamma_-(t)$.
However, {note that} at some discrete points in time the time local-generator is ill-defined, as the $\gamma_-(t)$ diverges.
Interestingly, for certain choices of the initial environmental state all three rates can diverge, see Fig. \ref{fig:JC} for an example with $\rho_E(0)=\ket{1}\bra{1}$. This happens at times when one of the functions $\eta_{\bot }(t)$, $\eta_{\parallel }(t)$ vanish.

The ill defined  time-local master equation does not generally have a unique solution \cite{PhysRevA.86.042107}. Nevertheless, typically one can assign a \textit{physically well-behaved} Hamiltonian (i.e. whose change in time is not too rapid in comparison to the timescale of the reduced system) to only a single solution.
A possible approach to bypass such singularities  derives a corresponding higher order, well defined, evolution equation \cite{hegde2021open}. Surprisingly, the corresponding higher order equation contains a different operatorial structure relative to the original master equation.


\subsection{Spin-star model}
\label{sec:spin_star}

The spin star model describes the dynamics of a central spin residing within a hot bath of environmental spins \cite{gaudin1976diagonalisation,breuer2004non,ferraro2008non,arshed2010channel}. Such a system may represent a two-level system in a bulk, as a Nitrogen Vacancy (NV) center within a diamond at room temperature \cite{hall2014analytic}, or an electron spin qubit coupled to a nuclear spin bath in a GaAs quantum dot \cite{barnes2012nonperturbative}.  The model is both tractable and shows  strong non-Markovian behavior, hence constitutes a natural choice in studying non-Markovian dynamics. In the asymptotic long time limit, the populations of the energy states of the central spin exhibit a complete relaxation to equilibrium, while the coherences only partially decohere.

To be concrete, we consider $K+1$ localized spin $\f{1}{2}$ particles with an identical transition frequency. 
The central spin 
interacts with $K$ environmental spins by a Heisenberg $XY$ type interaction. The joint dynamics are generated by 
the composite Hamiltonian
\begin{equation}
    \hat{H}=\hbar \omega \hat{\sigma}_z+2\hbar g\b{\hat{\sigma}_{+}\hat{J}_{-}+\hat{\sigma}_{-}\hat{J}_{+}}+\hbar\omega\sum_{k=1}^K\hat{\sigma}^{\b{k}}_z~~,
    \label{eq:spinstar_Ham}
\end{equation}
where $\hat{\sigma}_i$ and $\hat{\sigma}_i^{\b{k}}$, $i=x,y,z$ are Pauli 
operators of the central and environmental spins, correspondingly, $\omega$ is the transition frequency, and $g$ is the coupling constant. The total spin angular momentum of the environment is denoted by $\v{J}=\f{1}{2}\sum_{k=1}^K \v{\sigma}^{\b k}$, with associated creation anhilation operators
$\hat{J}_{\pm}=\sum_{k=1}^K\hat{\sigma}^{\b{k}}_{\pm}$, where $\v{\sigma}$ is the vector of Pauli operators, $\hat \sigma_{\pm}^{\b{k}}=\b{\hat{\sigma}_x^{\b k}\pm i\hat{\sigma}_y^{\b{k}}}/2$ and similarly for the central spins. We assume the environment is initially in a fully mixed state $\hat{\rho}_E\b t=\hat{I}_E/2^{-K}$.

The present model complies with the two thermodynamics postulates: Due to the resonance condition, the interaction term  of Eq. \eqref{eq:spinstar_Ham} commutes with the free Hamiltonian $\hbar \omega \b{\hat{\sigma}_z+\sum_{k=1}^K\hat{\sigma}_z^{\b{k}}}$, and the initial environment state is stationary under the free environment dynamics. This stationary state represents a thermal bath in the high temperature regime $k_B T\gg\hbar \omega$, serving a suitable approximation for room temperature experiments (energy scale of THz) on NV-centers (GHz) and GaAs quantum dots (MHz). Overall, these conditions and the previous analysis infers that the master equation should be of the form Eq. \eqref{eq:L_time_independent}.

We verify this by comparing the predicted structure with the explicit solution.
Due to the high symmetry of $\hat{H}$ one can derive an exact solution for the reduced central spin dynamics, given by 
\begin{equation}
    \hat{\rho}_S \b {t}= \f{\hat{I}_S}{2}+
    \f{r_z\b{t}}{2}\hat{\sigma}_z+{r}_+\b{t}\hat{\sigma}_+ +{r}_-\b{t}\hat{\sigma}_-~~,
    \label{eq:spin_star_sol1}
\end{equation}
where 
\begin{gather}
\label{eq:55l}
    r_z\b{t}=\kappa_z\b{t}r_z\b{0} 
     \\
    r_{\pm}\b{t}=e^{\pm i\omega t}\kappa\b{t}r_{\pm}\b{0}~~,
    \nonumber
\end{gather}
with
\begin{gather}
      \kappa_z{\b t}=\sum_{j,m}\f{d\b{j}}{2^K}\cos\b{4h\b{j,m}gt}
    \label{eq:spin_star_sol2}\\  \kappa\b{t}=\sum_{j,m}\f{d\b{j}}{2^K}\cos\b{2h\b{j,m}gt}\cos\b{2h\b{j,-m}gt}~~.
\nonumber
\end{gather}
Here, $j\leq K/2$ and $-j\leq m\leq j$ are the angular momentum quantum numbers and the $\b{j,m}$ dependent functions are given by 
\begin{gather}
    d\b{j} =\b{\begin{array}{c}
K\\
K/2-j
\end{array}}-\b{\begin{array}{c}
K\\
K/2-j-1
\end{array}}
\label{eqap:37}\\
h\b{j,m}={\sqrt{j\b{j+1}-m\b{m-1}}}~~.
\nonumber
\end{gather}

Given the exact solution, it is straightforward to deduce the associated dynamical generator in the interaction picture relative to the free dynamics (see details in Appendix \ref{apsec:spin_star_model})
\begin{equation}
\tilde{\cal L}\b{t}\sb{\bullet}=\eta _{-}\b{t}{\cal{D}}_{-}\sb{\bullet}+\eta_{+} \b t\cal{D}_{+}\sb{\bullet}~~,
\label{eq:exact_ME}
\end{equation}
with $
    {\cal D}_{\pm}\sb{\bullet}=\hat{\sigma}_{\pm}\bullet\hat{\sigma}_\mp-\f{1}{2}\{\hat{\sigma}_{\mp}\hat{\sigma}_{\pm},\bullet\}$
and 
\begin{equation}
    \eta_{\pm}\b t=\frac{\xi_\pm\b t}{2\kappa\b t\b{r_{x}\b 0-r_{y}\b 0}}
\end{equation}
\begin{multline}
    \xi_\pm\b t =\mp2\dot{\kappa}r_{x}\b 0\b{r_{z}\b t\pm1}\\+2\dot{\kappa}r_{y}\b 0\b{r_{z}\b t\pm1}+\kappa\dot{r}_{z}\b 0\b{r_{x}\b 0-r_{y}\b 0}~~,
    \nonumber
\end{multline}
where  ${r}_x \b 0 = r_+ \b 0+{r}_ - \b 0$ and $r_y\b 0 = i \b{ r_+\b 0 - r_- \b 0}$.

Equation \eqref{eq:exact_ME} constitutes the exact dynamical generator of the central spin when the kinetic coefficients $\eta_\pm\b{t}$ are well defined. 
When these coefficients obtain negative values the map violates CP-divisibility, which indicates that the dynamics are non-Markovian. In Fig. \ref{fig:spin_star_coeff} we present the kinetic coefficients in the weak coupling regime, clearly demonstrating that the dynamics of the central spin are non-Markovian. 

In comparison to the exact expression, the Lindblad jump operators of the general structure Eq. \eqref{eq:L_time_independent} constitute the eigenoperators of the free propagator $\hat{U}_S\b{t}=\exp\b{-i\omega\hat{\sigma}_z t}$. 
Accordingly, the non-invariant unitary eigenoperator are the raising and lowering operators $\hat{F}_\pm=\hat{\sigma}_\pm$, and the invariant subspace is spanned by $\hat{\sigma}_z$ and the identity. These identifications verify that the exact solution Eq. \eqref{eq:exact_ME} complies with the general structure Eq. \eqref{eq:L_time_independent}.

  \subsubsection*{Comparison of the approximate and exact master equations}
 \label{subsec:compare}
The kinetic coefficients of the master equation can be evaluated with the perturbative treatment of Sec. \ref{sec:kinetic coefficients}.  We demonstrate this approach by calculating the approximate master equation of the spin star model, and compare it to the exact solution.

\begin{figure}
\centering
\includegraphics[width=5.5cm]{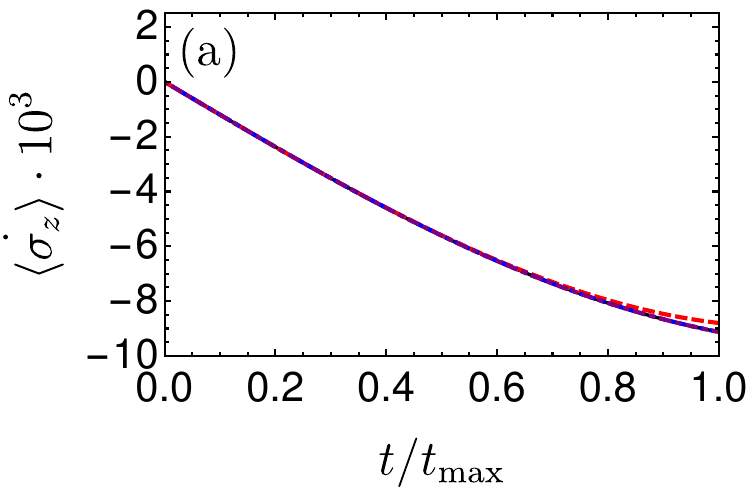}
\includegraphics[width=5.5cm]{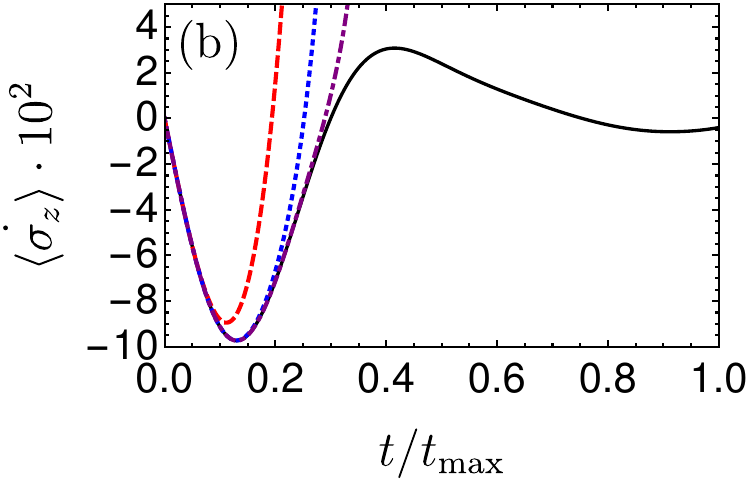}
\includegraphics[width=5.5cm]{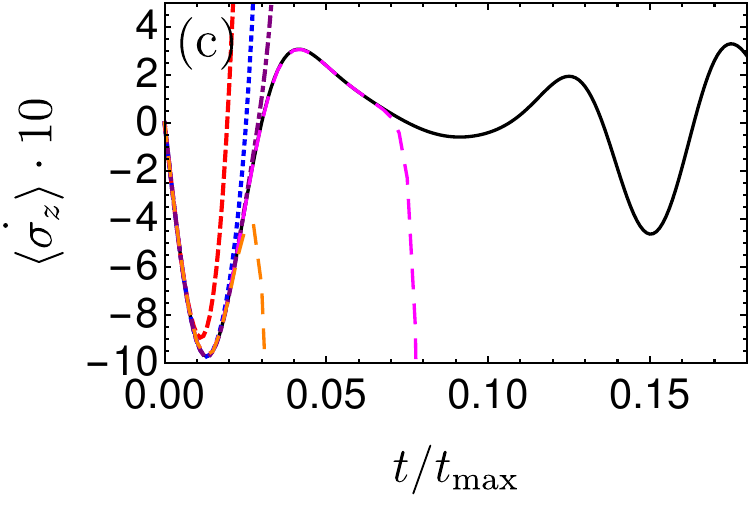}
\caption{Time derivative of $\mean{\hat{\sigma}_z}=\tr\b{\hat{\sigma}_z\hat{\rho}_S\b t}$ as a function of time for the exact solution (black continuous line) and approximate solution with various expansion orders $M$. Colored lines correspond to:  Red dashed - $M=2$, blue dotted - $M=6$ and purple dot-dashed line - $M=10$ for the Maclaurin series. A solution using a Chebychev series of order $M=10$ (orange dashed) and $M=38$ (dashed magenta) are represented for the strong coupling case. The different panels correspond to different interaction strengths (a) Weak interaction regime: $g=0.01$ (b) 
intermediate interaction: $g=0.1$ (c) Strong interaction $g=1$. The quality of the Maclaurin series is determined by the coupling strength and expansion order. For very weak interaction strength (a) small expansion order is sufficient, however, with increase of $g$, Panels (b) and (c), $M$ must be increased as well to insure accurate kinetic coefficients. The Chebychev series remains precise over its entire validity range, and deviates rapidly for large $t$.  The behaviour of $\f{d}{dt}{\mean{{\hat{\sigma}_x}}}$ is  qualitatively similar to the presented result, while $\f{d}{dt}{\mean{\hat{\sigma}_y}}$ vanishes due to the chosen initial central spin state.
Model parameters (in arbitrary units): $\omega = 2$, $r_z\b 0 = 0.3$, $r_-\b{0}=r_+\b{0}=0.1$, $K=10$ and $t_{\text{max}}=10$.}
\label{fig:spin_star}
\end{figure}

\begin{figure}
\centering
\includegraphics[width=5.5cm]{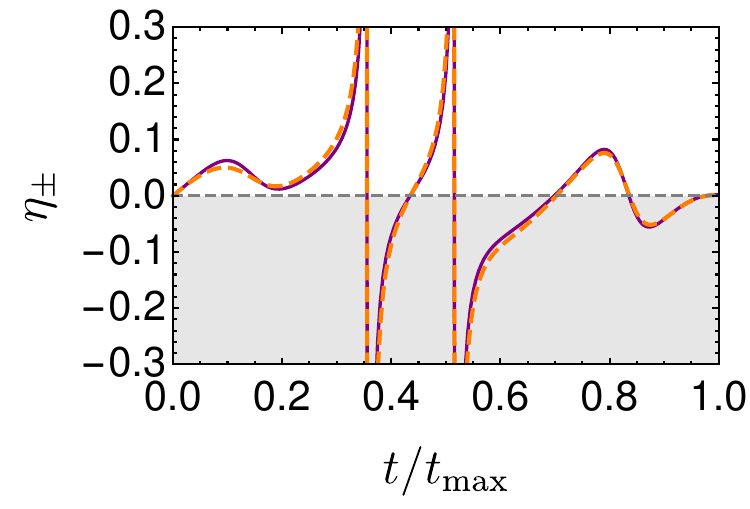}
\caption{ Exact kinetic coefficients of the spin star model as a function of the normalized time in the weak interaction regime. The continuous purple and dashed orange lines correspond to the kinetic coefficients $\eta_-$ and $\eta_+$, respectively. The coefficients characterize the dissipation and pure dephasing rates. The approximate kinetic coefficients obtained from the Maclaurin and Chebychev series, follow a similar behaviour in the regime where the approximation holds.  Model parameters are the same as in Fig. \ref{fig:spin_star}, with $g=0.01$ and $t_{\text{max}} = 200$. }
\label{fig:spin_star_coeff}
\end{figure}

As expected, the Maclaurin series produces accurate results in the vicinity of the origin $t=0$, see Fig. \ref{fig:spin_star}. The accuracy improves with the reduction of the interaction strength $g$. Nevertheless, for sufficient polynomial order $M$, the master equation can capture the dynamics under strong interactions.

In comparison, the Chebychev series enjoys a `global' accuracy, within its convergence range, see Fig. \ref{fig:spin_star} Panel (c).

\subsection{Spin-boson bath model conserving the total number of excitations}
\label{sec:num_cons_example}
The spin-boson model is a central model in the field of open quantum systems. It showcases both Markovian and non-Markovian open system dynamics in the short and long time regimes. In the following section, we demonstrate how the kinetic coefficients of Eq. \eqref{eq:num_gen} can be determined, allowing to completely characterize the spin dynamics in the weak coupling limit under the conservation of the total number of excitations. The demonstration involves a slight adjustment of a recent work of Rivas \cite{rivas2017refined}. 

We consider a spin interacting with a bosonic bath via an excitation conserving interaction. Such a scenario is represented by the following Hamiltonian 
\begin{equation}
    \hat{H}^{\b{N}} = \f{\hbar \omega_0}{2}\hat{\sigma}_z  +\sum_k \hbar\omega_k \hat{b}^\dagger_k\hat{b}_k + \sum_k g_k\b{ \hat{\sigma}_-\hat{b}^\dagger_k+\hat{\sigma}_+\hat{b}_k}~~.
    \label{eq:68num}
\end{equation}
Note, that this Hamiltonian can be also obtained by applying the RWA to the standard dipole approximation interaction \cite{scully1999quantum}.

The reduced dynamics can generally be expressed in terms of a cumulant expansion
\begin{equation}
    \tilde{\rho}_S\b t = \Lambda\sb{\hat{\rho}_S\b 0} = e^{{\cal Z}\b t}\sb{\hat{\rho}_S\b 0}~~,
    \label{eq:69num}
\end{equation}
where the exponent can be written as a sum of cumulants ${\cal Z}\b t=\sum_{i=1}^{\infty}K^{\b{i}}\b t$, where $K^{\b{i}}$ is of $i$-th order in interaction strength \cite{alicki1989master}. For a detailed derivation of the open system cumulants see Ref. \cite{majenz2013coarse} Sec. II B.
The symmetry associated with the conservation of the number of particles ($\sb{\hat{N},\hat{H}^{\b N}}=0$) restricts the form of the map and the exponent ${\cal Z}\b t$. In section \ref{sec:generalization to other symmetries} we showed that representation of the open system map in Liouville space must acquire a block diagonal form, in the operator basis $\{\ket{\eps_n}\bra{\eps_m}\}$, where each block is related to a different value of the excitation number $l$. The representation of the exponent ${\cal Z}$ has an identical structure, implying that ${\cal Z}$ is of the form of ${\cal L}^{\b N}$, Eq. \eqref{eq:num_gen}.
Namely, for the spin case this form has the same operatorial structure as the strict energy conservation case (Eq. \eqref{eq:phcov})
\begin{multline}
{\cal Z}\b{t}\sb{\bullet}= - \f{i}{\hbar} \sb{\bar{H}^{\b{N}},\bullet}\\+ \trr{\Gamma_+^{\b{N}}}\b{t}{\cal{D}}_+\sb{\bullet} +\trr{\Gamma_-^{\b{N}}}(t){\cal{D}}_-\sb{\bullet}+\trr{\Gamma_z^{\b{N}}}(t){\cal D}_z\sb{\bullet}~~,
\label{eq:rho70}
\end{multline}
where the kinetic coefficients $\{\Gamma^{\b{N}}\}$ are real (due to the hermitiacy preserving property of the map), and $\bar{H}^{\b{N}}$ is proportionate to $\hat{\sigma}_z$ from the trace preserving property of the map. We emphasis that the operatorial structure of Eq. \eqref{eq:rho70} is valid for arbitrary system-environment coupling and bath size, it describes the exact reduced dynamics generated by Hamiltoninan Eq. \eqref{eq:68num}  for an initial stationary environment state of arbitrary size. 

To derive the form of the dynamical generator we express the dynamical generator in the interaction picture in terms of the map's exponent
\begin{multline}
    \tilde{\cal L}^{\b {N}}\b t\sb{\tilde{\rho}_S\b t} \equiv
    \f{d}{dt}\tilde{\rho}_S\b t\\=\b{\f{d}{dt}e^{{\cal Z}\b t}}\tilde{\rho}_S\b 0\\ = \sb{\b{\f{d}{dt}e^{{\cal Z}\b t}}e^{-{\cal Z}\b t}}\tilde{\rho}_S\b t~~.
\end{multline}
This relation identifies the generator with the term in square brackets. Finally, by utilizing the identity $\f d{dt}\sb{e^{{\cal Z}\b t}}=\int_0^1{ds\,e^{s{\cal Z}\b t}\sb{\f{d{\cal Z}\b t}{dt}}e^{\b{1-s}{\cal Z}\b t}ds} $ \cite{snider1964perturbation,wilcox1967exponential} the generator becomes  
\begin{equation}
\tilde{\cal L}^{\b N}\b t=\int_0^1{ds\,e^{s{\cal Z}\b t}\sb{\f{d{\cal Z}\b t}{dt}}e^{-s{\cal Z}\b t}ds}
\label{eq:72}
\end{equation}
For a closed algebra of super-operators the expression can be solved explicitly.

In order evaluate the kinetic coefficients of the master equation, we consider a weak system-environment coupling ($g_k\ll 1$). This assumption allows truncating the cumulant expansion after second order, leading to 
\begin{multline}
    {\cal Z}\b t\approx K^{\b{2}}\b t \\=-\f 12{\cal T}\int_0^t{\int_0^t{dt_{1}}dt_{2}}\,\text{tr}_{E}\b{\sb{\tilde{H}_I\b{t_{1}},\sb{\tilde{H}_I\b{t_{2}},\hat{\rho}\b 0}}}~~,
    \label{eq:uuu}
\end{multline}
where $\cal T$ is anti-chronological time-ordering operator, $\hat{H}_I$ is the interaction term of Eq. \eqref{eq:68num}. 
Here the first order cumulant vanishes due to the $\text{tr}_E\b{\hat{b}_k\hat{\rho}_E\b 0}=0$ for a stationary initial state (this is also true for any odd  order environment correlation function, implying the present result is accurate up to fourth order in the coupling strength).

\begin{figure}
\centering
\includegraphics[width=7.5cm]{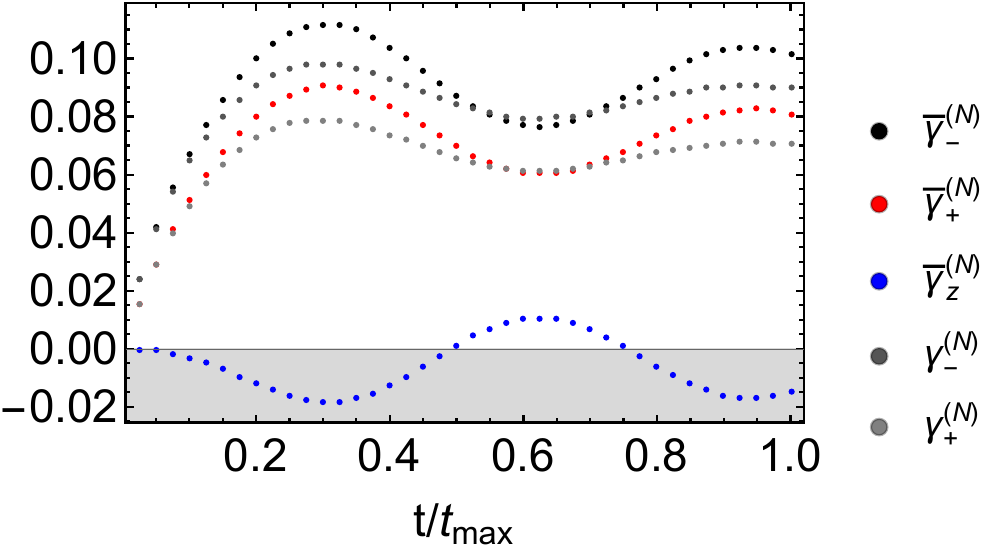}
\caption{ \trr{Kinetic coefficients of the spin boson model
as a function of the normalized time. The bath is taken to be thermal with an Ohmic spectral density function $J\b{\omega}=\alpha\omega e^{-\omega/\omega_{c.o}}$, where $\omega_{c.o}$ is the cut-off frequency (see Eqs. \eqref{eq:H7} and \eqref{eqap:2225} in Appendix \ref{apsec:num_model} for explicit expressions of the kinetic coefficients). The black, red, blue, dark grey and grey dotes correspond to the kinetic coefficients $\bar\gamma^{\b{N}}_-$, $\bar\gamma^{\b{N}}_+$,  $\bar\gamma^{\b{N}}_z$,
$\gamma^{\b{N}}_-$ and $\gamma^{\b{N}}_+$, respectively, as defined in Eq. \eqref{eq:rho4Order} and Eq. \eqref{eq:rho73}. Model parameters are: $\omega_{0} = 0.5$, $\omega_{c.o} = 1$, $\alpha=0.01$, $t_{\text{max}}=20$ and the reservoir bath is at $T=2 /{k_B}$.}}
\label{fig:spin_boson_coeff}
\end{figure} 

Schaller and Brandes \cite{PhysRevA.78.022106}, showed that  $K^{\b{2}}\b t$ has a GKLS form for all $t$, and accordingly, the approximated dynamical map, Eq. \eqref{eq:69num} is CPTP for all $t$ (also for arbitrary small $t$). This is the strength of this approach, since in many other derivations in the weak coupling limit,  at short times the positivity of the map is lost. A well known example of such a case is the Redfield equation \cite{Duemcke}, which nevertheless, does not affect its accuracy at later times \cite{PhysRevA.101.012103}. 
Another notable exception to the evolution equation retaining positivity of the evolved density operator for all times in weak-coupling regime has been obtained by employing a correlation picture approach \cite{alipour2020correlation}.

We proceed by substituting the explicit form of the interaction into Eq. \eqref{eq:uuu} and employ the symmetry considerations to obtain the second order approximations for $\trr{\{\Gamma^{\b N}\}}$ in Eq. \eqref{eq:rho70}. The kinetic coefficient of ${\cal D }_z$ vanishes, $\trr{\Gamma_z^{\b N}} = 0$, and $\trr{\Gamma_\pm^{\b N}}$ are given in Appendix {\ref{apsec:num_model}}.

Finally, by substituting the explicit form of ${\cal Z}\b t$ into Eq. \eqref{eq:72} and utilizing the commutation relations $\sb{{\cal{D}}_+,{\cal D}_-}={\cal D}_+-{\cal D}_- $, $\sb{{\cal{D}}_z',{\cal D}_\pm}=0$, where ${\cal{D}}_z'\sb{\bullet}=\sb{\hat{\sigma}_z,\bullet}$,  we obtain the non-Markovian generator
\begin{multline}
\tilde{\cal L}^{\b{N}}\b{t}\sb{\bullet}= - \f{i}{\hbar} \sb{\Upsilon^{\b N} \b t,\bullet}\\+ \gamma_+^{\b{N}}\b{t}{\cal{D}}_+\sb{\bullet} +\gamma_-^{\b{N}}(t){\cal{D}}_-\sb{\bullet}~~,
\label{eq:rho73}
\end{multline}
where $\Upsilon^{\b N}\b t$ and $\gamma_\pm^{\b{N}}\b t$ are given explicitly in the Appendix {\ref{apsec:num_model}}. Figure \ref{fig:spin_boson_coeff} presents the kinetic coefficients for an Ohmic thermal bath. 

\trr{We emphasis that beyond the weak coupling regime the generator maintains a similar structure which is still compatible with the symmetry restriction. Interestingly, a non-vanishing pure dephasing term ${\cal D }_z$ emerges, see Eq. \eqref{eq:rho70}. In addition, kinetic coefficients are slightly modified by the higher order correction
\begin{multline}
\tilde{\cal L}^{\b{N}}\b{t}\sb{\bullet}= - \f{i}{\hbar} \sb{\bar\Upsilon^{\b N} \b t,\bullet}+\bar\gamma_z^{\b{N}}(t){\cal{D}}_z\sb{\bullet}\\+ \bar\gamma_+^{\b{N}}\b{t}{\cal{D}}_+\sb{\bullet} +\bar\gamma_-^{\b{N}}(t){\cal{D}}_-\sb{\bullet}~~,
\label{eq:rho4Order}
\end{multline}
see Figure \ref{fig:spin_boson_coeff} for an example.
Further details regarding the cumulant expansion up to fourth order in the  interaction strength appears in Appendix \ref{asec:4Order}. Note, that the a similar emergence of pure dephasing at higher orders can be observed by directly expanding the generator \cite{smirne2010nakajima}}.

Imposing particle conservation on an exactly solvable model of
a qubit couple to a spin chain \cite{yao2020probing} will lead to the same structure of  Eq. \eqref{eq:rho73} with different kinetic coefficients. In both cases
$\tilde{\cal L}^{\b{N}}\b{t}$ exhibits highly non-Markovian behaviour at short times, while reproducing the Markvoian result in the long time limit, see Appendix \ref{asec:Mlimit}. Such a transition  is completely incorporated in the time-dependence of the kinetic coefficients, while the operatorial structure remains unmodified.

 

\section{Discussion}
\label{sec:discussion}
In the last decade the use of dynamical equations for open systems has increased significantly. 
Advancements in quantum technology has led to the development of novel experimental platforms, designed to minimize noise and decoherence. The design of these experiments requires accurate theoretical modelling which includes the environmental influence. Such simulations commonly rely on master equations.

Precise characterization of the evolution of an open quantum system is a hard task.
Many approaches have been perused in order to construct the desired dynamical equations of motion. These approaches may at times produce varying results and conflicting physical predictions, especially when the underlying assumptions and approximations are not critically questioned for the particular system. Such inconsistency can cause confusion and uphold the scientific advancement. We chose to study the general structure of the dynamical equations employing an axiomatic treatment and symmetry considerations to layout a clear picture.

The analysis framework considers a macroscopic viewpoint which assumes the dynamics of the entire universe is unitary. In addition, the evolution is generated by a constant total Hamiltonian, which leads to time-reversal symmetric evolution and conservation of all the energy moments.

In order to achieve a local description, the isolated universe is partitioned to the system of interest and environment. At initial time, it is assumed that the environment is in a stationary state and uncorrelated with the system.
These assumptions allow formulating a general formal form for the dynamical map, however, the detailed reduced description still remains out of reach.
In order to proceed we introduce a thermodynamically motivated symmetry consideration, and consider a system-environment interaction that satisfies strict energy conservation. This property implies time-translation symmetry of the reduced system dynamics.  
The crucial symmetry allowed developing the general structure of the reduced dynamical equations, without imposing the Markovian (CP-divisibility/semi-group property) or weak coupling
conditions. 

The open system dynamical symmetric structure, Eq. \eqref{eq:L_time_independent}, is similar to the GKLS Markovian master equation. The two equations differ by their kinetic coefficients, which are time-dependent and may be negative in the non-Markovian case. 
Interestingly, the operative form of the two master equations is the same, as the dynamical symmetry considerations dictate in both cases that the jump operators constitute eigenoperators of the free dynamical map \cite{dann2021open,dann2020thermodynamically}. The different roles associated with the two parts of the master equation, the operative structure and kinetic coefficients, motivate the following physical interpretation of
 Eq. \eqref{eq:L_time_independent}. 
 The operator structure is associated with the symmetry of the dynamics, dictating the possible transitions, which is related to thermodynamic considerations. Conversely, the kinetic coefficients contain all the details related to specific timescales of the system and environment, and therefore are associated with the kinematics.

The influence of the state's history is the main distinction between Markovian and non-Markovian dynamics. The memory is related to the non-instantaneous response of the environment to the interaction with the system. 
One of the main consequences of the formal structure is that it reveals that memory effects can be completely captured in terms of scalar functions, the kinetic coefficients. Moreover, these coefficients and corresponding memory effects can be classified according to the associated system's Bohr frequency.


In order to evaluate the dynamical symmetric structure in the proper context, it is beneficial to compare it to other known methods.
There are a number possible techniques to obtain the master equation. 
A stringent experimental procedure applies process tomography to fully characterize the open system dynamical map \cite{chuang1997prescription,poyatos1997complete,mitchell2003diagnosis,o2004quantum}. This procedure produces accurate results and incorporates all significant noisy effects. However, the 
method is limited due a typical scaling of $\sim N^4$, where $N$ is the Hilbert space size.

A theoretical alternative employs a first principle derivation to obtain the master equation \cite{davies1974markovian,nakajima1958quantum,zwanzig1960ensemble,wangsness1953dynamical,redfield1957theory,fano1957description,winczewski2021bypassing,yao2020probing, alipour2020correlation}. The 
drawback of this approach is that only rarely can one solve for the equation
under realistic experimental conditions. The difficulty is that the environment requires an explicit description. This can be obtained for idealized cases, such as, a linear boson bath and certain spin baths. Moreover, different derivations may lead to contradicting physical predictions.

A third pragmatic popular approach is to guess a master equation, based on simpler building blocks. The most widely used methodology employs the GKLS framework, adopting the Lindbladian to the specific physical scenario. This approach is simple and modular which explains its popularity in building models of open quantum systems.
The main drawback is that the ensued dynamics can violate physical principles \cite{hofer2017markovian,levy2014local}.
In contrast, the approach we present is motivated by thermodynamic principles which can be cast in terms of symmetry relations. This allows obtaining a dynamical description in modular fashion, while maintaining the symmetry considerations.  The analysis leading to the symmetric structure can be used to justify the practical heuristic approaches utilized to model experiments.
Alternatively, the framework can be inverted, obtaining from the experimental data the kinetic coefficients and high order environment correlation functions.

The current derivation is general and therefore can be extended to other dynamical symmetries,
which are associated to other conservation laws \cite{noether1971invariant}, see Sec. \ref{sec:generalization to other symmetries}. 
We studied various dynamical symmetries for driven and controlled systems in Sec. \ref{sec:td_Hamiltonians}. In this case the generator can be obtained even when the control timescale is comparable to the environment dynamical timescale. The approach can enable quantum open system control beyond the Markovian limit. Dynamical decoupling methods fall within this operational regime.

The present paper is in the line of generating a dynamical theory of quantum thermodynamics. Despite its name, traditional thermodynamics is not concerned with dynamics. It is a theory that classifies the possible transitions between equilibrium states \cite{giles2016mathematical,lieb1999physics}. Quantum thermodynamic resource theory extends this approach to the quantum regime, obtaining a partial order between single quantum states and the possible transitions \cite{horodecki2003reversible,horodecki2013fundamental,lostaglio2019introductory}. In analogy with the resource theory approach, we develop a thermodynamically motivated axiomatic approach to quantum dynamics of open systems. 
One of the major elements of both formulations, are the strict partitions between subsystems. Such partition is manifested by the strict energy conservation or the time-translation symmetry. In the present study, we show that non-Markovian dynamical equations can be derived by employing similar initial postulates. Therefore, the present approach can be considered as a dynamical extension to quantum resource theory.

In the thermodynamic context, the dynamical framework allows explicitly evaluating the power and heat flow. This may give a new insight on the quantum dynamical version of the first law, beyond the adiabatic or Markovian limits. We find
that in the presence of an external drive different partitions associated with different dynamical symmetries lead to varying master equations, see
Sec. \ref{sec:td_Hamiltonians}, and as a consequence, to different decompositions of the first law of thermodynamics \cite{dann2021unification}.

\begin{acknowledgements}
We thank Peter Salamon, James Nulton, Justin Sheek,
Erez Boukobza, Raam Uzdin, Anton Trushechkin and Andrea Smirne for insightful discussions. RD and RK acknowledge the support of the Adams Fellowship  Program of the Israel Academy of Sciences and Humanities, the National Science Foundation under Grant No. NSF PHY-1748958 and The Israel Science Foundation Grant No.  2244/14. NM acknowledges funding by the Alexander von Humboldt Foundation in form of a Feodor-Lynen Fellowship and by the UniMi Transition Grant H2020.
\end{acknowledgements}
 
 \appendix
 
\section{Commutation properties of dynamical maps and generators }

\label{apsec:proofA}
The dynamical map associated with the composite dynamics and the map of the isolated system dynamics satisfy the following theorem (a similar proof was given in Ref. \cite{dann2021open} we present it here for the sake of completeness).
\paragraph*{Theorem 1} 
Let $\hat{H}$ be the time-independent Hamiltonian of the composite system Eq. \eqref{eq:hamil},
with $\sb{\hat{H}_{SE},\hat{H}_{S}+\hat{H}_{E}}=0$, and let the initial state $\hat{\rho}_E\b 0$ be a stationary state of  $\hat{H}_E$
 then  the dynamical maps $\Lambda$, Eq. \eqref{eq:dynamical_map} and  ${\cal U}_{S}\sb{\hat{\rho}_{S}\b 0}=\hat{U}_{S}\b{t,0}\hat{\rho}_{S}\b 0\hat{U}_{S}^{\dagger}\b{t,0}$ commute, where $\hat{U}_S\b{t,0}=e^{-i \hat{H}_S t/\hbar}$ is the free propagator of the system and $\hat{U}\b{t,0}=e^{-i\hat{H}t/\hbar}$.
\paragraph*{Proof} 
We first introduce some notations: The free propagators of the environment and composite (uncoupled) system are given by $\hat{U}_E\b{t,0}=e^{-i\hat{H}_E t/\hbar}$ and 
$\hat{U}_{SE}\b{t,0}=e^{-i\b{\hat{H}_S+\hat{H}_E} t/\hbar}$, moreover, the spectral decomposition of the environment Hamiltonian reads $\hat{H}_E =\sum_i c_i \ket{\chi_i} \bra{\chi_i}$.
Since the initial state of the environment is stationary with respect to $\hat{H}_E$, it can also be expressed as $\hat{\rho}_E\b 0=\sum_{i}\lam_{i}\ket{\chi_{i}}\bra{\chi_{i}}$. To simplify the notation, in this proof we emit the time-dependence of the propagators and maps, nevertheless, it should be clear that they induce a time translation from initial time ($t'=0$) to time $t'=t$.

Utilizing the spectral decomposition of the environment's initial state any quantum dynamical map can be expressed in a Kraus form \cite{kraus1971general}
\begin{equation}
    \hat \rho_{S}\b t=\sum_{ij}\hat{K}_{ij}\hat \rho_{S}\b 0 \hat K_{ij}^{\dagger}~~, 
    \label{eq:Kraus_form}
\end{equation}
where $\hat{K}_{ij}=\sqrt{\lam_{i}}\bra{\chi_{j}}\hat U\b{t,0}\ket{\chi_{i}}$ with $\sum_{ij}\hat{K}_{ij}^\dagger \hat{K}_{ij}=\hat{I}_S$. In the Heisenberg representation the dynamical map becomes $\hat{O}^H_S\b t ={\Lambda}^*\sb{\hat{O}_S} = \sum_{ij}\hat{K}_{ij}^{\dagger}\hat O_{S}\b 0 \hat K_{ij} $, where the superscript $H$ and asterisk designate operators and superoperators (dynamical maps) in the Heisenberg representation and $\hat{O}_S$ is a system operator. 

Using the Kraus representation the product of dynamical maps is explicitly expressed as
\begin{multline}
    {\cal U}_{S}^*\sb{{{\Lambda}}^{*}\sb{\hat O_{S}}}=\hat U_{S}^{\dagger}\b{\sum_{ij}\hat K_{ij}^{\dagger}\hat O_{S}\hat K_{ij}}\hat U_{S}\\=\sum_{i}\lam_{i}\bra{\chi_{i}}\hat U_{S}^{\dagger}\hat U^{\dagger}\hat O_{S}\sum_{j}\ket{\chi_{j}}\bra{\chi_{j}}\hat U \hat U_{S}\ket{\chi_{i}}\\=\sum_{i}\lam_{i}\bra{\chi_{i}}\hat U_{S}^{\dagger}\hat U^{\dagger}\hat O_{S}\hat U \hat U_{S}\ket{\chi_{i}}~~,
\end{multline}
where the second equality is achieved by identifying the environment identity operator $\hat{I}_E = \sum_{j}\ket{\chi_{j}}\bra{\chi_{j}}$. Inserting the identity operator $\hat{U}_E\hat U_E^\dagger = \hat{I}_E $ twice, we obtain 
\begin{multline}
     {\cal U}_{S}^*\sb{ {\Lambda}^{*}\sb{\hat O_{S}}}=\sum_{i}\lam_{i}\bra{\chi_{i}}\hat U_{E}\hat{U}_{SE}^{\dagger}\hat U^{\dagger}\hat O_{S}\hat U\hat U_{SE}\hat U_{E}^{\dagger}\ket{\chi_{i}}\\
     =\sum_{i}\lam_{i}\bra{\chi_{i}}\hat{U}_{SE}^{\dagger}\hat U^{\dagger}\hat O_{S}\hat U\hat U_{SE}\ket{\chi_{i}}~~.
\end{multline}
 The second equality is obtained by utilizing the eigenvalue equation  $\hat{U}_E\ket{\chi_i} = e^{-i c_i t/\hbar}\ket{\chi_i}$ for the eigenstates $\{\ket{\chi_i}\}$. Next, strict energy conservation implies that $\sb{\hat{H},\hat{H}_S+\hat{H}_E}=0$, which in turn suggests that the associated propagators satisfy $\sb{\hat{U},\hat{U}_{SE}}=0$. This relation leads to the final form 
 \begin{equation}
     {\cal U}_{S}^*\sb{{ {\Lambda}^{*}}\sb{\hat O_{S}}}  =\sum_{i}\lam_{i}\bra{\chi_{i}}\hat U^{\dagger}\hat{U}_{SE}^{\dagger}\hat O_{S}\hat U_{SE}\hat U\ket{\chi_{i}}~~.
     \label{eq:direct_product}
 \end{equation}
Following a similar derivation the product in reverse order of the dynamical maps gives
\begin{multline}
    {\Lambda}^{*}\sb{{{\cal U}}_{S}^*\sb{\hat O_{S}}}=\sum_{ij}\hat K_{ij}^{\dagger}\hat U_{S}^{\dagger}\hat O_{S}\hat U_{S}\hat K_{ij}\\ =\sum_{i}\lam_{i}\bra{\chi_{i}}\hat U^{\dagger}\hat U_{SE}^{\dagger}\hat U_{E}\hat O_{S}\hat U_{E}^{\dagger}\hat U_{SE}U\ket{\chi_{i}}\\ =\sum_{i}\lam_{i}\bra{\chi_{i}}\hat U^{\dagger}\hat U_{SE}^{\dagger}\hat O_{S}\hat U_{SE}U\ket{\chi_{i}}~~,
    \label{eq:reverse_product}
\end{multline}
where the last equality stems from the commutativity of local operators of the  system and environment $\sb{\hat{U}_E,\hat{O}_S} = 0$. 

Finally, Eqs. \eqref{eq:direct_product} and \eqref{eq:reverse_product} imply the desired result 
\begin{equation}
    {\cal U}_{S}^*\sb{{{\Lambda}^{*}}\sb{\hat O_{S}}}={\Lambda}^{*}\sb{{\cal {\cal U}}_{S}^*\sb{\hat O_{S}}}~~_\blacksquare
\end{equation}
From the equivalence of the Schr\"odinger and Heisenberg representations we can infer that $\Lambda$ and ${\cal{U}}_S$ commute.

 \section{Structure of the dynamical generator}
 \label{apsec:dynamical_map_properties}
 In Sec. \ref{sec:gen_form} we utilize Lemma 2.2 of Ref. \cite{gorini1976completely}, the lemma states that:\\
 {\emph{Lemma}} Let $\cal L$ be a linear operator $M\b{N}\ra M\b N$, where $M\b N$ denotes the $C^{*}$ algebra of the $N\times N$ complex matrices,
and let $\{\hat{V}_{\alpha}\}$ with $\alpha=1,2,\dots,N^{2}$ be a complete orthonormal set in $M\b N$, viz. 
$\b{\hat{V}_{\alpha},\hat{V}_{\beta}}\equiv\text{tr}\b{\hat{V}_{\alpha}^{*}\hat{V}_{\beta}}=\delta_{\alpha\beta}$. Then 
$\cal L$ can be uniquely written in the form 
\begin{equation}
\sum_{\alpha,\beta=1}^{N^{2}}v_{\alpha\beta}\hat{V}_{\alpha}\hat A\hat{V}_{\beta}^{\dagger}  ~~, 
\end{equation}
where $\hat{A}\in M\b N$. In addition, if ${\cal{L}}\b{\hat A^{\dagger}}=\b{{\cal L} \hat A}^{\dagger}$, then $v_{\alpha\beta}=v_{\beta\alpha}^*$.
 
 \section{Explicit derivation of unitary invariant and non-invariant conditions}
 \label{apsec:conditions}
 In the following section we derive the restricted form of the master equation which complies with unitary invariant and non-invariant relations (Eqs. \eqref{eq:non-invariant_eig} and \eqref{eq:invariant_eig})
 \begin{equation}
     \Lambda\b {t}\sb{\hat{F}_{\alpha}} = \lam_{\alpha}\b t\hat{F}_{\alpha}~~
    \label{apeq:non-invariant_eig}
\end{equation}
 and
 \begin{equation}
      \Lambda\b t\sb{\hat{\Pi}_j} = \sum_i \mu_{ji}\b{t} \hat{\Pi}_{i}.~~
         \label{apeq:invariant_eig}
\end{equation}
Since $\Lambda$ is assumed to be invertable, Eq. \eqref{eq:lambda_diff} implies that the dynamical generator $\cal L$ satisfies analogous conditions.

Consider the general form of a linear map (Eq. \eqref{eq:linear_decomp_2}) in the $\{\hat{S}\}$  operators basis (Eq. \eqref{eq:S_basis})
\begin{equation}
    {\cal L}\b{t}\sb{\bullet}=\sum_{i,j,k,l = 1}^{N} c_{ijkl}\b{t}\ket{i}\bra{j} \bullet \ket {k}\bra{l}~~,
    \label{apeq:linear_decomp_2}
\end{equation}
where we expressed the basis operators explicitly in terms of system's energy eigenstates and used a double index notation for the coefficients ($\alpha\ra i,j$ and $\beta\ra l,k $).
Condition Eq. \eqref{apeq:linear_decomp_2} implies that  \ref{apeq:non-invariant_eig} now infers that
\begin{equation}
    \text{tr}\b{\hat{F}_{n'm'}^\dagger{\cal{L}}\sb{\hat{F}_{nm}}}\propto\delta_{nn'}\delta_{mm'} ~~, 
\end{equation}
which implies that the coefficients of the mixed terms $c_{iikl}=c_{klii}=0$ for $i,k,l=1,\dots N$ and $k\neq l$. Similarly, condition \eqref{apeq:invariant_eig} infers that the coefficients $c_{ikkl}=0$ for $i\neq l$.
These conditions leads to the following structure (equivalent to Eq. \eqref{eq:17}) 
 \begin{multline}
    {\cal L}\b{t} \sb{\ket{n}\bra{m}} = \\
   \sum_{a\neq b}^{N} c_{abba}\b{t} \ket{a}\langle{b}|n\rangle\langle{m}|b\rangle\bra{a}
    +\sum_{i,j=1}^{N}  c_{iijj}\b{t} \ket{i}\bra{i} n\rangle \bra{m}j\rangle\bra{j}~~.
    \label{eq:C5t}
\end{multline}

 
  
For $n\neq m$, $\ket{n}\bra{m}=\hat{F}_{nm}$ is a unitary non-invariant eigenoperator, and Eq. \eqref{eq:C5t} reduces to
\begin{multline}\label{eq:Cond1}
   {\cal L}\b{t} \sb{\hat{F}_{nm}}=\\ \sum_{a\neq b}^{N}c_{abba}\b{\delta_{bn}\delta_{mb}\ket a\bra a}+\sum_{k,l=1}^{N}d_{kkll}\delta_{kn}\delta_{ml}\ket k\bra l\\=d_{nnmm}\ket n\bra m=d_{nnmm}\hat{F}_{nm}~~.
\end{multline}
Thus, the non-invariant condition, ${\cal L}\b{t} \sb{\hat{F}_{nm}}\propto \hat{F}_{nm}$, is satisfied.
For $n=m$, $\ket{n}\bra{n}=\hat{\Pi}_{n}$ and we obtain 
\begin{multline}\label{eq:Cond2}
    {\cal L}\b{t}\sb{\hat{\Pi}_{n}}\\=\sum_{a\neq b}^{N}c_{abba}\delta_{bn}\ket a\bra a+\sum_{k,l=1}^{N}d_{kkll}\ket k\bra l\delta_{kn}\delta_{ln}\\=\sum_{a\neq n}^{N}c_{anna}\ket a\bra a+d_{nnnn}\ket n\bra n\\=\sum_{a\neq n}^{N}c_{anna}\hat{\Pi}_{a}+{d_{nnnn}}\hat{\Pi}_{n}~~,
\end{multline}
which demonstrates that the invariant condition, ${\cal L}\b{t}\sb{\hat{\Pi}_j} = \sum_{i=1}^N b_{ji} \hat{\Pi}_{i}$ for $b_{ij}\in\mathbb{C}$, holds.

 \section{Trace preserving condition}
 \label{apsec:invariant_term}
The proposed structure for the dynamical generator, Eq. \eqref{eq:17}, is simplified in Sec. \ref{sec:gen_form} by imposing the trace preserving property. This leads to the final form for ${\cal L}\b{t}$, given in Eq. \eqref{eq:L_time_independent}. Here we provide an explicit derivation for this result. This derivation follows a similar line as Lemma 2.3 of Ref. \cite{gorini1976completely}.

We begin by introducing a new operator basis $\{\hat{P}_i\}$ for the invariant subspace (linear combinations of $\{\hat{\Pi}_i\}$), satisfying $\hat{P}_N=\hat{I}/N$, while the rest of the operators are traceless operators. In this basis the source-drain term becomes
\begin{equation}
    \sum_{i,j=1}^{N}  p_{ij}\b{t} \hat{\Pi}_{i} \bullet\hat{\Pi}_{j}\ra \sum_{ij=1}^{N}d_{ij}\hat{P}_{i}\bullet \hat{P}_{j}~~,
\end{equation}
where the matrices $\sb{p_{ij}}$ and $\sb{d_{kl}}$ are related by a unitary transformation. 
In terms the operator basis 
\begin{equation}
\{\hat{T}\}\equiv\{\hat{F}_1,\dots, \hat{F}_{N\b{N-1}},\hat{P}_{1},\dots,
\hat{P}_N\}~~,
\end{equation}
the dynamical generator (Eq. \eqref{eq:17}) becomes 
\begin{multline}
    {\cal L}\b{t} \sb{\bullet} = \\
   \sum_{\alpha=1}^{N\b{N-1}} c_{\alpha\alpha}\b{t} \hat{F}_{\alpha} \bullet\hat{F}_{\alpha}^\dagger 
    +\sum_{i,j=1}^{N}  d_{ij}\b{t} \hat{P}_{i} \bullet\hat{P}_{j}\\
    \equiv \sum_{i,j=1}^{N^2} r_{ij} \hat{T}_i\bullet \hat{T}_j~~.
    \label{apeq:D3t}
\end{multline}
Note that $r_{ij}$ does not vanish only for $i,j\in{\sb{1,N\b{N-1}}}$ or $i,j\in\sb{N\b{N-1},N^2}$.

For any operator in $\hat{A}\in M\b N$ the of the dynamical generator expressed as 
\begin{multline}
{\cal L}\sb{\hat{A}}=\f{1}{N} r_{N^2N^2} \hat A\\
+\f 1{N}\sum_{i=1}^{N^2-1}\b{r_{iN^2}\hat T_{i}\hat{A}+r_{N^2i}\hat{A}\hat{T}_{i}}+\sum_{i,j=1}^{N^2-1}r_{ij}\hat T_{i}\hat A\hat T_{j}\\=-\f{i}{\hbar}\sb{\bar{H},\hat{A}}+\{\bar{G},\hat{A}\}+\sum_{i,j=1}^{N^{2}-1}r_{ij}\hat{T}_{i}\hat{A}\hat{T}_{j}~~,
\end{multline}
where
\begin{gather}
    \bar{H}=\f {\hbar}{2i}\b{\hat{T}^{\dagger}-\hat T}\\
    \bar{G}=\f 1{2N}r_{N^2N^2}\hat{I}+\f 12\b{\hat{T}^{\dagger}+\hat T}~,
    \nonumber
\end{gather}
with
\begin{equation}
\hat{T}=\f{1}{N}\sum_{i=1}^{N^2-1}r_{i N^2}\hat{T}_{i}=\f{1}{N}\sum_{1=1}^{N}d_{iN}\hat{P}_i~~.
\end{equation}
The trace preserving property implies that
\begin{equation}
\text{tr}\b{{\cal L}\b{t}\hat{A}}=\text{tr}\sb{\b{2\bar{G}+\sum_{i,j=1}^{N^2-1}r_{ij}\hat{T}_{j}^{\dagger}\hat T_{i}}\hat{A}}=0~~,  
\end{equation}
 for all $\hat{A}\in M\b N$.  This infers that $\bar{G}=-\f 12\sum_{i,j=1}^{N^2-1}r_{ij}\hat{P}_{j}^{\dagger}\hat{P}_{i}$, which leads to the final form
 \begin{multline}
 {\cal D}\sb{\hat{A}}=-\f{i}{\hbar}\sb{\bar{H},\hat{A}}\\+\sum_{i,j=1}^{N-1}r_{ij}\b{\hat{T}_{i}\hat A\hat T_{j}^{\dagger}-\f 12\{\hat T_{j}^{\dagger}\hat T_{i},\hat A\}}~~.
 \end{multline}
 By expressing the ${\hat{T}}$ basis in terms of $\{\hat{F}\}$ and $\{\hat{P}\}$ (Eq. \eqref{apeq:D3t}) the relation between $r_{ij}$ to $c_{\alpha\alpha}$ and $d_{ij}$ coefficients, we obtain Eq. \eqref{eq:L_time_independent}
 \begin{multline}
    {\cal L}\b{t} \sb{\bullet} = -\f{i}{\hbar}\sb{\bar{H}(t),\bullet}\\
   +\sum_{\alpha=1}^{N\b{N-1}} c_{\alpha\alpha}\b{t} \left( \hat{F}_{\alpha} \bullet\hat{F}_{\alpha}^\dagger -\f{1}{2}\{\hat{F}_{\alpha}^\dagger\hat{F}_{\alpha},\bullet\}\right)\\
    +\sum_{i,j=1}^{N-1}d_{ij}\b{t}\b{\hat{P}_{i}\bullet\hat{P}_{j}^{\dagger}-\f 12\{\hat{P}_{j}^{\dagger}\hat{P}_{i},\bullet\}}~~,
\end{multline}

 \section{Chebychev expansion}
 \label{apsec:chebychev}
 The Chebychev expansion is utilized in Sec. \ref{sec:spin_star} to approximate the dynamical map of the spin star and obtain accurate kinetic coefficients. 
We first define a normalized generator so the associated eigenvalues are contained  with in the convergence range of the Chebychev polynomial
\begin{equation}
    {\cal O} = 2\f{\tilde{\cal{L}}-\lam_{\text{min}}}{\lam_\text{max}-\lam_{min}}-{\cal I}\,\,\in\sb{1,-1}
\end{equation}
where $\lam_\text{max}$ and $\lam_\text{min}$ are the maximum and minimum eigenvalues of $\cal{L}$, and $\cal I$ is the identity. This definition gives
\begin{equation}
    e^{\tilde{\cal L}t}=e^{s} e^{r\tilde{\cal O}}~~,
\end{equation}
where $s=t\b{\lam_\text{max}+\lam_\text{min}}/2$ and $r=t\b{\lam_\text{max}-\lam_\text{min}}/2$. The Chebychev series for $e^{i r\tilde{\cal O}}$ is obtained by expanding the function $e^{i x}$ with $x\in\sb{-1,1}$, utilizing the orthgonality condition
\begin{equation}
    \int_{-1}^{1}T_n\b{x}T_m\b{x}\f{dx}{\sqrt{1-x^2}}=\Bigg\{\begin{array}{c}
0\,\,\text{if}\,\,n\neq m\\
\pi\,\,\text{if}\,\,n=m=0\\
\f{\pi}2\,\,\text{if}\,\,n=m\neq0
\end{array}~~.
\end{equation}
Leading to 
\begin{equation}
    \tilde{\rho}_S\b t = e^s\sum_m a_m\b{r} \text{tr}_E\b{T_m\b{\tilde{\cal O}\sb{\hat \rho\b 0}}}~~,
\end{equation}
where the expansion coefficients are given by \begin{equation}
a_m\b{r} = \b{2i^m-\delta_{m0}}J_m\b{r}~~,
\end{equation} 
and $J_m$ is the $m$'th Bessel $J$ function.

The coefficients of the expansion for $\tilde{\cal L}$, Eq. \eqref{eq:Chebychev}, are given by 
\begin{equation}
    w_m\b{r\b t}=\f{d}{dt}a_m\b{r\b t}~~.
\end{equation}

 \section{Error bound associated with removal of a degeneracy}
 \label{apsec:bound}
 Consider two quantum channels $\Lambda^U\b{t}$ and $\Lambda^V\b t$ of the form of Eq. \eqref{eq:dynamical_map}, with corresponding unitaries $\hat{U}$ and $\hat{V}$, an initial system-environment state $\hat{\rho}\b{0}=\sum_i p_i\ket{\psi_i}\bra{\psi_i}$, expressed in terms of the orthonormal basis $\{\ket{\psi_i}\}$, and an element  $\hat M$ of an arbitrary POVM.
The difference in the probabilities $P^U$ and $P^{V}$ (associated with the reduced dynamics $\Lambda^{U}\b{t}$ and $\Lambda^{V}\b{t}$) of obtaining a certain outcome related to the element $\hat M$ can be bounded. Following Ref. \cite{nielsen2002quantum} (pg. 195)  we introduce the state $\ket{\Delta_{i}}\equiv\b{U-V}\ket{\psi_{i}}$ and employ the Cauchy-Schwarz inequality 
 \begin{multline}
     |P_U-P_V|=\Big|\text{tr}\b{\b{\Lambda^U\b{t}\sb{\hat{\rho}\b 0}- \Lambda^V\b t\sb{\hat{\rho}\b 0}}\hat{M}}\Big|
     \\=\sum_{i}p_{i}\Big|\bra{\psi_{i}}{\hat{U}^{\dagger}\hat M\hat U\ket{\psi_{i}}}-\bra{\psi_{i}}{\hat{V}^{\dagger}\hat{M}\hat{V}\ket{\psi_{i}}}\Big|\\=\sum_{i}p_{i}|\bra{\psi_{i}}U^{\dagger}M\ket{\Delta_{i}}+\bra{\Delta_{i}}MV\ket{\psi_{i}}|\\\leq\sum_{i}p_{i}\b{|\bra{\psi_{i}}U^{\dagger}M\ket{\Delta_{i}}|+|\bra{\Delta_{i}}MV\ket{\psi_{i}}|}\\\leq\sum_{i}p_{i}\b{||\ket{\Delta_{i}}||+||\ket{\Delta_{i}}||}\\\leq  2E\b{\hat{U},{\hat{V}}}~~.
  \end{multline}
with
 \begin{equation}
     E\b{\hat{U},{\hat{V}}}=  \text{max}_{\ket{\psi}}\Big|\Big|\b{\hat{U}-\hat{V}}\ket{\psi}\Big|\Big|~~,
 \end{equation}
 and the norm is defined as $||\ket{\psi}||=\sqrt{\mean{\psi|\psi}}$.

 For $\hat{U}\b t=e^{-i\hat{H}t/\hbar}$, $\hat{V}=e^{-i\hat{H}'t/\hbar}$ and $E\b{\hat{H},\hat{H}'}=\eps$, we get
\begin{multline}
 |P_U-P_V|\leq \text{max}_{\ket{\psi}}\Big|\Big|\b{\hat{I}-e^{i{\hat{H}'}t/\hbar}e^{-i{\hat{H}}t/\hbar}}\ket{\psi}\Big|\Big|\\
     =\Big|\Big|\sb{i\b{\hat{H}'-\hat{H}}t/\hbar+O\b{\eps^2}}\ket{\psi}\Big|\Big|\\\leq\eps t+O\b{\eps^{2}}~~.
 \end{multline}
 The connection to the error between the degenerate and non-degenerate maps, Eq. \eqref{eq:error} Sec. \ref{sec:degeneracy}, is achieved by identifying $\hat{H}$ and ${\hat{H}'}$ with the joint Hamiltonian (of the system and environment), generating the unitary dynamics associated with $\hat{\Lambda}^{deg}\b t$ and $\hat{\Lambda}^{\eps}\b t$. 
 
 \section{Thermomajorization condition}
 \label{apsec:thermomaj}
 Thermomajorization is a mathematical condition involving two vectors $\v x, \v y\in {\mathbb{R}}^n$ and an associated Hamiltonian $\hat{H}$. In the present context the vectors are constructed from the populations of two density operators $\hat{\rho}^x$ and $\hat{\rho}^y$ in the energy eigenbasis of the associated Hamiltonian $\hat{H}$. To evaluate whether $\v x$ thermomajorizes $\v y$, one first defines the so-called $\beta$ ordered vectors of $\v x$ and $\v y$, with indices $x_i^{\downarrow \beta} =x_{\pi\b{i}}$ where $\pi\b{i}$ is the permutation ensuring that $x_{\pi\b 1}e^{\beta E_1}\geq x_{\pi\b 2}e^{\beta E_2}\geq\cdots x_{\pi\b n}e^{\beta E_n}$ and similarly for $y^{\downarrow\beta}$. The $\beta$-ordered  vectors are next utilized to define the thermomajorization-curves, these are piece-wise linear curves joining the origin and points $\b{\sum_{i=1}^k e^{\beta E_{\pi\b{i}}},\sum_{i=1}^k x_{\pi{i}}}$ for $k=1,\dots,n$, and similarly for $\v y$. The vector $\v x$ thermomajorizes $\v y$ iff the  thermomajorization curve associated with $\v x$ does not lie below the thermomajorization curve of $\v y$.
 
\section{Dynamical symmetry associated with the conservation of the number of excitations}
 \label{apsec:number_conservation}
 Conservation of the total number of excitations is manifested by the commutation of the total Hamiltonian with the number operators ${\hat{N}}=\hat{N}_S+\hat{N}_E$. In the following we should that the conservation law, along with an initial stationary environment state, implies the dynamical symmetry ${\cal U}_{N,S}\circ\Lambda=\Lambda\circ{\cal U}_{N,S}$.
 
 \paragraph*{Proof}
 We first as introduce a number of notions: The spectral decomposition of the environment Hamiltonian is given by $\hat{H}_E = \sum_i c_i\ket{\chi_i}\bra{\chi_i}$.
The environment is initially assumed to be in a stationary state with respect to the free dynamics, allowing to express it as $\hat{\rho}_E\b{0} = \sum_i\lam_i\ket{\chi_i}\bra{\chi_i}$. ${\cal{U}}_{N,j}=\hat{U}_{N,j}\bullet\hat{U}_{N,j}^\dagger$, with $\hat{U}_{N,j}=e^{i\hat{N}_j}$ and $j=S,E$, and $\hat{U}_{N,SE}=e^{i\b{\hat{N}_{S}+\hat{N}_E}}$. 

The reduced system dynamics are given by 
 \begin{equation}
 \hat{\rho}_{S}\b t=\text{tr}_{E}\b{\hat{U}\b{t,0}\rho_{S}\b 0\otimes \hat \rho_{E}\b 0\hat{U}^{\dagger}\b{t,0}}    
 \end{equation}
Substituting the spectral decomposition of the initial environment state leads to the Kraus form \cite{kraus1971general}
\begin{equation}
    \hat{\rho}_S\b t = \sum_{ij} \hat{K}_{ij}\hat{\rho}_S\b 0 \hat{K}_{ij}^{\dagger}
\end{equation}
where $\hat{K}_{ij} = \sqrt{\lam_i}\bra{\chi_j}\hat{U}\b{t,0}\ket{\chi_i}$ with $\sum_{ij}\hat{K}_{ij}^\dagger \hat{K}_{ij}=\hat{I}_S$.  
In the Heisenberg representation the dynamical map obtains the form $\hat{O}_S^{H}\b{t} = \Lambda^\ddagger\sb{\hat{O}_S}=\sum_{ij}\hat{K}_{ij}^\dagger \hat{O}_S\hat{K}_{ij}$ where $\hat{O}_S$ is a general system operator and the superscript $H$ designates that the operator is in the Heisenberg representation.
Similarly, ${\cal U}_{N,S}$ in the Heisenberg representation becomes ${\cal U}_{N,S}^\ddagger\sb{\bullet} = \hat{U}_{N,S}^\dagger\bullet \hat{U}_{N,S} $.

The product of maps can be now expressed as 
\begin{multline}
    {\cal U}_{N,S}^{\ddagger}\sb{\Lambda^{\ddagger}\sb{\hat O_{S}}}=\hat U_{N,S}^{\dagger}\b{\sum_{ij} \hat K_{ij}^{\dagger}\hat O_{S}\hat K_{ij}}\hat U_{N,S}\\
=\sum_{i}\lam_{i}\bra{\chi_{i}}\hat U_{N,S}^{\dagger}\hat U^{\dagger}\hat O_{S}\sum_{j}\ket{\chi_{j}}\bra{\chi_{j}}\hat U\hat U_{N,S}\ket{\chi_{i}}\\
=\lam_{i}\bra{\chi_{i}}\hat U_{N,S}^{\dagger}U^{\dagger}\hat O_{S}\hat U\hat U_{N,S}\ket{\chi_{i}}~~.
\nonumber
\end{multline}
Utilizing the relations $\hat{U}_{N,E}\ket{\chi_k} = e^{ik}\ket{\chi_k}$ we get
\begin{multline}
    =\lam_{i}\bra{\chi_{i}}\hat U_{N,E}^{\dagger}\hat U_{N,S}^{\dagger}\hat U^{\dagger}\hat O_{S}\hat U\hat U_{N,S}\hat U_{N,E}\ket{\chi_{i}}\\=\lam_{i}\bra{\chi_{i}}\hat U_{N,SE}^{\dagger}\hat U^{\dagger}O_{S}\hat U\hat U_{N,SE}\ket{\chi_{i}}\\=\lam_{i}\bra{\chi_{i}}\hat U^{\dagger}\hat U_{N,SE}^{\dagger}\hat O_{S}\hat U_{N,SE}\hat U\ket{\chi_{i}}~~,
    \label{eq:F3}
\end{multline}
where in the last equality with utilize the commutation relation $\sb{\hat{N}_S+\hat{N}_E,\hat{H}}$.

The reverse product can be expressed as
\begin{multline}
\Lambda^{\ddagger}\sb{{\cal U}_{N,S}^{\ddagger}\sb{O_{S}}}=\sum_{ij}\hat K_{ij}^{\dagger}\hat U_{N,S}^{\dagger}\hat O_{S}\hat U_{N,S}\hat K_{ij}\\=\sum_{i}\lam_{i}\bra{\chi_{i}}U^{\dagger}\hat U_{N,SE}^{\dagger}\hat U_{N,E}\hat O_{S}\hat U_{N,E}^{\dagger}\hat U_{N,SE}\hat U\ket{\chi_{i}}\\=\sum_{i}\lam_{i}\bra{\chi_{i}}\hat U^{\dagger}\hat U_{N,SE}^{\dagger}\hat O_{S}\hat U_{N,SE}\hat U\ket{\chi_{i}}~~,
\label{eq:F5}
\end{multline}
where in the second equality we inserted the identity $\hat{U}_{N,E}\hat{U}_{N,E}^\dagger =\hat{I}_E$ and utilized the fact that environment and system operators.

Finally, equations \eqref{eq:F3} and \eqref{eq:F5} infer the desired result
\begin{equation}
    \Lambda^{\ddagger}\sb{{\cal U}_{N}^{\ddagger}\sb{O_{S}}}={\cal U}_{N}^{\ddagger}\sb{\Lambda^{\ddagger}\sb{O_{S}}}~~_\blacksquare
\end{equation}

 \section{Dynamics of the spin-star model}
 \label{apsec:spin_star_model}
 The dynamical solution of the spin-star model (Sec. \ref{sec:spin_star}) was first presented in Ref. \cite{breuer2004non}. For the sake of completeness, we briefly discuss the solution and give a detailed derivation.  
 
 The connection between the present model, Eq. \eqref{eq:spinstar_Ham} and the derivation in \cite{breuer2004non} is obtained by transitioning to the interaction picture with respect to the free dynamics $\hat{H}_S+\hat{H}_E=\hbar \omega \b{\hat{\sigma}_z+\sum_{k=1}^K\hat{\sigma}_z^{\b{k}}}$. In the interaction picture, the Liouville von-Neumann equation becomes 
 \begin{equation}
     \f{d}{dt}\tilde{\rho}\b t=\tilde{\cal L}^{\b{SE}}\sb{\tilde{\rho}\b t}=-\f{i}{\hbar}\sb{\tilde{H},\tilde{\rho}\b t}
     \label{eq:VN_spin_star}
 \end{equation}
 with
 \begin{equation}
     \tilde{H}=2g\b{\hat \sigma_{+}\hat J_{-}+\hat \sigma_{-}\hat J_{+}}~~,
 \end{equation}
 where operators in the interaction picture are designated by a superscript tilde, and $\hat{\rho}\b t$ is the joint density operator. The solution for the joint dynamics can be formally expressed in terms of the generator of Eq. \eqref{eq:VN_spin_star}:  $\hat{\rho}\b{t}=e^{g\tilde{ \cal L}^{\b{SE}} t}\sb{\hat{\rho}\b{0}}$. This relation allows writing the central spin evolution in terms of a power series of the generator 
 \begin{multline}
     \tilde{\rho}_{S}\b t=\text{tr}_{E}
     \b{e^{g{\tilde{\cal L}^{\b{SE}}}t}\sb{\hat \rho_{S}\b{0}\otimes\hat \rho_{E}\b {0}}}\\=
     \sum_{k=0}^{\infty}\f{\b{g t}^{k}}{k!}\text{tr}_{E}\b{\b{\tilde{\cal L}^{\b{SE}}}^{k}\sb{\hat{\rho}_{S}\b{0}\otimes\hat\rho_{E}\b{0}}}~~,    
     \label{apeq:d3}
 \end{multline}
 where 
 \begin{equation}
     \b{\tilde{\cal L}^{\b{SE}}}^{k}\sb{\bullet}=i^{k}\sum_{l=0}^{k}\b{-1}^{l}\b{\begin{array}{c}
k\\
l
\end{array}}\tilde{H}^{l}\bullet \tilde{H}^{k-l}~~.
\label{apeq:d4}
 \end{equation}
The symmetric structure of $\tilde{H}$ leads to a simple relation for the moments
\begin{gather}
    \tilde{H}^{2n}=4^{n}\b{\hat{\sigma}_{+}\hat{\sigma}_{-}\b{\hat{J}_{-}\hat{J}_{+}}^{n}+\hat{\sigma}_{-}\hat{\sigma}_{+}\b{\hat{J}_{+}\hat{J}_{-}}^{n}}\\
    \tilde{H}^{2n+1}=2\cdot4^{n}\b{\hat \sigma_{-}\hat J_{+}\b{\hat J_{-}\hat J_{+}}^{n}+\hat \sigma_{+}\hat J_{-}\b{\hat J_{+}\hat J_{-}}^{n}}~~.
    \nonumber
\end{gather}
 For an odd $k$, $\text{tr}_E\b{\tilde{H}^{l}\hat{\rho}_S\b t\otimes\hat{\rho}_E \tilde{H}^{k-l}}=0$, due to an odd power of $\hat{J}_+$ or $\hat{J}_-$ with $\hat{\rho}_E\b {0}\propto \hat{I}_E$. 
 As a result, only the even powers $k=2n$ contribute to the infinite sum of Eq. \eqref{apeq:d3}. 
 
 We separate the treatment to two cases, even or odd values of $l$. For an even $l=2m$ one obtains 
 \begin{multline}
     \tilde{H}^{l}\hat{\rho}\b{0} \tilde{H}^{k-l}=\\\f {1}{2^{K}}\b{4g^2}^{n}\b{\hat \sigma_{+}\hat \sigma_{-}\hat\rho_{S}\b{0}\hat \sigma_{-}\hat \sigma_{+}\b{\hat J_{-}\hat J_{+}}^{m}\b{\hat J_{+}\hat J_{-}}^{\b{n-m}}+\text{h.c}}~~,
 \end{multline}
 which contributes
\begin{multline}
    \f{1}{2^{K}}\b{-g^2}^{n}\sum_{m=0}^{n}\b{-1}^{2m}\b{\begin{array}{c}
2n\\
2m
\end{array}}4^{n}\\\times\b{\hat \sigma_{+}\hat \sigma_{-}\hat \rho_{S}\b{0}\hat \sigma_{-}\hat \sigma_{+}\Big \langle \b{\hat J_{-}\hat J_{+}}^{m}\b{\hat J_{+}\hat J_{-}}^{n-m}\Big \rangle_E+\text{h.c}}\\=\b{-4g^2}^{n}\sb{\sum_{m=0}^{n}\b{\begin{array}{c}
2n\\
2m
\end{array}}R_{m}^{n-m}}\b{r_{+}\b {0}\hat \sigma_{+}+r_{-}\b{0}\hat \sigma_{-}}
\label{apeq:d7}
\end{multline}
to Eq. \eqref{apeq:d4}, where 
\begin{equation}
R_{m}^{n-m}=\f{1}{2^{K}}\Big \langle\b{\hat J_{+}\hat J_{-}}^{n-m}\b{\hat J_{-}\hat J_{+}}^{m}\Big \rangle~~.  
\end{equation}

For an odd $l=2m+1$ we obtain
\begin{multline}
   \tilde{H}^{l}\hat{\rho}\b{0} \tilde{H}^{k-l}\\ =\b{4g^2}^{n}\sb{\hat \sigma_{-}\hat \rho_{S}\b{0}\hat \sigma_{+}\b{\hat J_{+}\hat J_{-}}^{n}+\text{h.c}}\\=\b{4g^2}^{n}\sb{\f{\hat I_{S}}2-\f{r_{z}\b{0} }{2}\hat \sigma_{z}}\b{\hat J_{+}\hat J_{-}}^{n}~~,
\end{multline}

which contributes a term
\begin{multline}
    \b{-16g^2}^{n}{\f{r_{z}\b {0}}{2}\hat{\sigma}_{z}}\Big \langle\b{{\hat J_{+}\hat J_{-}}^{n}}\Big \rangle
    \label{apeq:d10}
\end{multline}
to Eq. \eqref{apeq:d4}. In the last transition we utilized the relation 
\begin{multline}
    \sum_{m=0}^{n-1}\b{-1}^{2m+1}\b{\begin{array}{c}
2n\\
2m+1
\end{array}}\\=\sum_{m=0}^{2n}\b{\begin{array}{c}
2n\\
m
\end{array}}\b{-1}^{m}1^{2n-m}-\sum_{m=0}^{2n}\b{\begin{array}{c}
2n\\
l
\end{array}}1^{m}1^{2n-m}\\=\b{1-1}^n-\b{1+1}^n=-4^n~~.
\end{multline}

Gathering Eqs. \eqref{apeq:d7} and \eqref{apeq:d10} we get 
\begin{multline}
    \text{tr}_{E}\b{\b{{\cal L}^{\b{SE}}}^{2n}\hat\rho_{S}\b {0}\otimes2^{-K}\hat I_{E}}\\=\b{-16g^2}^{n}Q_{n}\f{r_{z}\b {0}}{2}\hat\sigma_{z}+\b{-4g^2}^{n}\sb{\sum_{m=0}^{n}\b{\begin{array}{c}
2n\\
2m
\end{array}}R_{m}^{n-m}}\\\times \b{r_{+}\b{0}\hat\sigma_{+}+r_{-}\b{0}\hat\sigma_{-}}~~,
\label{eqap:e12}
\end{multline}
with
\begin{equation}
    Q_{n}=\f 1{2^{K}}\Big\langle{\b{\hat J_{+}\hat J_{-}}^{n}}\Big\rangle~~.
\end{equation}
We can now recognize that corresponding time-dependent coefficients of $\hat{\rho}_S\b{t}$ (Eq. \eqref{eq:spin_star_sol1}) are 
\begin{gather}
    r_z\b{t}=\kappa_z\b{t}r_z\b{0} 
     \label{apeq:f14}\\
    r_{\pm}\b{t}=e^{\pm i\omega t}\kappa\b{t}r_{\pm}\b{t_0}~~,
    \nonumber
\end{gather}
where
\begin{multline}
    \kappa_{z}\b t=\sum_{k=0}^{\infty}\f{\b{g t}^{2k}}{2k!}\b{-16}^{k}Q_{k}\\=\f 1{2^{K}}\text{tr}_{E}\b{\sum_{k=0}^{\infty}\f{1}{2k!}\b{4ig\sqrt{\hat J_{+}\hat J_{-}}t}^{2k}}\\=\f 1{2^{K}}\text{tr}_{
    E}\b{\cos\b{4g \sqrt{\hat J_{+}\hat J_{-}}t}}~~,
\end{multline}
and
 \begin{multline}
     \kappa\b t=\sum_{2k=0}^{\infty}\f{\b{g t}^{2k}}{2k!}\b{-4}^{k}\sb{\sum_{l=0}^{k}\b{\begin{array}{c}
2k\\
2l
\end{array}}R_{l}^{k-l}}\\=\f 1{2^{N}}\text{tr}_{E}\bigg(\sum_{k=0}^{\infty}\f 1{2k!}\sum_{l=0}^{k}\b{\begin{array}{c}
2k\\
2l
\end{array}}\\ \times\sb{\b{2ig\sqrt{\hat J_{+}\hat J_{-}} t}^{2k-2l}\b{2ig\sqrt{\hat J_{-}\hat J_{+}} t}^{2l}}\bigg)\\=\f 1{2^{K}}\cos\b{2g\sqrt{\hat J_{+}\hat J_{-}} t}\cos\b{2g\sqrt{\hat J_{-}\hat J_{+}} t}~~.
 \end{multline}
 Here we utilized the relation
 \begin{multline}
     \cos\b x\cos\b y=\f 12\b{\cos\b{x+y}+\cos\b{x-y}}\\=\f 12\b{\sum_{k=0}^{\infty}\f{\b{x+y}^{2k}}{2k!}+\sum_{k=0}^{\infty}\f{\b{x-y}^{2k}}{2k!}}\\=\sum_{k=0}^{\infty}\f 1{2k!}\b{\sum_{l=0}^{k}\b{\begin{array}{c}
2k\\
2l
\end{array}}x^{2l}y^{2k-2l}}~~.
 \end{multline}
 
 We next introduce a basis for the environment's Hilbert space, consisting of simultaneous eigenstates of $\v{J}^2$ and $\hat{J}_z$: $\{\ket{j,m,\nu}\}$, where the index $\nu$ label eigenstates which correspond to the same $\b{j,m}$ quantum numbers. This allows simplifying the time-dependent coefficients $r_z\b{t}$ and $r_{\pm}\b{t}$ by utilizing the degeneracy of states with the same $\b{j,m}$ quantum numbers: $d\b{j,m}$, and the relation $\sqrt{\hat{J}_{+}\hat{J}_{-}}\ket{j,m,\nu}=h\b{j,m}\ket{j,m,\nu}$, as defined in Eq. \eqref{eqap:37}.
 Finally, the final form of the time-dependent coefficients determine the exact reduced dynamics of the central spin, given in Eqs. \eqref{eq:spin_star_sol1} and \eqref{eq:spin_star_sol2}.
 
 The approximate kinetic coefficients of Sec. \ref{subsec:compare} (relying on the Maclaurin and Chebychev) can be calculated using  Eq. \eqref{eqap:e12} and the relation
 \begin{multline}
     \text{tr}_{E}\b{\b{\tilde{\cal L}^{\b{SE}}}^{n}\hat\rho\b {t_0}} \\=\b{-\f{i}{\hbar}}^n\text{tr}_E\b{{\sb{\hat H_{SE},\sb{...\sb{\hat H_{SE},\hat{\rho}\b{t_0}}}}}}~~,
 \end{multline}
 including $n$ commutation relations on the RHS.

\subsection*{Derivation of the dynamical generator}
The dynamical generator associated with the exact dynamical map can be obtained by the following procedure. 
We work in the interaction picture relative to the free dynamics ($\hat{H}_0=\hbar \omega \hat{\sigma}_z$). In this picture the coefficients of the density matrix evolve according to (see Eq. \ref{eq:spin_star_sol1})
\begin{gather}
\label{apeq:2041}
    \tilde{r}_z\b t =  r_z\b t=\kappa_z\b t r_z \b 0\\
    \tilde{r}_\pm \b 0 = \kappa\b t r_\pm \b 0
    \nonumber
\end{gather}
Alternatively, the density matrix can be expressed in terms of the Pauli operators
\begin{equation}
    \tilde{\rho}_S\b t = \f{1}{2}\b{\hat{I}_S +\tilde{r}_x\b t \hat{\sigma}_x+\tilde r_y\b t \hat{\sigma}_y+\tilde r_z \hat{\sigma}_z}~~,
\end{equation}
where
\begin{gather}
 \label{eqap:2050}
 \tilde{r}_x \b t = \tilde r_+ \b t+\tilde{r}_- \b t\\
 \tilde r_y\b t = i \b{\tilde r_+\b t- \tilde r_- \b t}~~.
 \nonumber
\end{gather}
Now the action of the dynamical generator in the interaction picture gives 
\begin{equation}
    \tilde{\cal{L}}\b t\sb{\tilde{\rho}_S\b t} =\f{d}{dt}\tilde{\rho }_S\b t =  \dot{\tilde{r}}_x\b t \hat{\sigma}_x+\dot{\tilde r}_y\b t \hat{\sigma}_y+\dot{\tilde r}_z \hat{\sigma}_z~~.
    \label{apeq:2055}
\end{equation}
In addition, the most general structure of the master equation for a two-level system which complies with the strict energy conservation is of the following form (a specific case of Eq. \eqref{eq:L_time_independent}) 
\begin{multline}
    \tilde{\cal{L}}\sb{\bullet} = -i\eta_0\sb{\hat{\sigma}_z,\bullet}+ \eta_-\b t {\cal D}_-\b t\sb{\bullet} +\eta_+\b t {\cal D}_+ \sb{\bullet}\\+ \eta_z\b t {\cal D}_z\sb{\bullet}~~,
    \label{apeq:2059}
\end{multline}
where the superoperators ${\cal{D}}_\pm$ are defined bellow Eq. \eqref{eq:exact_ME}, $   {\cal D}_z\sb{\bullet}=\hat{\sigma}_z\bullet\hat{\sigma}_z-\bullet$, and
${\eta_+,\eta_-,\eta_z,\eta_0}$ are real time-dependent variables. First, we identify that the dynamics of $\tilde{r}_\pm\b t$ in Eq. \eqref{apeq:2041} is determined by a real function $\kappa\b t$. Since the commutation relation of the lowering and raising operators satisfy $\sb{\hat{\sigma}_z,\hat{\sigma}_\pm}=\pm\hat{\sigma}_\pm$ the unitary term in the interaction picture (the term proportionate to $i$) vanishes, i.e. $\eta_0=0$.

Equating $\text{tr}\b{\tilde{\cal{L}}\b t\sb{\tilde{\rho}_S\b t}\hat{\sigma_i}}$ for Eqs. \eqref{apeq:2055} and \eqref{apeq:2059} for $i=x,y,z$, leads to a set of linear equations connecting the kinetic coefficients to
$\{\tilde{r}_i\}$ and $\{\dot{\tilde{r}}_i\b t\}$. The relation can be summarized by
\begin{equation}
    \sb{\begin{array}{ccc}
-\f 12\tilde r_{x} & -\f 12 \tilde r_{x} & -2\\
-\f 12 \tilde r_{y} & -\f 12 \tilde r_{y} & -2\\
1-\tilde r_{z} & -\b{1+\tilde r_{z}} & 0
\end{array}}\sb{\begin{array}{c}
\eta_{+}\\
\eta_{-}\\
\eta_{z}
\end{array}}=\sb{\begin{array}{c}
\dot{\tilde r}_{x}\\
\dot{\tilde r}_{y}\\
\dot{\tilde r}_{z}
\end{array}}~~.
\end{equation}
By solving for the vector of kinetic coefficients we obtain
\begin{gather}
    \eta_{+}=\frac{-2\dot{\tilde{r}}_{x}\b{1+\tilde{r}_{z}}+2\dot{\tilde{r}_{y}}\b{1+\tilde{r}_{z}}+\dot{\tilde{r}}_{z}\b{\tilde{r}_{x}-\tilde{r}_{y}}}{2\b{\tilde{r}_{x}-\tilde{r}_{y}}} \nonumber\\ \label{apeq:2092}
    \eta_{-}=\frac{2\dot{\tilde{r}}_{x}\b{\tilde{r}_{z}-1}-2\dot{\tilde{r}}_{y}\b{\tilde{r}_{z}-1}+\dot{\tilde{r}}_{z}\b{\tilde{r}_{x}-\tilde{r}_{y}}}{2\b{\tilde{r}_{x}-\tilde{r}_{y}}}
    \\\eta_{z}=\frac{\dot{\tilde{r}}_{x}\tilde{r}_{y}-\dot{\tilde{r}}_{y}\tilde{r}_{x}}{2\b{\tilde{r}_{x}-\tilde{r}_{y}}}~~.
    \nonumber
\end{gather}
Substituting relations \eqref{eqap:2050} and \eqref{apeq:2041} into Eq. \eqref{apeq:2092} leads the final form of the kinetic coefficients, $\eta_z$ vanishes 
and 
\begin{equation}
    \eta_{\pm}\b t=\\\frac{\xi_\pm\b t}{2\kappa\b t\b{r_{x}\b 0-r_{y}\b 0}}
\end{equation}
with
\begin{multline}
    \xi_\pm\b t =\mp2\dot{\kappa}r_{x}\b 0\b{r_{z}\b t\pm1}\\+2\dot{\kappa}r_{y}\b 0\b{r_{z}\b t\pm1}+\kappa\dot{r}_{z}\b 0\b{r_{x}\b 0-r_{y}\b 0}~~.
\end{multline}
These kinetic coefficients obtain negative values (see Fig. \ref{fig:spin_star_coeff}), indicating that the dynamical map violates CP-divisiblity, which signifies that the dynamics are non-Markovian.

\section{Kinetic coefficients of the spin-boson bath model under conservation of the total number of excitations}
\label{apsec:num_model}

We derive the master equation of a spin coupled weakly to a bosonic thermal bath, under dynamics satisfying the the conservation of the number of excitations.

In the interaction picture the total dynamics (Eq. \eqref{eq:68num}) are governed by the interaction Hamiltonian
\begin{equation}
    \tilde{H}_I \b t= \sum_k g_k\b{\hat{\sigma}_-b_k^\dagger e^{-i\b{\omega_0-\omega_k}t}+\text{h.c}}~~,
\end{equation}
where $\omega_0,\omega_k>0$. 
We substitute the interaction Hamiltonian into the second cumulant Eq. \eqref{eq:uuu}, and utilize the known relations  for a thermal bosonic bath: $\mean{\hat{b}_k\hat{b}_{k'}}=0$, $\mean{\hat{b}_k\hat{b}_{k'}^\dagger}=\delta_{kk'}\b{1+\bar{n}_T\b{\omega_k}}$, $\mean{\hat{b}_k^\dagger\hat{b}_{k'}}=\delta_{kk'}{\bar{n}_T\b{\omega_k}}$, were $\bar{n}_T\b{\omega_k}$ is the Bose-Einstein distribution at temperature $T$. These ansatzs lead to 
\begin{multline}
    {\cal Z}\b t \approx K^{\b{2}}\b t = 
    \hat{\rho}_{S}\b 0-\f{i}{\hbar}\sb{\bar{H}^{\b{N}}\b t,\hat \rho_{S}\b 0} \\
    +\trr{\Gamma_{-}^{\b N}}\b t{\cal{D}}_-\sb{\hat{\rho}_S\b 0}+\trr{\Gamma_{+}^{\b N}}\b t{\cal{D}}_+\sb{\hat{\rho}_S\b 0}~~,
\end{multline}
where ${\cal D}_\pm$ are defined bellow Eq. \eqref{eq:exact_ME} and the kinetic coefficients are considered to be correct only up to second order in spin-bath interaction. 
The second order contribution to $\hat{H}^{\b{N}}$ (Eq. \eqref{eq:rho70}) becomes (for more details see Appendix B of Ref \cite{rivas2017refined})
\begin{multline}
    \bar{H}^{\b{N}}\b t =\f{1}{2i}\int_0^{t}dt_1 \int_0^{t}dt_2 \text{sgn}\b{t_1-t_2 }\\\times\text{tr}_E\b{\tilde{H}_I\b{t_1}\tilde{H}_I\b{t_2}\hat{\rho}_E\b 0}\\=
     \Xi_{+}\b t\hat{\sigma}_{+}\hat \sigma_{-}+\Xi_{-}\b t\hat\sigma_{-}\hat \sigma_{+},
     \label{eq:H3}
\end{multline}
with 
\begin{multline}
    \Xi_{\mp}\b t=\sum_{k}g_{k}^{2}\f 1{2i}\int _0^t dt_{1}\int_ 0^t dt_{2}\\
    \text{sgn}\b{t_{1}-t_{2}}e^{\mp i\b{\omega_0-\omega_k}\b{t_{1}-t_{2}}}\mean{\hat b_{\mp k}b_{\mp k}^{\dagger}}_E~~.
     \label{eq:H4}
\end{multline}
and
\begin{multline}\label{eq:Gamma1}
    \trr{\Gamma_{\pm}^{\b N}}\b{t}=\sum_kg_k^2\int_0^t{dt_{1}}\int _0^t dt_{2}e^{\mp i{\omega_0\b{ t_{1}-t_{2}}}}\\\times\text{tr}\b{\tilde{b}_{\mp k}\b{t_{1}}\tilde{b}_{\mp k}^{\dagger}\b{t_{2}}\hat{\rho}_{E}\b 0}\\
    =\sum_kg_k^2\int_0^t{dt_{1}}\int _0^t dt_{2}e^{\mp i\b{\omega_0-\omega_k}\b{ t_{1}-t_{2}}}\\\times\mean{\hat{b}_{\mp k}\hat{b}_{\mp k}^{\dagger}}_E~~,
\end{multline}
where $\hat{b}_{- k}=\hat{b}_{k}^\dagger$.

Taking the continuum limit and defining the bath spectral density function $J\b{\omega}\sim\sum_{k}g_{k}^{2}\delta\b{\omega-\omega_{k}}$, we get 
\begin{multline}
    \trr{\Gamma_{\pm}^{\b N}}\b{t}=\\ \int_{0}^{\infty}{d\omega_k}J\b{\omega_k}\mean{\hat{b}_{\mp k}\hat{b}_{\mp k}^{\dagger}}_E t^{2}\text{sinc}^{2}\b{\b{\omega_{0}-\omega_k}t/2}~~,
\end{multline}
In explicit form, the kinetic coefficients become
\begin{gather}
     \trr{\Gamma_{+}^{\b N}}\b{t}= \int_{0}^{\infty}{d\omega_k}J\b{\omega_k}\bar{n}_T\b{\omega_k} t^{2}\text{sinc}^{2}\b{\b{\omega_{0}-\omega_k}t/2}~~,\nonumber\\
     \trr{\Gamma_{-}^{\b N}}\b{t}= \int_{0}^{\infty}{d\omega_k}J\b{\omega_k}\b{\bar{n}_T\b{\omega_k}+1} t^{2}\text{sinc}^{2}\b{\b{\omega_{0}-\omega_k}t/2}~~.
     \label{eq:H7}
\end{gather}

The integral of the Lamb-shift  term, Eqs. \eqref{eq:H3} and \eqref{eq:H4}, can be simplified (see Appendix B of Ref. \cite{rivas2017refined})
\begin{multline}
    \Xi_{\mp}\b t=\f 1{2i}\int_{-\infty}^{\infty}d\varphi\int _0^t dt_{1}\int_ 0^t dt_{2}\\
    e^{ i\b{\pm\omega_0-\varphi}\b{t_{1}-t_{2}}}{\cal P}\int_0^{\infty}d\omega_k\, J\b{\omega_k}\f{\mean{\hat b_{\mp k}\hat{b}_{\mp k}^{\dagger}}_E}{\varphi-\omega_k}~~.
\end{multline}
where $\cal P$ denotes the Cauchy principle value of the integral. This leads to
\begin{multline}
    \Xi_{\mp}\b t=\f 1{2\pi}\int_{-\infty}^{\infty}{d\omega}\,t^{2}\text{sinc}^{2}\b{\f{\omega\mp\omega_{0}}2}\\{\cal P}\sb{\int _0^{\infty}{d\omega_{k}\,J\b{\omega_{k}}\f{\mean{\hat b_{\mp k}\hat b_{\mp k}^{\dagger}}_{E}}{\omega-\omega_{k}}}}~~.
\end{multline}
Next, we substitute $\hat{\sigma}_+\hat{\sigma}_- = \b{\hat{{I}}+\hat \sigma_z}/2$ and $\hat{\sigma}_-\hat{\sigma}_+ = \b{\hat{{I}}-\hat \sigma_z}/2$ into Eq. \eqref{eq:H3} which gives the simple form
\begin{equation}
    \bar{H}^{\b{N}} = \Xi\b t \hat{\sigma_z}~~,
\end{equation}
where 
\begin{multline}
    \Xi\b t  = \b{\Xi_+\b t-\Xi_-\b t}/2\\=
    \f 1{4\pi}\int_{-\infty}^{\infty}{d\omega}\,t^{2}\\\Bigg\{\text{sinc}^{2}\b{\f{\b{\omega-\omega_{0}}t}2}{\cal P}\sb{\int _0^{\infty}{d\omega_{k}J\b{\omega_{k}}\f{\bar n_{T}\b{\omega_{k}}+1}{\omega-\omega_{k}}}}\\-\text{sinc}^{2}\b{\f{\b{\omega+\omega_{0}}t}2}{\cal P}\sb{\int _0^{\infty}{d\omega_{k}J\b{\omega_{k}}\f{\bar n_{T}\b{\omega_{k}}}{\omega-\omega_{k}}}}\Bigg\}~~.
\end{multline}

Finally,  ${\cal Z}\b t\approx K^{\b{2}}\b t$ is substituted into Eq. \eqref{eq:72} and algebra of the super operators $\sb{{\cal{D}}_+,{\cal D}_-}={\cal D}_+-{\cal D}_- $, $\sb{{\cal{D}}_z',{\cal D}_\pm}=0$ is employed to obtain the non-Markovian generator (Eq. \eqref{eq:rho73})
\begin{multline}
\tilde{\cal L}^{\b{N}}\b{t}\sb{\bullet}= - \f{i}{\hbar} \sb{\Upsilon^{\b N} \b t,\bullet}\\+ \gamma_+^{\b{N}}\b{t}{\cal{D}}_+\sb{\bullet} +\gamma_-^{\b{N}}(t){\cal{D}}_-\sb{\bullet}~~,
\end{multline}
where the kinetic coefficients are given by (Appendix A Ref. \cite{rivas2017refined})
\begin{multline}
    \gamma_{\pm}^{\b N}=\f 1{\b{\trr{\Gamma}_{+}^{\b N}+\trr{\Gamma}_{-}^{\b N}}^{2}}\times\\\bigg\{\b{e^{-\b{\trr{\Gamma}_{+}^{\b N}+\trr{\Gamma}_{-}^{\b N}}}-1}\b{\dot{\trr{\Gamma}}_{\mp}^{\b N}\trr{\Gamma}_{\pm}^{\b N}-\dot{\trr{\Gamma}}_{\pm}^{\b N}\trr{\Gamma}_{\mp}^{\b N}}\\+\b{\dot{\trr{\Gamma}}_{+}^{\b N}+\dot{\trr{\Gamma}}_{-}^{\b N}}\b{\b{\trr{\Gamma}_{\pm}^{\b N}}^{2}+\trr{\Gamma}_{+}^{\b N}\trr{\Gamma}_{-}^{\b N}}\bigg\}~~,
    \label{eqap:2225}
\end{multline}
and the Lamb-shift by
\begin{equation}\label{eq:Lamb}
    \Upsilon^{\b{N}}=\dot{\Xi}~~,
\end{equation}
where we left out the explicit time-dependence for the sake of conciseness.
\subsection{Markovian limit}
\label{asec:Mlimit}
In the Markovian limit, the coefficients of the second cumulant become time-independent, and the master equation converges to the standard Markovian result.
We denote the long time of a general variable $x\b t$
\begin{equation}
    x^{\b{\infty}}\equiv \text{lim}_{t\ra\infty} x\b t~~.
\end{equation}
In the long time regime the sinc functions can be approximated by delta functions, and the kinetic coefficients converge to (Eq. \eqref{eq:H7})  (see for example Appendix D of Ref. \cite{winczewski2021bypassing})
\begin{multline}
      \gamma_- ^{M\b{\infty}}=e^{\hbar\omega_0/k_B T}\gamma_{+}^{M\b{\infty}}\\=\text{lim}_{t\ra\infty} \gamma_-^M\b t/t\\= 2\pi J\b{\omega_0}\b{\bar{n}_T\b{\omega_0}+1} ~~,
\end{multline}
where $k_B$ is the Boltzmann constant, and the Lamb-shift term is given by 
\begin{equation}
    \Upsilon^{M\b{\infty}} =\f{1}{2}{\cal P}\sb{\int _0^{\infty}{d\omega_{k}J\b{\omega_{k}}\b{\f{\bar n_{T}\b{\omega_{k}+1}}{\omega_0-\omega_{k}}+\f{\bar n_{T}\b{\omega_{k}}}{\omega_0+\omega_{k}}}}}~~.
\end{equation}
This implies  that the second order cumulant becomes ${\cal Z}\b t=\tilde{\cal L}^{M} t$ ($M$ signifies the Markovian limit), where 
\begin{multline}
\tilde{\cal L}^{M}\sb{\bullet}= - \f{i}{\hbar} \sb{\Upsilon^{M\b{\infty}} ,\bullet}\\+ \gamma_+^{M\b{\infty}}{\cal{D}}_+\sb{\bullet} +\gamma_-^{M\b{\infty}}{\cal{D}}_-\sb{\bullet}~~,
\end{multline}
and the dynamical generator (Eq. \eqref{eq:72}) converges to $\tilde{\cal L}^{\b N}\sb{\hat{\rho}_S\b t}={\tilde{\cal{L}}^{M}\sb{\hat{\rho}_S\b t}}$.

\subsection{Derivation of the fourth order cumulant}
\label{asec:4Order}

\trr{To obtain the fourth order cumulant one expands the dynamical map up to fourth order in interaction strength, obtaining
\begin{align*}
\tilde\rho_S(t)\approx \tilde\rho_S(0) - \int\limits_0^t \hspace{-0.1cm}\int\limits_0^{t_1}\hspace{-0.1cm} d\tilde{t} \text{tr}_E([\tilde H_I(t_1),[\tilde H_I(t_2),\tilde\rho_S(0) \otimes \rho_\beta]])+ \\
\int\limits_0^t \hspace{-0.1cm} \int\limits_0^{t_1}\hspace{-0.1cm}\int\limits_0^{t_2} \hspace{-0.1cm} \int\limits_0^{t_3}\hspace{-0.1cm} d\tilde{t} \text{tr}_E([\tilde H_I(t_1),[\tilde H_I(t_2)[\tilde H_I(t_3),[\tilde H_I(t_4),\tilde\rho_S(0) \otimes \rho_\beta]]])\\:=
\left(1 - \int\limits_0^t \hspace{-0.1cm}\int\limits_0^{t_1}\hspace{-0.1cm} d\tilde{t} c_2(t_1,t_2)+
\int\limits_0^t \hspace{-0.1cm} \int\limits_0^{t_1}\hspace{-0.1cm}\int\limits_0^{t_2} \hspace{-0.1cm} \int\limits_0^{t_3}\hspace{-0.1cm} d\tilde{t} c_4(t_1,t_2,t_3,t_4)\right)\tilde\rho_S(0),
\end{align*}}
\trr{where we have used the fact that the initial state of the environment is assumed to be thermal state and, consequently, only expectation values consisting of the same numbers of creation/annihilation operators are non-zero. Additionally, we have introduced the notation $d\tilde t = dt_1 dt_2...$ and
\begin{gather}
  c_2\b{t_1,t_2} \boldsymbol{\cdot} = \text{tr}_E([\tilde H_I(t_1),[\tilde H_I(t_2),\boldsymbol{\cdot} \otimes \rho_\beta]])\\
    c_4\b{t_1,t_2,t_3,t_4}\boldsymbol{\cdot}  =\nonumber\\ \text{tr}_E([\tilde H_I(t_1),[\tilde H_I(t_2)[\tilde H_I(t_3),[\tilde H_I(t_4),\boldsymbol{\cdot}\otimes \rho_\beta]]]]).
    \nonumber
\end{gather}}


\trr{On the other hand, we write the reduced density operator at time $t$ in terms of cumulants
\begin{align*}
\tilde\rho_S(t)= e^{{\cal Z} (t)}  \tilde\rho_S(0) \approx e^{\f{1}{2}{\cal Z}_{2} (t)+ \f{1}{24} {\cal Z}_4 (t)}  \tilde\rho_S(0).
\end{align*}
Comparing the two above equations, we identify the relation
\begin{align*}
{\cal Z}_2=& -2\int\limits_0^t \hspace{-0.1cm}\int\limits_0^{t_1}\hspace{-0.1cm} d\tilde{t} c_2(t_1,t_2),\\
{\cal Z}_4=& 12\int\limits_0^t \hspace{-0.1cm} \int\limits_0^{t_1}\hspace{-0.1cm}\int\limits_0^{t_2} \hspace{-0.1cm} \int\limits_0^{t_3}\hspace{-0.1cm} d\tilde{t}(2 c_4(t_1,t_2,t_3,t_4) - C_{22}(t_1,t_2,t_3,t_4))~~,
\end{align*}
where a change of limits in the time integrals in the second term leads to 
\begin{align*}
C_{22}(t_1,t_2,t_3,t_4)&=c_2(t_3,t_4)c_2(t_1,t_2) 
+c_2(t_2,t_4)c_2(t_1,t_3)\\
&+c_2(t_2,t_3)c_2(t_1,t_4)  +  c_2(t_1,t_4)c_2(t_2,t_3)\\ 
&+ c_2(t_1,t_3)c_2(t_2,t_4) + c_2(t_1,t_2)c_2(t_3,t_4).
\end{align*}
\\
For ease of notation we introduce functions $f(t)$ and $g(t)$ 
\begin{align*}
   g(t):=\sum_k |g_k|^2 n_k e^{-i(\omega_0-\omega_k) t}, \\  
f(t):= \sum_k |g_k|^2 (n_k+1) e^{i(\omega_0-\omega_k) t}.
\end{align*}}
\trr{Employing the notation $f_{12}^R=\Re(f(t_1-t_2))$, etc., after a lengthy calculation we obtain
\begin{align*}
2c_4(t_1,t_2,t_3,t_4)-C_{22}(t_1,t_2,t_3,t_4)=-2\times\nonumber \\ 
{\cal{D}}_z(2g^R_{14}f^R_{23}+2f^R_{14}g^R_{23}+g^R_{13}f^R_{24}+f^R_{13}g^R_{24}+g^I_{13}f^I_{24}+f^I_{13}g^I_{24} )\\
+2{\cal{D}}_- (2f^R_{12}g^R_{34}+2f^R_{13}(f^R_{24}+g^R_{24})+2f^R_{23}(f^R_{14}+g^R_{14})\\-2f^R_{34}g^R_{12}-2f^I_{14}(f^I_{23}-g^I_{23})-2f^I_{24}(2f^I_{13}-g^I_{13})) \nonumber \\ 
+2{\cal{D}}_+(2g^R_{12}f^R_{34}+2g^R_{13}(g^R_{24}+f^R_{24})+2g^R_{23}(g^R_{14}+f^R_{14})\\-2g^R_{34}f^R_{12}-2g^I_{14}(g^I_{23}-f^I_{23})-2g^I_{24}(2g^I_{13}-f^I_{13}))\nonumber\\
-\f{i}{\hbar}{\cal{D}}_z'\Im(g_{14}(2f_{23}+f_{23}^*-g_{23})-f_{14}(2g_{23}+g_{23}^* - f_{23})\\-g_{13}(g_{24}-f_{24}^*)+f_{13}(f_{24}-g_{24}^*))~~,
\end{align*}}
\trr{which determine the higher order
corrections to the kinetic coefficients $\bar\Gamma_{+}^{\b N}\b{t}$, $\bar\Gamma_{-}^{\b N}\b{t}$, $\bar\Upsilon^{\b N} \b t$, and lead to the appearance of a pure dephasing term, which strength is determined by the kinetic coefficient $\bar\Gamma_{z}^{\b N}\b{t}$. The connections between these kinetic coefficients and the one occurring in the generator, Eq.\eqref{eq:rho4Order}, are the mathematically identical to the second order case and are given by Eq. \eqref{eqap:2225} and Eq. \eqref{eq:Lamb}. Additionally, due to commutation of ${\cal{D}}_z$ with ${\cal{D}}_-$, ${\cal{D}}_+$ and  ${\cal{D}}_z'$ one gets
\begin{align*}
    \bar\gamma_z^{\b N}\b{t}=\dot{\bar\Gamma}_{z}^{\b N}\b{t}~~.
\end{align*}}



\end{document}